# ConQuer-92

Revised report on the conceptual query language LISA-D

*Confidential*

Asymetrix Report 94-5

H.A. Proper
Asymetrix Research Laboratory
Department of Computer Science
University of Queensland
Australia 4072
E.Proper@acm.org

**Abstract**

In this report the conceptual query language ConQuer-92 is introduced. This query language serves as the backbone of InfoAssistant's query facilities. Furthermore, this language can also be used for the specification of derivation rules (e.g. subtype defining rules) and textual constraints in InfoModeler.

This report is solely concerned with a formal definition, and the explanation thereof, of ConQuer-92. The implementation of ConQuer-92 in SQL-92 will be treated in a separate report.

## 1 Introduction

In this report we introduce the conceptual query language which will serve as the backbone of InfoAssistant's query facilities, and which can also be used in InfoModeler for the specification of derivation rules and constraints. The definition of this language is a restriction, and a slight extension at the same time, of the existing language LISA-D ([HPW93], [HPW97], [HPW94]).

In ConQuer, a central role is played by the so-called *path-expressions*. In its most elementary form, a path-expression describes a path through the conceptual schema.

The most important extension with respect to LISA-D is the ability to deal (and name) intermediate results of path-expressions. In operational terms this means that we can select *any* intermediate result of a path-expression to become part of the final query result. Later we will see some convincing examples of this.

Furthermore, a few minor changes in the definitions have been made to bridge the gap between LISA-D and FORML. LISA-D has been restricted in the sense that certain restrictions had to be made to ensure that the language can be implemented on top of SQL-92. The proposed name of the new language is therefore ConQuer-92, which is an acronym for *Con*ceptual *Quer*ies. For IA it has been agreed to simply refer to the language as ConQuer, but for internal purposes it still makes sense to make the difference between ConQuer-92. The resulting language can later be extended further when SQL-3 can be used as a target platform, leading to ConQuer-3. Using SQL-3 as the target platform will in particular allow us to define recursive queries. The definition of ConQuer-92 as provided in the report also includes many of the aspects that are present in the FORML ([Hal95]) language for the actual definition of derivation rules, subtype defining rules, and textual constraints. In the LISA-D articles no special syntax for this purpose was introduced. This report is solely concerned with a formal definition of ConQuer-92. The implementation of ConQuer-92 in SQL-92 will be treated in a separate report.

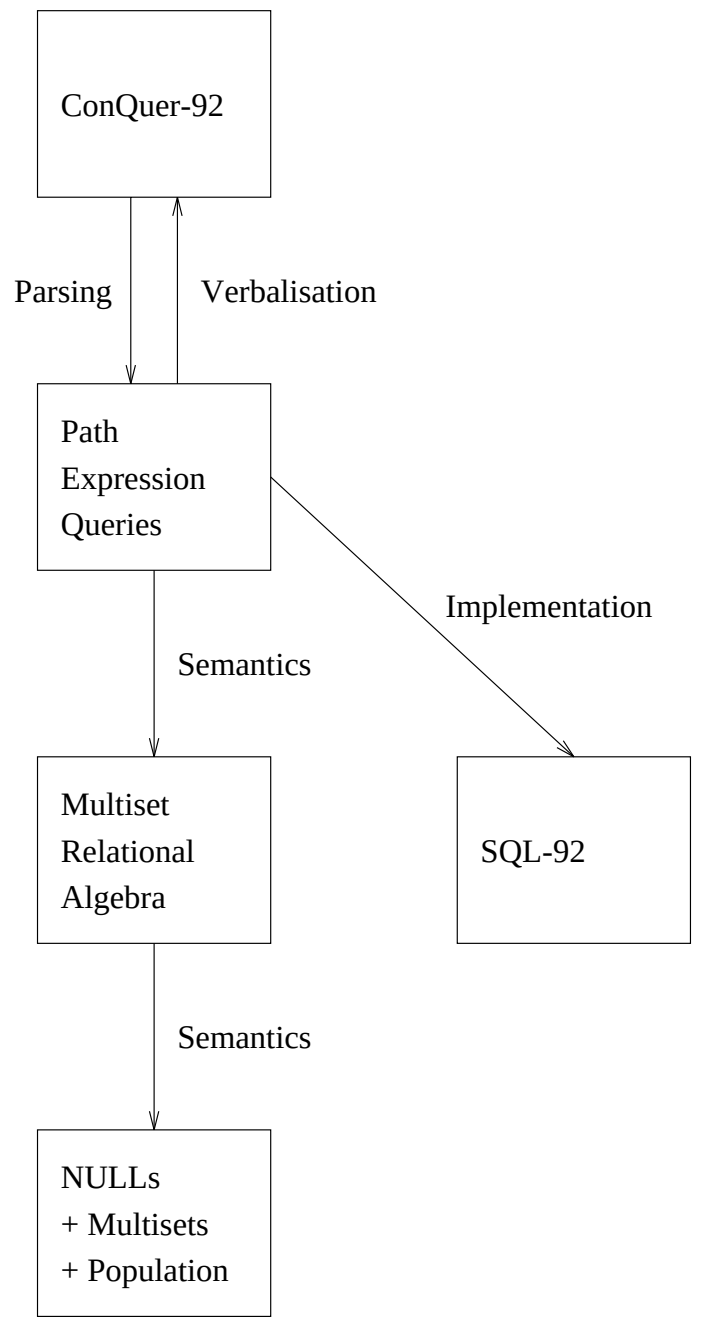

Figure 1: Setup of the ConQuer-92 definition

ConQuer-92 is introduced using a layered approach. Each layer represents one level of abstraction. In figure 1 we have illustrated the multi-layered definition of ConQuer-92. The bottom layer of the specification is formed by NULLs, multisets and populations of ORM schemas. The difference between a set and a multiset is that elements in a set can occur only once, whereas a multiset allows for multiple occurrences of the same element.

The next level of abstraction introduces an extended form of relational algebra which features some extra operations, and is based on multisets (or bags) rather than traditional sets. The relational algebra can be regarded as forming the formal foundation of SQL. A key difference between traditional relational algebra and SQL is, however, that SQL is multiset based whereas relational algebra is set based. This effectively means that the result of an SQL query may contain multiple occurrences of the same value. This deviation between RA and SQL is based on efficiency. Removing multiple occurrences in queries is expensive from an implementational point of view, as it involves ordering the entire result, and searching for multiple occurrences of the same values.

On the next level up are the path-expression queries. This is an abstract language which will not be seen by a user, but is used for the internal storage of queries, constraints, and derivation rules, in the fact base. The largest part of this language is formed by the path-expressions. Therefore, we will simply refer to this level as the *path-expression level* in the remainder.

Finally, the highest level of abstraction is formed by ConQuer-92 itself. The difference between the ConQuer level and the path-expression level is nill with respect to the expressiveness of the languages. However, the ConQuer level is the language that is used to communicate path-expression queries to users, and vice versa. While the path-expression level is not intended for ’human consumption’, the ConQuer level is only intended for ’human consumption’.

In a next report, a mapping from path-expressions to SQL-92 will be provided. The definition of the semantics of path-expressions in terms of relational algebra can thus be regarded as a specification for this mapping to SQL. With this mapping, the relation between ConQuer-92, path-expression queries, and SQL-92 can be illustrated using the diagram depicted in figure 2. The path-expression query is the query that is actually manipulated (and stored) by the IA tool. ConQuer can be seen as an external representation of a path-expression. When we change the verbalisation of object types, predicates, etc. in the conceptual schema, then the path-expression does not change, but the verbalisation of the path-expression may change. Conversely, when we change the mapping from a conceptual schema to a logical schema, the SQL statement generated from a given path-expression will have to change as well.

The existing query formulation techniques: Query by Navigation (QBN), Query by Construction (QBC), Query by Outline (QBO), Spider Queries (SQ), Point to Point Queries (PPQ), and natural-languages based query formulation using NUQL, all use path-expression queries as a common format for storing (partial) queries. This is illustrated in figure 3.

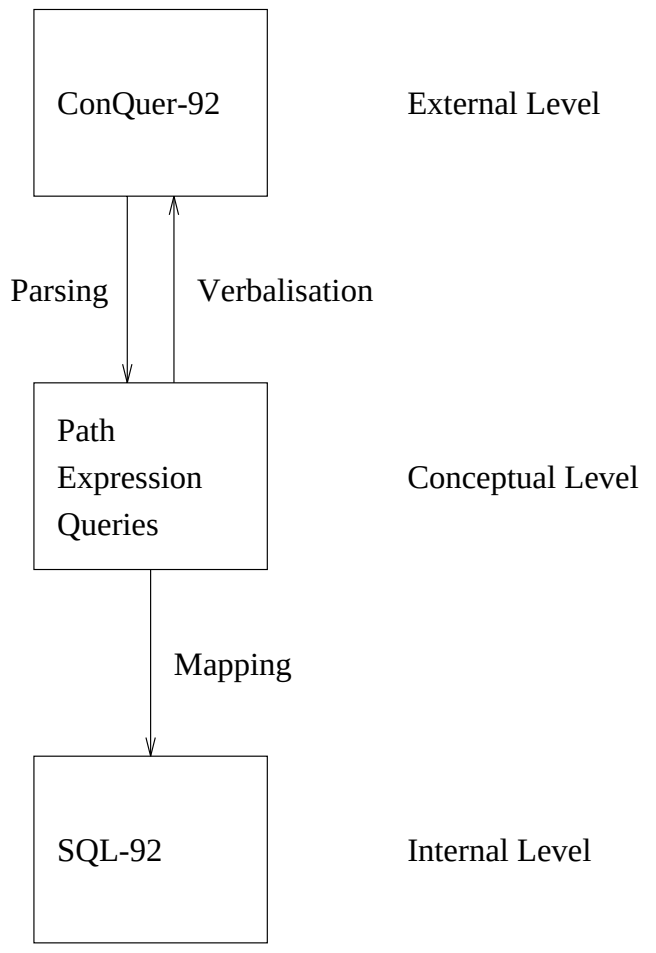

Figure 2: The relation between SQL, path-expressions queries, and ConQuer

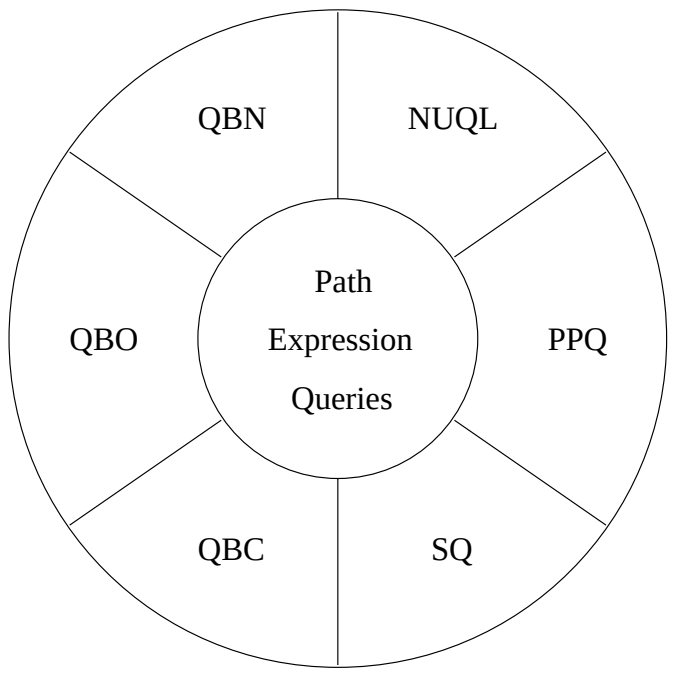

Figure 3: Interplay between query formulation tools

The structure of the report is as follows. In section 2 we provide a brief formal definition of an ORM schema. In section 3 we briefly discuss some of the problems concerning the treatment of NULL values. A brief overview of the definition of multisets is provided in section 4. To make the mapping to SQL-92 easier to define, the semantics of the path-expressions have been defined in an extended version of the relational algebra. This relational algebra is introduced in section 5. The definition of path-expression queries is provided in section 6. Finally, that part of ConQuer-92 that will be visible for ordinary users is discussed in section 7 and section 8. The BNF grammars of both the path expression level and the ConQuer level are discussed in separate appendices.

# 2 ORM Models

In this section we provide a brief discussion of the formalisation of ORM we use in this report. We have setup the formalisation in such a way that it does not rely to much on the details of the ORM meta-model which is in use for InfoModeler. So any changes, in particular the addition of new modelling concepts, will not lead to dramatic changes in the formalisation presented here.

A conceptual schema is presumed to consist of a set of types $\mathcal{TP}$. This set of types is split into two pairs of two subsets; based on two dichotomies. Firstly, a distinction is made based on the underlying structure of types. This results in the set of relationship types $\mathcal{RL}$, and the set of ordinary object types $\mathcal{OB}$. Note that this is not a disjunctive dichotomy as nested relationship types are in both classes.

The second dichotomy is based on the denotability of instances of types. Types which are directly denotable are referred to as value types $\mathcal{VL}$, and correspond to the types whose instances have a direct denotation such as strings, numbers, etc. The types which cannot be directly denoted are the non-value types. This double dichotomy is discussed in more detail in [HP95].

Relationship types are build from roles. Let $\mathcal{RO}$ be the set of all such roles in the conceptual schema. The fabric of the conceptual schema is then captured by two functions and two predicates. The set of roles associated to a relationship type are provided by the partition: $\mathsf{Roles} : \mathcal{RL} \to \wp(\mathcal{RO})$. Using this partition, we can define the function $\mathsf{Rel}$ which returns for each role the relationship type in which it is involved: $\mathsf{Rel}(r) = f \iff r \in \mathsf{Roles}(f)$. Every role has an object type at its base called the player of the role, which is provided by the function: $\mathsf{Player} : \mathcal{RO} \to \mathcal{OB}$. As an example, consider the schema depicted in figure 4. In this schema we have marked each role of the relationship type $f$ with a letter $(p, q, r)$. In this schema we have: $\mathsf{Roles}(f) = \{p, q, r\}$ and $\mathsf{Player}(p) = A$, $\mathsf{Player}(q) = B$, $\mathsf{Player}(r) = C$.

We presume the existence of a relation $\sim \subseteq \mathcal{OB} \times \mathcal{OB}$ providing us with the types that are type related, i.e. types that may share instances. Typical examples of type related object types are subtypes. In the currently used version of ORM for InfoModeler, two types are type related if they are in the same subtype hierarchy, and it does not follow from a disjunction constraint that they are disjunctive. For a definition of this relationship in more basic concepts of an ORM schema, please refer to [HPW93] or [HP95].

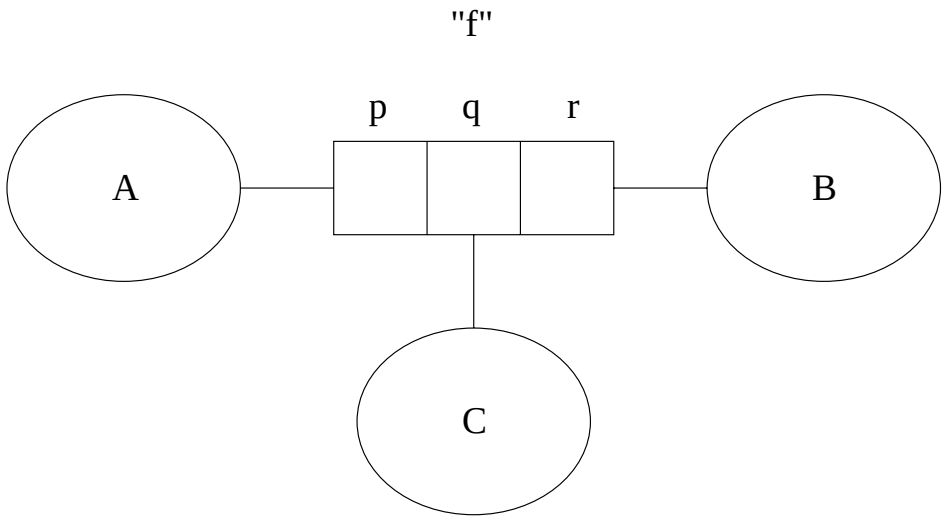

Figure 4: An example conceptual schema

In ORM some object types may be specialised or be polymorphic (in the current version of InfoModeler only specialisation applies). Sometimes we need to access the so-called *root types* of a given object type. A root type of an object type $x$ is a type $y$ which itself is not a specialisation or polymorphic type, such that $x$ is a specialisation or polymorphism of $y$. The roots of a type are formally provided by: $\mathsf{RootsOf} : \mathcal{TP} \to \wp(\mathcal{TP})$. Due to possible multi-rooted specialisation hierarchies and the use of polymorphism, this function yields a set of types. In the current version of ORM supported by InfoModeler, all subtyping hierarchies have only one root. Therefore this function will always lead to singleton sets (sets with one element only) for ORM models developed in InfoModeler. Again, for a more detailed definition of such a function in terms of more basic concepts of an ORM schema, please refer to [HPW93] or [HP95].

Finally, since all instances of the population of a type must be identifiable in terms of a combination of values, we presume that each non value type has associated an identification scheme (or reference scheme). This identification schema either consists of a sequence of either role pairs or single roles (relationship types). It is provided by: $\mathsf{Idf} : (\mathcal{TP} - \mathcal{VL}) \to (\mathcal{RO} \times \mathcal{RO})^+ \cup \mathcal{RO}^+$. Note that when dealing with disjunctive identification schemes this becomes: $\mathsf{Idf} : (\mathcal{TP} - \mathcal{VL}) \to ((\mathcal{RO} \times \mathcal{RO})^+ \cup \mathcal{RO}^+)^+$. This is, however, beyond the currently supported version of ORM. These are all the assumptions we have to make on the underlying ORM data model.

# 3 NULL Values

In the new definition of the path-expression level, it has become essential to be able to deal with NULL values, since we now want to deal with intermediate results as well. In this report we do not explicitly concern ourselves with a proper definition of the semantics of NULL values as there are various ways of dealing with NULL values. The basic idea of this section is to identify this as a couse of possible problems. In practice we will stick as much as possible to the standards dictated by SQL-92. In the remainder of this report we will, wherever appropriate, return to the NULL value issue.

The central issues are the behaviour of NULLs in arithmetic operations, logic expressions, and equality. As an illustration of the choices involved, we provide some examples of different choices:

$$
\begin{array}{rcl}
|A \cup \{\mathrm{NULL}\}| & = & |A| + 1 \text{ if NULL} \not\in A \\
|A \cup \{\mathrm{NULL}\}| & = & |A| + 0 \\
|A \cup \{\mathrm{NULL}\}| & = & \mathrm{NULL} \\
\mathrm{NULL} \in \{\mathrm{NULL}\} & \Longleftrightarrow & \mathit{true} \\
\mathrm{NULL} \in \{\mathrm{NULL}\} & \Longleftrightarrow & \mathit{false} \\
\mathrm{NULL} \in \{\mathrm{NULL}\} & \Longleftrightarrow & \mathit{unknown} \\
a + \mathrm{NULL} & = & a \\
a + \mathrm{NULL} & = & \mathrm{NULL} \\
\mathrm{NULL} = \mathrm{NULL} & \Longleftrightarrow & \mathit{true} \\
\mathrm{NULL} = \mathrm{NULL} & \Longleftrightarrow & \mathit{false} \\
\mathrm{NULL} = \mathrm{NULL} & \Longleftrightarrow & \text{unknown}
\end{array}
$$

# 4 Multisets

In this section, which is completely based on a section in [HPW93], the concept of *multiset* is introduced formally. *Multisets* ([Lew85]), also known as *multiple membership sets* ([Lev79]), or *bags* ([Par90]), differ from ordinary sets in that a multiset may contain an element more than once.

As an example, to illustrate the difference between sets and multisets, consider the set $\{a, b, c, c\}$. This is a set with only *three elements*. These elements are: $a, b, c$. Although the element $c$ was written *twice* within the set enumeration: $\{a, b, c, c\}$, it only occurs *once* in the set. So for sets we have: $\{a, b, c, c\} = \{a, c, b\} = \{a, b, a, c\}$, etc. In a multiset, elements can occur multiple times. However, in a multiset, just like sets, elements do not have an order. We will denote a multiset in the following format: $\{\!\![a, b, c, c]\!\!\}$. In this multiset, the elements $a$ and $b$ occur once, and $c$ occurs twice. For multisets we have: $\{\!\![a, b, c, c]\!\!\} \neq \{\!\![a, b, c]\!\!\}$, but we do have: $\{\!\![a, b, c, c]\!\!\} = \{\!\![a, c, b, c]\!\!\}$, etc. The relation between lists (or sequences), bags, and sets can be summarised as follows:

| | List | Bag | Set |
|---|---|---|---|
| Order | Yes | No | No |
| Frequency | Yes | Yes | No |

Multisets over an underlying domain $X$ are elegantly introduced as functions: $X \to \mathbb{N}$, assigning to each $x \in X$ its frequency. So if we use as underlying domain all letters in the alphabet, an example of such a function is: $f = \{a : 2, b : 3, z : 1\}$ (ignoring all 0 occurring elements). The $f$ is now a function where $f(a) = 2$, $f(b) = 3$, and $f(z) = 1$. This $f$ corresponds to the multiset: $\{\!\![a, a, b, b, b, z]\!\!\}$.

In the definitions of the operations on multisets, the $\lambda$-calculus notation provided by [Bar84], will be employed. This notation is nothing more than a mathematical equivalence of a function declaration. For instance $\lambda x.x^2$ is the polynomial function assigning $x^2$ to each $x$-value. The statement $Sqr = \lambda x.x^2$ corresponds to the following definition in Modula-2:

```
FUNCTION Sqr (x :REAL) :REAL
  BEGIN
    RETURN(x * x);
  END;
```

The lambda calculus allows us to reason about anonymous functions. For example $\lambda x.x^2$ is the function yielding the square of the parameter to the function. This would correspond to:

```
FUNCTION (x :REAL) :REAL
  BEGIN
    RETURN(x * x);
  END;
```

This is an anonymous function. Some more advanced programming languages, like Algol-68 allow for anonymous functions.

In the theory of multisets, like in set theory, $\varnothing$ denotes the empty multiset. The definition of the empty multiset is: $\lambda x.0$. As a last example of what this looks like in Modula-2:

```
FUNCTION (x :REAL) :REAL
  BEGIN
    RETURN(0);
  END;
```

Sets can be defined in a variety of ways. One way which we shall use very often is the set comprehension schema. For example $\{x \in \mathbb{N} \mid x > 10\}$ denotes the set of natural numbers $x \in \mathbb{N}$ such that they are larger than 10 ($x > 10$). This could be illustrated using the following piece of pseudo code:

```
FOR EACH x ∈ IN DO
  IF x > 10 THEN
    Add x to the result;
  END IF;
END FOR;
```

Quite often we will omit the $\in \mathbb{N}$ part if from the condition part it is clear what the domain of $x$ is. If $<$ would only be defined for natural numbers, then we would write: $\{x \mid x > 10\}$. Another alternative is to lift the $\in \mathbb{N}$ part over the middle bar '—' (which is pronounced as 'such that'), leading to: $\{x \mid x \in \mathbb{N} \wedge x > 10\}$. This latter format is actually the preferred format for the comprehension schema for multisets.

The comprehension schema for a multiset has the format: $\{\!\![e{\uparrow}^n \mid C(e, n)]\!\!\}$. The $C$ is some kind of condition where $e$ is the element and $n$ is the frequency in which the element should occur in the resulting multiset. So $C(e, n)$ should be a binary predicate such that $C(e, n_1) \wedge C(e, n_2) \Rightarrow n_1 = n_2$. The notation $e{\uparrow}^n$ is used to indicate that $e$ should indeed occur $n$-times in the resulting multiset. Sometimes we will allow us to write $e{\uparrow}^{n+m}$, or more general $e{\uparrow}^{f(n_1,\ldots,n_l)}$ as long as $n + m$ and $f(n_1, \ldots, n_l)$ return positive natural numbers. For example $\{\!\![x{\uparrow}^3 \mid x \in \{1, 2, 3\}]\!\!\}$ leads to a multiset: $\{\!\![1, 1, 1, 2, 2, 2, 3, 3, 3]\!\!\}$.

With respect to NULL values, special care has to be taken when $C(e, n)$ evaluates to *unknown*. Without NULL values, $C(e, n)$ will evaluate to *true* or *false*, but allowing for NULL values means that it might lead to *unknown* as well. In the case of SQL the *unknown* is treated as 'insufficient proof', so the element $e$ will *not* occur in the result set.

The traditional union, intersection, and difference operations from set theory are defined for multisets as follows:

$$
\begin{aligned}
N \cup M &\triangleq \lambda x. N(x) + M(x) \\
N \cap M &\triangleq \lambda x. \min(N(x), M(x)) \\
N - M &\triangleq \lambda x. \max(N(x) - M(x), 0)
\end{aligned}
$$

Some examples are:

$$
\begin{aligned}
&\{\!\![a, b, a]\!\!\} \cup \{\!\![b, c, c]\!\!\} = \{\!\![a, a, b, b, c, c]\!\!\} \\
&\{\!\![a, b, a]\!\!\} \cap \{\!\![b, b, a, c]\!\!\} = \{\!\![a, b]\!\!\} \\
&\{\!\![a, b, b, a, c]\!\!\} - \{\!\![a, b, b, b \mid =]\!\!\} \{\!\![a, c]\!\!\}
\end{aligned}
$$

Bag comprehension can be used for intentional denotations of multisets. Extentional denotations are defined by: $\{\!\![a]\!\!\} \triangleq \{\!\![a{\uparrow}^q \mid q = 1]\!\!\}$ and $\{\!\![a_1, \ldots, a_n]\!\!\} \triangleq \{\!\![a_1]\!\!\} \cup \ldots \cup \{\!\![a_n]\!\!\}$.

In the case of sets we write $e \in E$ to say '$e$ is an element of the set $E$'. For multisets we write $e \in^n M$, with the formal meaning $M(e) = n$. Sometimes, when we are not interested in the frequency, we will write $e \in M$ for $M(e) > 0$.

The comparison operator $N \subseteq M$ for multisets is defined as: $\forall_x [N(x) \leq M(x)]$. For example $\{\!\{a, b, c, c\}\!\} \subseteq \{\!\{a, b, b, c, c, c\}\!\}$, but not $\{\!\{a, b, c, c\}\!\} \subseteq \{\!\{a, b, c\}\!\}$. From this operator, the $\subset$ comparison is derived in the usual way: $N \subseteq M \wedge N \neq M$. This allows for the definition of the powerset of a multiset: $\mathbb{M}(X) = \{\!\{ Y\uparrow^1 \mid Y \subseteq X \}\!\}$. For example, the powerset of $\{\!\{a, b, b\}\!\}$ is:

$$\{\!\{\varnothing, \{\!\{a\}\!\}, \{\!\{b\}\!\}, \{\!\{a, b\}\!\}, \{\!\{b, b\}\!\}, \{\!\{a, b, b\}\!\}\}\!\}$$

Coercions from multiset to set and vice versa are defined by the following functions:

$$\begin{aligned} \mathsf{Set}(N) &\triangleq \{ x \mid x \in N \} \\ \mathsf{Multi}(S) &\triangleq \{\!\{ x\uparrow^1 \mid x \in S \}\!\} \end{aligned}$$

For example: $\mathsf{Set}(\{\!\{a, a, b, b, b\}\!\}) = \{a, b\}$, and $\mathsf{Multi}(\{a, b\}) = \{\!\{a, b\}\!\}$. The number of elements in a multiset is counted by $|N| \triangleq \sum_{x \in X} N(x)$, so $\left|\{\!\{a, b, c, c, a\}\!\}\right| = 5$. With respect to NULL values, the exact result of this expression depends on the way the $\sum$ (the $+$) operation handles NULLs.

By making assumptions on the underlying domains $X$ we can introduce some more interesting operations. If $N$ is a multiset over an arithmetic domain $X$, then the following operations can be defined:

$$\begin{aligned} \max(N) &\triangleq \max(\mathsf{Set}(N)) \\ \min(N) &\triangleq \min(\mathsf{Set}(N)) \\ \mathrm{sum}(N) &\triangleq \textstyle\sum_{x \in X} x \times N(x) \end{aligned}$$

These operations return the maximum, minimum and sum of the values in the multiset. Some examples are: $\max(\{\!\{1, 3, 9, 9, 1\}\!\}) = 9$, $\min(\{\!\{1, 3, 9, 9, 1\}\!\}) = 1$, and $\mathrm{sum}(\{\!\{1, 3, 9, 9, 1\}\!\}) = 23$. Again the exact result of $\mathrm{sum}(N)$ in the case of NULL values depends on the treatment of NULLs in arithmetic operations.

# 5 Relational Algebra

As stated before, the semantics of path-expressions is defined in terms of relational algebra expressions. However, since path-expressions are based on multisets, the relational algebra we use here needs to be defined in such a way that it can deal with multisets (and NULL values). The style of relational algebra we use here is one which is widely used in academic papers. We could have chosen a style which would make some definitions simpler, but the used algebra is certainly closer to what is needed for a mapping to SQL than any of the other relational algebra styles.

## 5.1 Preliminaries

A relational algebra expression effectively defines a derived relationship type. For such expressions two functions play a central role. The first function (Sch) is concerned with the set of columns (attribute names) in this derived relationship. This function provides the underlying structure (c.f. a sequence of roles in a normal relationship type) of the derived relationship type resulting from the relational algebra expression. The second function (Val) with the actual population of the derived relationship type. The formal signatures of these functions are:

$$\mathsf{Sch} : \mathcal{RA} \rightarrow \wp(\mathsf{Attrs})$$
$$\mathsf{Val} : \mathcal{RA} \times \mathsf{POP} \times (\mathsf{Attrs} \rightarrowtail \Omega) \rightarrow \wp(\mathsf{Attrs} \rightarrow \Omega)$$

In these definitions, $\mathcal{RA}$ is the set of relational algebra expressions, Attrs the set of attributes names (column names) that can be used in the relational algebra expressions, and $\Omega$ the set of all instances that are allowed in the result of a relational algebra expression. Note that we presume the existence of an explicit NULL element in $\Omega$.

The signature of the first function (Sch) tells us that this function takes a relational algebra expression as a parameter and returns a set of attribute names. These attribute names are the column names of the resulting table. From a theoretical point of view it is more elegant and more conceptual to see this as a set of attribute names rather than a sequence. If we would model this as a list of attribute names, than the relational schemas $[a,b,c]$ would not be the same as the relational schema $[b,c,a]$! In general, however, we want to regard these to be the same! Therefore we model the header of a relational algebra table as a *set* of attribute names.

The signature of the second function (Val) is a bit more complex. This function should return the population of the table that can be associated to a relational algebra expression. The relational algebra expression derives this population from the population of the underlying conceptual schema. Therefore it needs a relational algebra expression and the current database population (an element from POP) as input parameters. However, a third parameter is required. We allow for the use of correlated subqueries. These correlated subqueries, like in SQL, need to be evaluated in the context of one tuple from the surrounding query. For example, consider the SQL query:

```
SELECT department, name, pay
FROM employees x
WHERE pay >
  (SELECT AVG(pay)
  FROM employees
  WHERE department = x.department)
```

When evaluating the subquery, the reference to x.department refers to a tuple in the result from the surrounding query, i.c. a tuple from employees. The third parameter to the Val function allows us to pass on such a tuple to the evaluation of a subquery. In our formalisation a tuple from a relational algebra table is sees as a partial(!) function from the set of attribute names to the possible instances.

In the relational algebra expressions we use in the context of ConQuer-92, we allowed the roles from the underlying ORM schema to be used as attribute names. This is not just allowed because we are trying to be friendly, but it is highly essential as we need to be able to refer to the population of the conceptual schema! When generating an SQL statement, these roles should be replaced by proper attribute names that are derived from the names of the roles and object types. These names will indeed correspond to the attribute names used for in the logical representation of the conceptual schema. For the relational algebra level we effectively have: $\mathcal{RO} \subseteq \mathsf{Attrs}$ (but it does *not* mean that all attributes are roles).

## 5.2 Denotational semantics

The semantics of relational algebra (and later the semantics of path-expressions and ConQuer) are defined conform the style of denotational semantics (see e.g. [Sto77]). In denotational semantics, the semantics of each syntactical construct (of the language which's semantics are being defined) is defined in terms of other syntactical constructs, and ultimately in terms of isome underlying semantical domain. In our case the underlying semantical domain are the multisets and the populations of conceptual schemas. An important role in denotational semantics is played by the environment, which can be seen as representing the state of a program. In our case the environment of the first semantics function is empty, but the environment of the second semantics function consists of two parts. In the latter case, a relational algebra expression is evaluated in the context (environment) of a conceptual database population and possibly a tuple from a surrounding query.

In the style of denotational semantics, the construct which's semantics are defined is put between double brackets: $[\![P]\!]$ and the environment between the traditional parenthesis: $(p,t)$. Conform this style of defining semantics we write $\mathsf{Sch}[\![P]\!]$ and $\mathsf{Val}[\![P]\!]\,(p,t)$, where $P$ is the algebra expression which's semantics are to be defined, $p$ is a population, and $t$ is a tuple.

The actual semantics of a relational algebra expression $P$, which is not a nested subquery, can now be expression as: $\mathsf{Val}[\![P]\!]\,(p,\varnothing)$, where $p$ is the conceptual database population.

## 5.3 Projection

The first relational algebra operation we introduce is projection. In a projection the columns of a relation can be (re)defined by means of expressions in terms of other attributes. Therefore, a projection must provide a function $R$ : $\mathsf{Attrs} \rightarrowtail \mathtt{RAScalarExpr}$ assigning expressions to attributes. RAScalarExpr is a set of arithmetic-expressions of which the semantics is supposed to be provided by a function:

$$\mathsf{Expr} : \mathtt{RAScalarExpr} \times \mathsf{POP} \times (\mathsf{Attrs} \rightarrowtail \Omega) \to \Omega$$

This function takes as input parameters an expression, the current conceptual database population, and a tuple from any surrounding query. Later on in this section we define this function in more detail.

The semantics of the projection operator can now be defined formally as:

$$\begin{aligned} \mathsf{Sch}[\![\pi_R\, P]\!] &\triangleq \mathsf{dom}(R) \\ \mathsf{Val}[\![\pi_R\, P]\!]\,(p,t) &\triangleq \bigcup_{u \in^n \mathsf{Val}[\![P]\!]\,(p,t)} \{\![(\lambda_{a \in \mathsf{dom}(R)} \cdot \mathsf{Expr}[\![R(a)]\!]\,(p, t \ltimes u)) \uparrow^n]\!\} \end{aligned}$$

In this definition we can see the tuple $t$ for subqueries in actual use. The tuple passing mechanism is needed since the expressions in $R$ could quite well contain entire subqueries! The expression $t \ltimes u$ is the updated tuple $t$ in the context of relational algebra expression $P$. The original tuple $t$ was passed on by the surrounding query (if any) of $\pi_R\, P$. Since $P$ may contain attribute names that are the same as in the surrounding query, these local names must overwrite the global names (simple scoping rules). Therefore we pass the tuple $t \ltimes u$ parameter on to the evaluation of the expression. The formal definition of the overwrite ($\ltimes$) function is:

$$f \ltimes g \triangleq \{\langle x, y\rangle \in f \mid x \notin \mathsf{dom}(g)\} \cup g$$

Before we continue, we will give a part for part breakdown of the formal definition of the projection. For each tuple $u$ in the result of $P$, we need to evaluate the expressions contained in $R$. Therefore we unite these evaluations by

$$\bigcup_{u \in^n \mathsf{Val}[\![P]\!]\,(p,t)}$$

The actual evaluation of the expressions in $R$ leads to a single tuple, where the values of the tuples are each evaluated in the context of the population $p$ and tuple $t \ltimes u$:

$$\{\![(\lambda_{a \in \mathsf{dom}(R)} \cdot \mathsf{Expr}[\![R(a)]\!]\,(p, t \ltimes u)) \uparrow^n]\!\}$$

As a simple example consider the projection: $\pi_{a:=x+y;b:=z}\, P$ where we presume that $P$ is the following table (this table will be used as a running example in the remainder of this section):

| x | y | z |
|---|---|---|
| 1 | 2 | 'a' |
| 2 | 4 | 'b' |

This projection would lead to the table:

| a | b |
|---|---|
| 3 | 'a' |
| 6 | 'b' |

Based on the projection operator three derived operators can be defined. An existing relation can be extended with additional columns by means of the following operation:

$$\alpha_R P \triangleq \pi_{X \ltimes R} P$$

where $R : \mathsf{Attrs} \rightarrowtail \mathtt{RAScalarExpr}$ provides the definition of the additional columns, the existing columns of $O$ are copied by $X = \mathsf{DefMap}(\mathsf{Sch}[\![P]\!])$, and $\mathsf{DefMap}(A) \triangleq \{\langle a, a\rangle \mid a \in A\}$. The DefMap function is used to generate the proper expressions needed to leave the existing columns in-tact. Existing column definitions in $X$ are overwritten by the ones provided in $T$ using the $\ltimes$ function. For example $\alpha_{a:=x+y;z:=y-x} P$ would lead to:

| x | y | z | a |
|---|---|---|---|
| 1 | 2 | 1 | 3 |
| 2 | 4 | 2 | 6 |

Renaming existing attributes can be done by the following operation:

$$\rho_R P \triangleq \pi_{X \ltimes R} P$$

where $R : \mathsf{Attrs} \rightarrowtail \mathsf{Attrs}$ provides the renaming, and $X = \mathsf{DefMap}(\mathsf{Sch}[\![P]\!]) - \mathsf{ran}(R)$. An example is $\rho_{a:x,b:y} P$ leads to:

| a | b | z |
|---|---|---|
| 1 | 2 | 'a' |
| 2 | 4 | 'b' |

Sometimes a set of existing attributes $A$ of a relation need to be removed. For this purpose, the $\delta_A$ operation can be employed. The operation is identified by:

$$\delta_A P \triangleq \pi_X P$$

where $X = \mathsf{DefMap}(\mathsf{Sch}[\![P]\!] - A)$ For instance $\delta_x P$ results in:

| y | z |
|---|---|
| 2 | 'a' |
| 4 | 'b' |

## 5.4 Expressions

We now define the syntactical category `RAScalarExpr` in more detail. Let $c$ be a constant, $a$ an attribute, $p$ a role, $P$ be an existing relational algebra expression, $f$ an arithmetic operation, and $E_1, \ldots, E_n$ be arithmetic-expressions, than the arithmetic-expressions that can be used in the projection operations are:

$$\begin{aligned}
\mathsf{Expr}[\![c]\!]\,(p,t) &\triangleq c \\
\mathsf{Expr}[\![a]\!]\,(p,t) &\triangleq t(a) \\
\mathsf{Expr}[\![a.p]\!]\,(p,t) &\triangleq t(a)(p) \\
\mathsf{Expr}[\![\mathsf{Count}(P)]\!]\,(p,t) &\triangleq |\mathsf{Val}[\![P]\!]\,(p,t)| \\
\mathsf{Expr}[\![\mathsf{Sum}(P,a)]\!]\,(p,t) &\triangleq \Sigma_{t\in^n \mathsf{Val}[\![P]\!]\,(p,t)}(t(a) \times n) \\
\mathsf{Expr}[\![\mathsf{Min}(P,a)]\!]\,(p,t) &\triangleq \min\left\{t(a) \;\middle|\; t \in \mathsf{Val}[\![P]\!]\,(p,t)\right\} \\
\mathsf{Expr}[\![\mathsf{Max}(P,a)]\!]\,(p,t) &\triangleq \max\left\{t(a) \;\middle|\; t \in \mathsf{Val}[\![P]\!]\,(p,t)\right\} \\
\mathsf{Expr}[\![f(E_1,\ldots,E_n)]\!]\,(p,t) &\triangleq f(\mathsf{Expr}[\![E_1]\!]\,(p,t),\ldots,\mathsf{Expr}[\![E_n]\!]\,(p,t))
\end{aligned}$$

Furthermore, as an abbreviation we can define: $\mathsf{Avg}(P,a) = \mathsf{Sum}(P,a)/\,\mathsf{Count}(P)$. Note that $\mathsf{Expr}[\![a]\!]\,(p,t)$ only has a meaning if tuple $t$ is defined for $a$, while $\mathsf{Expr}[\![a.p]\!]\,(p,t)$ is defined only if tuple $t$ is defined for $a$ and the result is a tuple (i.e. a relationship instance) which is defined for role $p$. As an example, $\pi_{a:=\mathsf{Sum}(P,x),b:=\mathsf{Sum}(P,x)+y}\,P$ yields:

| a | b |
|---|---|
| 3 | 5 |
| 3 | 7 |

Note that the expressions with the functions such as Min, etc, are only allowed if the underlying domains allow for this.

The above defined expressions can be directly (without using a projection operation like in the example above) coerced into a relational algebra expression by:

$$\begin{aligned}
\mathsf{Sch}[\![\epsilon_x\,E]\!] &\triangleq \{x\} \\
\mathsf{Val}[\![\epsilon_x\,E]\!]\,(p,t) &\triangleq \{\!\!\{\langle x, \mathsf{Expr}[\![E]\!]\,(p,t)\rangle\}\!\!\}
\end{aligned}$$

## 5.5 Selection

The selection operation $\sigma_C\,P$ operates like a filter. It takes the tuples from the result of $P$, and returns them when condition $C$ is satisfied. The operation can be defined formally by:

$$\begin{aligned}
\mathsf{Sch}[\![\sigma_C\,P]\!] &\triangleq \mathsf{Sch}[\![P]\!] \\
\mathsf{Val}[\![\sigma_C\,P]\!]\,(p,t) &\triangleq \{\!\!\{u{\uparrow}^n \;\middle|\; u \in^n \mathsf{Val}[\![P]\!]\,(p,t) \wedge \mathsf{Cond}[\![C]\!]\,(p,t \ltimes u)\}\!\!\}
\end{aligned}$$

In this definition we can, again, see the tuple $t$ for the subqueries in actual use. The $t \ltimes u$ is the updated tuple $t$ in the context of relational algebra expression $P$. The condition $C$ is an element of syntactic category of conditions: $\mathtt{RAConditions}$. This is a set of conditions whose semantics is defined by the function: $\mathsf{Cond} : \mathtt{RAConditions} \times \mathsf{POP} \times (\mathsf{Attrs} \rightarrowtail \Omega) \to \mathbb{B}$ Each condition is evaluated in the context of a conceptual database population and a tuple passed on from any surrounding query.

## 5.6 Conditions

Using the above defined expressions we can define the syntactic category of conditions in more detail. Let R be a binary relation over the underlying domains such as $\mathrm{R} \in \{<, \leq, =, \neq, \geq, >\}$, let S be a binary relation on multisets such as $\mathrm{S} \in \{\subset, \subseteq, =, \neq, \supseteq, \supset\}$, let $\mathrm{L} \in \{\wedge, \vee, \underline{\vee}, \Rightarrow\}$, let $E_1, E_2$ be arithmetic-expressions, and let $C, C_1, C_2$ be conditions, then we have the following rules:

$$
\begin{aligned}
\mathsf{Cond}[\![E_1 \mathrel{\mathrm{R}} E_2]\!]\,(p,t) &\triangleq \mathsf{Expr}[\![E_1]\!]\,(p,t) \mathrel{\mathrm{R}} \mathsf{Expr}[\![E_n]\!]\,(p,t) \\
\mathsf{Cond}[\![P_1 \mathrel{\mathrm{S}} P_2]\!]\,(p,t) &\triangleq \mathsf{Val}[\![P_1]\!]\,(p,t) \mathrel{\mathrm{S}} \mathsf{Val}[\![P_n]\!]\,(p,t) \\
\mathsf{Cond}[\![E \in P(a)]\!]\,(p,t) &\triangleq \exists_{u \in \mathbb{P}[\![P]\!]\,(p)} \left[u(a) = \mathsf{Expr}[\![E]\!]\,(p,t)\right] \\
\mathsf{Cond}[\![\neg C]\!]\,(p,t) &\triangleq \neg\, \mathsf{Cond}[\![C]\!]\,(p,t) \\
\mathsf{Cond}[\![C_1 \mathrel{\mathrm{L}} C_2]\!]\,(p,t) &\triangleq \mathsf{Cond}[\![C_1]\!]\,(p,t) \mathrel{\mathrm{L}} \mathsf{Cond}[\![C_2]\!]\,(p,t)
\end{aligned}
$$

## 5.7 Connection to the Conceptual Schema

The careful reader might observe that we do not yet have a connection between the relational algebra expressions and the underlying conceptual schema. This connection is provided by the following operator, which allows us to introduce a type into a relational algebra expression. Let $x$ be an object type, and $a$ be an attribute name, then:

$$
\begin{aligned}
\mathsf{Sch}[\![\mathcal{T}_a\, x]\!] &\triangleq \{a\} \\
\mathsf{Val}[\![\mathcal{T}_a\, x]\!]\,(p,t) &\triangleq \{\!\!\{ \langle a, i\rangle{\uparrow}^1 \mid i \in \mathsf{Pop}(x) \}\!\!\}
\end{aligned}
$$

All this operation does is create a one-column table with column name $a$. Each element from the population of $x$ occurs only (and exactly) once in this table.

## 5.8 Advanced operations

As we are using multisets, the following operation is intended to ignore multiple occurrences when desired. It is the algebraic version of the SQL DISTINCT command.

$$
\begin{aligned}
\mathsf{Sch}[\![\mathsf{Ds}\, P]\!] &\triangleq \mathsf{Sch}[\![P]\!] \\
\mathsf{Val}[\![\mathsf{Ds}\, P]\!]\,(p,t) &\triangleq \mathsf{Multi}(\mathsf{Set}(\mathsf{Val}[\![P]\!]\,(p,t)))
\end{aligned}
$$

The algebraic equivalent of the SQL GROUP BY statement is the following grouping operator:

$$\begin{aligned}\mathsf{Sch}[\![\varphi_X\, P]\!] &\triangleq \mathsf{Sch}[\![P]\!]\\ \mathsf{Val}[\![\varphi_X\, P]\!]\,(p) &\triangleq \left\{\!\!\left[ (t \ltimes s)\!\uparrow^1 \,\middle|\, t \in VP \wedge s = \lambda_{a\in Y}. \bigcup_{u \in^n VP} \{\![ u(a)\!\uparrow^n \mid u[X] = t[X] ]\!\} \right]\!\!\right\}\end{aligned}$$

where $VP = \mathsf{Val}[\![P]\!]\,(p)$ and $Y = \mathsf{Sch}[\![P]\!] - X$. This operation does a similar grouping as in SQL on the attribute names provided in $X$. As an example, let $R$ be:

| a | b |
|---|---|
| 1 | 'a' |
| 2 | 'b' |
| 1 | 'c' |
| 2 | 'b' |

For $\varphi_X\, R$, this would lead to:

| a | b |
|---|---|
| 1 | $\{\![ 'a', 'c' ]\!\}$ |
| 2 | $\{\![ 'b', 'b' ]\!\}$ |

In our relational algebra we have the normal inner-join, as well as the left-join The inner-join is formally identified by:

$$\begin{aligned}\mathsf{Sch}[\![P \bowtie Q]\!] &\triangleq \mathsf{Sch}[\![P]\!] \cup \mathsf{Sch}[\![Q]\!]\\ \mathsf{Val}[\![P \bowtie Q]\!]\,(p) &\triangleq \{\![ t\!\uparrow^{n\times m} \mid t : X \to \Omega \;\wedge\; t[P] \in^n \mathsf{Val}[\![P]\!]\,(p) \wedge t[Q] \in^m \mathsf{Val}[\![Q]\!]\,(q) ]\!\}\end{aligned}$$

where $t[P]$ is an abbreviation for $t[\ \mathsf{Sch}[\![P]\!]\ ]$ and $X = \mathsf{Sch}[\![P \bowtie Q]\!]$. The left-join is defined as:

$$\begin{aligned}\mathsf{Sch}[\![P ⟕ Q]\!] &\triangleq \mathsf{Sch}[\![P]\!] \cup \mathsf{Sch}[\![Q]\!]\\ \mathsf{Val}[\![P ⟕ Q]\!]\,(p) &\triangleq \{\![ t\!\uparrow^{n\times m} \mid t : X \to \Omega \;\wedge\; t[P] \in^n \mathsf{Val}[\![P]\!]\,(p) \wedge t[Q_t] \in^m \mathsf{Val}[\![Q_t]\!]\,(q) ]\!\}\end{aligned}$$

where $X = \mathsf{Sch}[\![P ⟕ Q]\!]$, $Q_t = \pi_Y\, Q$ and $Y = \mathsf{DefMap}(\mathsf{Sch}[\![Q]\!] \cap \mathsf{dom}(t))$.

Finally, for relational algebra expressions $P$ and $Q$, where $\mathsf{Sch}[\![P]\!] = \mathsf{Sch}[\![Q]\!]$, we can define the following three operations which should be familiar from set theory:

$$\begin{aligned}\mathsf{Sch}[\![P \cup Q]\!] &\triangleq \mathsf{Sch}[\![P]\!]\\ \mathsf{Val}[\![P \cup Q]\!]\,(p) &\triangleq \mathsf{Val}[\![P]\!]\,(p) \cup \mathsf{Val}[\![Q]\!]\,(p)\end{aligned}$$

$$\begin{aligned}\mathsf{Sch}[\![P \cap Q]\!] &\triangleq \mathsf{Sch}[\![P]\!]\\ \mathsf{Val}[\![P \cap Q]\!]\,(p) &\triangleq \mathsf{Val}[\![P]\!]\,(p) \cap \mathsf{Val}[\![Q]\!]\,(p)\end{aligned}$$

$$\begin{aligned} \mathsf{Sch}[\![P - Q]\!] &\triangleq \mathsf{Sch}[\![P]\!] \\ \mathsf{Val}[\![P - Q]\!]\,(p) &\triangleq \mathsf{Val}[\![P]\!]\,(p) - \mathsf{Val}[\![Q]\!]\,(p) \end{aligned}$$

For this latter class of operations it is essential to have $\mathsf{Sch}[\![P]\!] = \mathsf{Sch}[\![Q]\!]$ because for these operations the tuples must 'fit' together with each other.

### 5.9 Non-value type instances

In the relational algebra definition as given in this section, instances of entity types (or objectified relationship types) are treated like instances of value types. However, when we translate the path expressions to SQL we should realise that we have to replace the references to the abstract entity instances to concrete references to value types based on the reference schemes.

For example, if Person is an object type with reference scheme $\langle \mathsf{Surname}, \mathsf{Firstname} \rangle$, then whenever we use instances of Person to make comparisons, we have to replace them by comparisons of the proper value types. Let Student and Co-Worker be overlapping subtypes of Person, and let $s$ be a student and $w$ be a co-worker. In the relational algebra we might write in a selection statement: $\neg(s = w)$. In SQL we should realise that we have to split the $s$ and $w$ attribute names in two. This would lead to: NOT ( s-Surname = w-Surname AND s-Firstname = w-Firstname ).

## 6 Path Expressions

Path-expressions are formal constructs for expressing derived relationship types by closely following the underlying information structure. Path-expressions can be constructed from elements of the information structure (roles, object types), constants and a number of operators. They are evaluated with respect to the current population of the information structure. In its elementary form, a path-expression corresponds to a path through the information structure, starting and ending in an object type.

In this section we discuss the abstract syntax of path expressions, the concrete semantics is provided in the appendix. The difference between abstract syntax and concrete syntax is that abstract syntax describes rules by which parse trees for the expressions in the defined language can be build. This necessarily means that issues like ambiguities in the parsing of the languages (an ambiguous grammar) are not relevant. A concrete syntax on the other hand, is also concerned with parsing issues. As a result, an abstract syntax will usually not contain disambiguating constructs like '(', ')'.

The set of path-expressions for a given information structure $\mathcal{IS}$, is denoted as `PathExpr`. The semantics for the path-expressions does not directly refer to a population itself. This is exactly what we want since we need to generate SQL without already evaluating the query! The semantics of path-expressions is therefore defined by translating the path-expressions to relational algebra. The final query result of a path-expression is then obtained by 'executing' the resulting relational algebra expression (or SQL statements) in the context of the current database population. The translation is done by the function:

$$\mathbb{P} : \texttt{PathExpr} \times (\mathsf{Attrs} \times \mathcal{TP}) \times \wp(\mathsf{Attrs}) \rightarrow \mathcal{RA}$$

We define this semantics, again, using the style of denotational semantics. The environment of the semantics function is a typing relationship ($T \subseteq \mathsf{Attrs} \times \mathcal{TP}$) and a set $B \in \wp(\mathsf{Attrs})$ indicating which attributes (variables on the ConQuer-92 level) have already been bound to an underlying type.

The relational algebra expressions used to express the semantics of path-expressions always have at least two columns. These are the $hd$ and $tl$ column. The first column represents the head (start) of the path-expression, and the second column represents the tail (end) of the path-expression. Intermediate results of a path-expression can be represented in a separate column by providing an explicit attribute name. Later in this section we see a construct to introduce such extra columns.

In a ConQuer-92 expression, all variables must be of some type. Just as in the following Pascal fragment the type of i is INTEGER, and j is of type REAL, variables in ConQuer-92 expressions are types as well:

```
VAR
  i: INTEGER;
  j: REAL;
BEGIN
  i := 1;
  j := 1 + i;
END
```

The reason for using a typing relationship rather than a typing function, which means that a variable may be of more than one type, will become clear in the remainder of this section.

Obviously, the typing function has to be determined before the actual translation of a path-expression to a relational algebra expression can be done. The typing is derived by using the function:

$$\mathbb{T} : \texttt{PathExpr} \rightarrow \wp(\mathsf{Attrs} \times \mathcal{TP})$$

This function is defined more formally in subsection 6.10. There it shall also become clear why the typing is provided as a relationship over $\mathsf{Attrs} \times \mathcal{TP}$. An obvious requirement on path-expressions with respect to typing is:

$$\langle a, x \rangle , \langle a, y \rangle \in \mathbb{T}[\![P]\!] \Rightarrow x \sim y$$

requiring attributes to be in one type relatedness class only. This means that we do allow variables to be of more than one type, but they must be of related types. Furthermore, all attributes occurring in a path should indeed be typed:

$$\pi_1\,\mathbb{T}[\![P]\!] \text{ is exactly the set of attributes in } P$$

As of now we use $\mathsf{Attr}[\![P]\!]$ as an abbreviation for $\pi_1\,\mathbb{T}[\![P]\!]$.

During the evaluation of a path-expression, an attribute must always be bound to its associated type. The notion of a *bound variable* stems from logic. For example, in the formula $\varphi(x)$ the variable $x$ is not bound to any domain. It can range over all individuals known in our *universe*. If $D$ is a set of individuals, then in $\forall_{x \in D}\,[\varphi(x)]$ the $x$ is said to be bound to the domain $D$. To make certain that all variables in ConQuer-92 are bound, we sometimes need to explicitly bind an attribute to its proper type. If $T$ is a typing, then this can converted into a function providing proper *bindings* of attributes/variables to their types by the function:

$$\begin{aligned} &\mathsf{Bind} : \wp(\mathsf{Attrs} \times \mathcal{TP}) \rightarrow (\mathsf{Attrs} \rightarrow \mathcal{RA}) \\ &\mathsf{Bind}(T) \triangleq \lambda_{a \in \pi_1 T}.\ \bigcup_{\langle a,x\rangle \in T}\ \bigcup_{y \in \mathsf{RootsOf}(x)} \mathcal{T}_a\,x \end{aligned}$$

The expression:

$$\bigcup_{\langle a,x\rangle \in T}\ \bigcup_{y \in \mathsf{RootsOf}(x)} \mathcal{T}_a\,x$$

is a relational algebra expression which forces the attribute $a$ to be limited to instances of the root types of the types of $a$.

Finally, we also introduce a function:

$$\mathbb{L} : \mathtt{PathExpr} \times (\mathsf{Attrs} \rightarrowtail \mathcal{TP}) \rightarrow \wp(\mathcal{TP} \times \mathcal{TP})$$

which is intended to administer the possible combinations of types for the start and end of a path-expression. As stated before, a path-expression basically corresponds to a path in the conceptual schema connecting types. More complex path expressions will correspond to sets of paths through the conceptual schema. As such, complex path expressions can have multiple start and end type combinations. The $\mathbb{L}$ functions provides us with these combinations. This function will allows us to make optimisations and remove ambiguities in verbalisations. For example, if for a certain path expression $P$ we have $\mathbb{L}[\![P]\!]\,(T) = \varnothing$, it means that $\mathbb{P}[\![P]\!]\,(T, B)$ evaluates to the empty set in any population.

## 6.1 Linear path-expressions

The first class of path-expressions we introduce are the linear path-expressions. These are the linear paths as they may result from a *point to point query* ([Pro94b]) or *query* *by navigation* ([Pro94a]). The linear path-expressions are called linear as they always correspond to a single path through the conceptual schema.

Linear path-expression have two elementary building blocks. Each type and role from the conceptual schema can be used as a linear path-expression. When a role $r$ used as a linear path-expression it becomes a so-called *role-entry* as it provides an entry to a fact type ($\mathsf{Rel}(r)$) from a type ($\mathsf{Player}(r)$) via a role ($r$).

These two basic building blocks lead to the following formal rules. If $x$ is a type, $p$ a role, $P$ an existing path-expression, and $a$ each time a fresh attribute, then we can define:

$$\begin{aligned}\mathbb{P}[\![x]\!]\,(T,B) &\triangleq \pi_{hd=a,tl=a}\,\tau_a\,x\\ \mathbb{P}[\![p]\!]\,(T,B) &\triangleq \pi_{hd=a.p,tl=a}\,\tau_a\,\mathsf{Rel}(p)\end{aligned}$$

If $x$ is a type with population $\mathsf{Pop}(x) = \{1,2,3\}$, then the path-expression $x$ leads to:

| $hd$ | $tl$ |
|---|---|
| 1 | 1 |
| 2 | 2 |
| 3 | 3 |

If $f$ is a fact type with $\mathsf{Roles}(f) = \{p,q\}$, and $\mathsf{Pop}(f) = \{\{p:1,q:a\},\{p:2,q:b\},\{p:3,q:c\}\}$, then for a path-expression $p$ we have:

| $hd$ | $tl$ |
|---|---|
| 1 | $\{p:1,q:a\}$ |
| 2 | $\{p:2,q:b\}$ |
| 3 | $\{p:3,q:c\}$ |

Note that while on the relational algebra level attributes and roles could be used interchangeably (as column names), on the path expression level they are to be treated separately. In path expressions, roles correspond to connections between the player of the role and the relationship type in which the role is involved, while attributes correspond to variables.

The possible combinations of head and tail types for the basic constructions are given by:

$$\begin{aligned}\mathbb{L}[\![x]\!]\,(T) &\triangleq \left\{\langle u,v\rangle \;\middle|\; u\sim x \wedge v \sim x\right\}\\ \mathbb{L}[\![p]\!]\,(T) &\triangleq \left\{\langle u,v\rangle \;\middle|\; u\sim \mathsf{Player}(p) \wedge v \sim \mathsf{Rel}(p)\right\}\end{aligned}$$

The first important complex construction on linear path-expressions allows us to reverse a (linear) path-expression. If we have a path from a type $x$ to $y$, then reversing it will lead to a path from $y$ to $x$. Let $P$ be a path, then we have:

$$\mathbb{P}[\![P^{\leftarrow}]\!](T,B) \triangleq \rho_{hd=tl,tl=hd}\,\mathbb{P}[\![P]\!](T,B)$$
$$\mathbb{L}[\![P^{\leftarrow}]\!](T) \triangleq \{\langle v,u\rangle \mid \langle u,v\rangle \in \mathbb{L}[\![P]\!](T)\}$$

A special case of a reversed path is a reversed role. If $r$ is a role, then $r$ is a path-expression (role-entry) as well. The reversal of $r$ leads to $r^{\leftarrow}$, and is referred to as a *role-exit*, as it provides a path from fact type $\mathsf{Rel}(r)$ to type $\mathsf{Player}(r)$. As an example, let $p$ be the role from the fact type $f$ as shown above, then $p^{\leftarrow}$ results in:

| *hd* | *tl* |
|---|---|
| $\{p:1,q:a\}$ | 1 |
| $\{p:2,q:b\}$ | 2 |
| $\{p:3,q:c\}$ | 3 |

Using concatenation, path-expressions may be combined into yet more complex expression. Let $P$ and $Q$ be (linear) path-expressions and $a$ a fresh attribute, then we can define:

$$\mathbb{P}[\![P \circ Q]\!](T,B) \triangleq \delta_a(\rho_{a=tl}\,P \bowtie \rho_{a=hd}\,Q)$$
$$\mathbb{L}[\![P \circ Q]\!](T) \triangleq \{\langle u,w\rangle \mid \langle u,v\rangle \in \mathbb{L}[\![P]\!](T) \wedge \langle v,w\rangle \in \mathbb{L}[\![Q]\!](T)\}$$

This operation corresponds to a head/tail concatenation of two existing path expressions. If $P$ and $Q$ are path-expressions with results:

$P:$

| *hd* | *tl* |
|---|---|
| 1 | $a$ |
| 2 | $b$ |
| 3 | $a$ |

$Q:$

| *hd* | *tl* |
|---|---|
| $a$ | $k$ |
| $c$ | $l$ |
| $b$ | $l$ |

then the path expression $P \circ Q$ leads to:

| *hd* | *tl* |
|---|---|
| 1 | $k$ |
| 2 | $l$ |
| 3 | $k$ |

The set of linear path-expressions is exactly defined as the set of path-expressions that can be build from the above constructions.

## 6.2 Complex operations

A (linear) path-expression corresponds to a path through the information structure. The front and tail elements of such paths play a central role in the concatenation of path-expressions. Sometimes we want to limit our interest to the front elements only. For this purpose we introduce the following operation:

$$\begin{aligned}\mathbb{P}[\![\mathsf{Fr}\,P]\!]\,(T,B) &\triangleq \alpha_{tl=hd}\,\mathbb{P}[\![P]\!]\,(T,B)\\ \mathbb{L}[\![\mathsf{Fr}\,P]\!]\,(T) &\triangleq \{\langle u,u\rangle \mid \langle u,v\rangle \in \mathbb{L}[\![P]\!]\,(T)\}\end{aligned}$$

As we are working with multisets, the distinct operation must be present on the path-expression level as well:

$$\begin{aligned}\mathbb{P}[\![\mathsf{Ds}\,P]\!]\,(T,B) &\triangleq \mathsf{Ds}(\mathbb{P}[\![P]\!]\,(T,B))\\ \mathbb{L}[\![\mathsf{Ds}\,P]\!]\,(T) &\triangleq \mathbb{L}[\![P]\!]\,(T)\end{aligned}$$

If $P$ returns the following table:

| $hd$ | $x$ | $tl$ |
|---|---|---|
| $a$ | 1 | 2 |
| $a$ | 3 | 4 |
| $b$ | 5 | 6 |
| $a$ | 3 | 4 |

we have:

$\mathsf{Fr}\,P$ :

| $hd$ | $x$ | $tl$ |
|---|---|---|
| $a$ | 1 | $a$ |
| $a$ | 3 | $a$ |
| $b$ | 5 | $b$ |
| $a$ | 3 | $a$ |

$\mathsf{Ds}\,P$ :

| $hd$ | $x$ | $tl$ |
|---|---|---|
| $a$ | 1 | 2 |
| $a$ | 3 | 4 |
| $b$ | 5 | 6 |

The cartesian product of path-expressions is identified by:

$$\mathbb{P}[\![P \times Q]\!]\,(T,B) \triangleq \pi_X(\delta_{tl}\,\mathbb{P}[\![P]\!]\,(T,B) \bowtie \delta_{hd}\,\rho_{tl=hd}\,\mathbb{P}[\![Q]\!]\,(T,B))$$

where $X = \mathsf{DefMap}(\mathsf{Sch}\cdot\mathbb{P}[\![P]\!]\,(T,B) \cup \mathsf{Sch}\cdot\mathbb{P}[\![Q]\!]\,(T,B))$

$$\mathbb{L}[\![P \times Q]\!]\,(T) \triangleq \{\langle u,x\rangle \mid \langle u,v\rangle \in \mathbb{L}[\![P]\!]\,(T) \wedge \langle w,x\rangle \in \mathbb{L}[\![Q]\!]\,(T)\}$$

If we have the following results of path expressions:

$P$ :

| $hd$ | $tl$ |
|---|---|
| 1 | 2 |
| 3 | 4 |

$Q$ :

| $hd$ | $x$ | $tl$ |
|---|---|---|
| 5 | $a$ | $c$ |
| 6 | $b$ | $d$ |

then we have:

$P \times Q$ :

| $hd$ | $x$ | $tl$ |
|---|---|---|
| 1 | $a$ | 5 |
| 1 | $b$ | 6 |
| 3 | $a$ | 5 |
| 3 | $b$ | 6 |

The following operations are taken from [HPW97]. Let $P$ and $Q$ be path-expressions, and $a$, $b$ be fresh attributes, then:

$$\mathbb{P}[\![P \subseteqq Q]\!]\,(T,B) \triangleq \pi_X\,\sigma_{\pi_{b=tl}(\sigma_{hd=a}\,\mathbb{P}[\![P]\!]\,(T,B)) \subseteqq \pi_{b=hd}\,\mathbb{P}[\![Q]\!]\,(T,B)}\,\alpha_{a=hd}\,P$$

$$\mathbb{P}[\![P \supseteqq Q]\!]\,(T,B) \triangleq \pi_X\,\sigma_{\pi_{b=tl}(\sigma_{hd=a}\,\mathbb{P}[\![P]\!]\,(T,B)) \supseteqq \pi_{b=hd}\,\mathbb{P}[\![Q]\!]\,(T,B)}\,\alpha_{a=hd}\,P$$

$$\mathbb{P}[\![P \equiv Q]\!]\,(T,B) \triangleq \pi_X\,\sigma_{\pi_{b=tl}(\sigma_{hd=a}\,\mathbb{P}[\![P]\!]\,(T,B)) = \pi_{b=hd}\,\mathbb{P}[\![Q]\!]\,(T,B)}\,\alpha_{a=hd}\,P$$

For $\mathrm{O} \in \{\subseteqq, \supseteqq, \equiv\}$ we have:

$$\mathbb{L}[\![P \;\mathrm{O}\; Q]\!]\,(T) \;\triangleq\; \{\langle u,w\rangle \mid \langle u,v\rangle \in \mathbb{L}[\![P]\!]\,(T) \wedge \langle v,w\rangle \in \mathbb{L}[\![Q]\!]\,(T)\}$$

Let the path expressions $P$ and $Q$ result in:

$P$:

| $hd$ | $tl$ |
|---|---|
| $a$ | 1 |
| $a$ | 2 |
| $b$ | 1 |
| $b$ | 2 |
| $b$ | 3 |
| $b$ | 4 |
| $c$ | 1 |
| $c$ | 2 |
| $c$ | 3 |

$Q$:

| $hd$ | $tl$ |
|---|---|
| 1 | $f$ |
| 2 | $g$ |
| 3 | $h$ |

For this we would for example have:

$P \subseteqq Q$:

| $hd$ | $tl$ |
|---|---|
| $a$ | 1 |
| $a$ | 2 |
| $c$ | 1 |
| $c$ | 2 |
| $c$ | 3 |

$P \supseteqq Q$:

| $hd$ | $tl$ |
|---|---|
| $b$ | 1 |
| $b$ | 2 |
| $b$ | 3 |
| $b$ | 4 |
| $c$ | 1 |
| $c$ | 2 |
| $c$ | 3 |

$P \equiv Q$:

| $hd$ | $tl$ |
|---|---|
| $c$ | 1 |
| $c$ | 2 |
| $c$ | 3 |

The $\subseteqq$ operation is used to select those head elements from $P$ such that all the tail elements associated to that head element occur as head of $Q$. An example is:

A president who has a hobby WHICH ARE ALL IN hobby of president: "Clinton"

which results in the presidents (and their hobbies) who have only hobbies that are also hobbies of president Clinton. The $\supseteqq$ operation, on the other hand, is used to select the head elements from $P$ which have associated a set of tail elements that includes all heads from $Q$. As an example consider:

A president who has a hobby THAT INCLUDES ALL hobby of president: "Clinton"

This expression results in those presidents who have at least all hobbies that president Clinton has. The $\underline{\subseteq}$ and $\underline{\supseteq}$ operation are combined by the $\underline{=}$ operation. For instance,

A president who has a hobby MATCHING ALL hobby of president: "Clinton"

leads to the presidents which have *exactly* the same set of hobbies as president Clinton.

With the $\underline{\otimes}$ operations we can select the head-tail combinations that are not returned by $\circ$:

$$\mathbb{P}[\![P \underline{\otimes} Q]\!]\,(T,B) \triangleq (\delta_{tl}\,P \bowtie \delta_{hd}\,Q) - \mathbb{P}[\![P \circ Q]\!]\,(T,B)$$

where each time $X = \mathsf{DefMap}(\mathsf{Sch} \cdot \mathbb{P}[\![P]\!]\,(T,B))$

$$\mathbb{L}[\![P \underline{\otimes} Q]\!]\,(T) \triangleq \{\langle u,x\rangle \mid \langle u,v\rangle \in \mathbb{L}[\![P]\!]\,(T) \wedge \langle w,x\rangle \in \mathbb{L}[\![Q]\!]\,(T)\}$$

As an example let $P$ and $Q$ lead to:

$P:$

| $hd$ | $tl$ |
|---|---|
| $a$ | $b$ |
| $c$ | $d$ |

$Q:$

| $hd$ | $tl$ |
|---|---|
| $b$ | 3 |
| $d$ | 4 |

then we have:

$P \underline{\otimes} Q:$

| $hd$ | $tl$ |
|---|---|
| $a$ | 4 |
| $c$ | 3 |

$P \circ Q:$

| $hd$ | $tl$ |
|---|---|
| $a$ | 3 |
| $c$ | 4 |

As a result, the $\underline{\otimes}$ operation allows us to select the connections between the head elements in $P$ and tail elements o f $Q$ that are not connected via $P \circ Q$. A concrete example is:

A president who has a hobby EXCLUDING hobby of president: "Clinton"

On path-expressions we also have the normal set theoretic operations like $\cap$, $\cup$, and $-$. However, in the case of path-expressions special care has to be taken for with differing attributes. In taking the union, intersection and difference of path-expressions, the path-expressions involved are first coerced such that they have the same set of attributes. After the union, intersection, or difference, has been calculated the information that was removed during the coercion is added again.

$$
\begin{array}{rcl}
\mathbb{P}[\![P \cap Q]\!](T,B) & \triangleq & (\pi_X\, \mathbb{P}[\![P]\!](T,B) \cap \pi_X\, \mathbb{P}[\![Q]\!](T,B)) \bowtie \mathsf{Ds}(P) \bowtie \mathsf{Ds}(Q) \\
\mathbb{T}[\![P \cap Q]\!](T) & \triangleq & \mathbb{T}[\![P]\!](T) \cap \mathbb{T}[\![Q]\!](T) \\
\mathbb{P}[\![P \cup Q]\!](T,B) & \triangleq & (\pi_X\, \mathbb{P}[\![P]\!](T,B) \cup \pi_X\, \mathbb{P}[\![Q]\!](T,B)) ⟕ \mathsf{Ds}(P) ⟕ \mathsf{Ds}(Q) \\
\mathbb{T}[\![P \cup Q]\!](T) & \triangleq & \mathbb{T}[\![P]\!](T) \cup \mathbb{T}[\![Q]\!](T) \\
\mathbb{P}[\![P - Q]\!](T,B) & \triangleq & (\pi_X\, \mathbb{P}[\![P]\!](T,B) - \pi_X\, \mathbb{P}[\![Q]\!](T,B)) \bowtie \mathsf{Ds}(P) \\
\mathbb{T}[\![P - Q]\!](T) & \triangleq & \mathbb{T}[\![P]\!](T)
\end{array}
$$

where each time $X = \mathsf{DefMap}(\mathsf{Sch} \cdot \mathbb{P}[\![P]\!](T,B) \cap \mathsf{Sch} \cdot \mathbb{P}[\![Q]\!](T,B))$

Note the use of the left-join in the case of the union. In practice, we will make much use of an intersection, union, and set difference. Therefore, we also introduce the following abbreviations:

$$
\begin{array}{rcl}
P_1 \Cup P_2 & \triangleq & \mathsf{Fr}(P_1) \cup \mathsf{Fr}(P_2) \\
P_1 \Cap P_2 & \triangleq & \mathsf{Fr}(P_1) \cap \mathsf{Fr}(P_2) \\
P_1 \not- P_2 & \triangleq & \mathsf{Fr}(P_1) - \mathsf{Fr}(P_2)
\end{array}
$$

We also introduce the traditional binary operations: $<, \leq, =, \neq, \geq, >$ into the path-expressions. Let in the following definition $R \in \{<, \leq, =, \neq, \geq, >\}$, let $a$, $b$ be fresh attributes, and let $P$ and $Q$ be path-expressions, then:

$$
\begin{array}{rcl}
\mathbb{P}[\![P \mathrel{\mathrm{R}} Q]\!](T,B) & \triangleq & \pi_X\, \sigma_{a \mathrel{\mathrm{R}} y}(\rho_{a=tl}\, \mathbb{P}[\![P]\!](T,B) \bowtie \rho_{b=hd}\, \mathbb{P}[\![Q]\!](T,B)) \\
\mathbb{L}[\![P \mathrel{\mathrm{R}} Q]\!](T) & \triangleq & \{\langle u,v \rangle \mid \langle u,v \rangle \in \mathbb{L}[\![P]\!](T) \wedge \langle v,w \rangle \in \mathbb{L}[\![Q]\!](T)\}
\end{array}
$$

where $X = \mathsf{DefMap}(\mathsf{Sch} \cdot \mathbb{P}[\![P]\!](T,B) \cup \mathsf{Sch} \cdot \mathbb{P}[\![Q]\!](T,B))$

If $P$ and $Q$ are path expressions with results:

$P:$

| $hd$ | $tl$ |
|---|---|
| $a$ | 100 |
| $b$ | 233 |
| $c$ | 250 |
| $d$ | 130 |

$Q:$

| $hd$ | $x$ | $tl$ |
|---|---|---|
| 50 | 50 | $k$ |
| 101 | 101 | $l$ |
| 200 | 200 | $m$ |

then we have:

$P < Q:$

| $hd$ | $x$ | $tl$ |
|---|---|---|
| $a$ | 101 | $l$ |
| $a$ | 200 | $m$ |
| $d$ | 200 | $m$ |

The above definition in itself would not allow us to write: Salary $< 1000$ if Salary is an entity type identified through a value type Amount. However, on the ConQuer-92 level we introduce an abbreviation mechanism which indeed allows us to write such path-expressions.

As the new generation of path-expressions allows for intermediate results, we now also introduce a path shuffle operation for path-expressions:

$$\begin{aligned}
\mathbb{P}[\![\mathsf{Path}(P, a_1, \ldots, a_n)]\!]\,(T, B) &\triangleq \rho_{hd=a_1, tl=a_n}\, \pi_{\mathsf{DefMap}(\{a_1, \ldots, a_n\})}\, \mathbb{P}[\![P]\!]\,(T, B - \{a_1, \ldots, a_n\}) \\
\mathbb{L}[\![\mathsf{Path}(P, a_1, \ldots, a_n)]\!]\,(T) &\triangleq \big\{\langle u, v\rangle \;\big|\; u \sim T(a_1) \wedge v \sim T(a_n)\big\}
\end{aligned}$$

This operation allows us to make projections and operates in a way similar to that of the SQL SELECT statement (therefore we shall not provide an elaborated example with tables). As an concrete example of how this may look like in ConQuer-92, inspired by [Hal95], consider:

THE PATH FROM $x$ TO $y$ OF
President $x$ who is married to a Spouse $y$ AND ALSO is member of the Party: "Republican"

resulting in the presidents and spouses of presidents who are a member of the republican party.

To support the use of mix-fix predicate verbalisations in ConQuer, we introduce the following construction (although it is basically an abbreviation) on path expressions. Let $p_1, \ldots, p_n$ be roles of the same fact types, and $P_2, \ldots, P_{l-1}$

$$\langle p_1, p_2 : \mathbb{I}[\![X_2]\!], \ldots, p_{l-1} : \mathbb{I}[\![X_{l-1}]\!], p_l\rangle \triangleq p_1 \circ \bigcap_{1<i<l} \mathsf{Ds}\,\mathsf{Fr}(p_i^{\leftarrow} \circ \mathbb{I}[\![P_i]\!]) \circ p_l$$

Two special operations are introduced that take care of coercing simply identified entity types to value types. This will allow us to write MoneyAmount $> 1000$ rather than MoneyAmount of AUD $> 1000$. The coercion operations, which are basically abbreviations, are identified by:

$$\mathsf{HdCoerce}(P) \triangleq \begin{cases} \mathsf{HdCoerce}(s \circ r^{\leftarrow} \circ P) & \text{if } \big\{t(hd) \;\big|\; t \in \mathbb{L}[\![P]\!]\big\} = \{x\} \wedge \mathsf{Idf}(X) = [\langle r, s\rangle] \\ P & \text{otherwise} \end{cases}$$

$$\mathsf{TlCoerce}(P) \triangleq \begin{cases} \mathsf{TlCoerce}(P \circ r \circ s^{\leftarrow}) & \text{if } \big\{t(tl) \;\big|\; t \in \mathbb{L}[\![P]\!]\big\} = \{x\} \wedge \mathsf{Idf}(X) = [\langle r, s\rangle] \\ P & \text{otherwise} \end{cases}$$

The HdCoerce and TlCoerce functions allow us to write: Salary $> 1000$, rather than Salary of \$Amount $> 1000$.

## 6.3 Scalar expressions

Scalar-expressions are regarded as a special kind of path-expression. However, we still do introduce a separate class `PEScalarExpr` $\subseteq$ `PathExpr` of scalar-expressions. Regarding scalar-expressions as a special class of path-expressions leads to an orthogonal language. The reason for introducing a special subclass is that this allows us to optimise the mapping to relational algebra (and SQL). We return to this issue when we introduce the semantic coercion rule from scalar-expressions to path-expressions.

The semantics of scalar-expressions are provided by:

$$\mathbb{E} : \texttt{PEScalarExpr} \times (\mathsf{Attrs} \times \mathcal{TP}) \times \wp(\mathsf{Attrs}) \rightarrow \mathcal{RA}$$

Let $c$ be a constant, and $P$ be path-expressions, then this class of path-expressions is provided as:

$$\begin{aligned}
\mathbb{E}[\![c]\!]\,(T,B) &\triangleq c\\
\mathbb{E}[\![\mathsf{Count}(P)]\!]\,(T,B) &\triangleq \mathsf{Count}(\mathbb{P}[\![P]\!]\,(T,B), hd)\\
\mathbb{E}[\![\mathsf{Sum}(P)]\!]\,(T,B) &\triangleq \mathsf{Sum}(\mathbb{P}[\![P]\!]\,(T,B), hd)\\
\mathbb{E}[\![\mathsf{Min}(P)]\!]\,(T,B) &\triangleq \mathsf{Min}(\mathbb{P}[\![P]\!]\,(T,B), hd)\\
\mathbb{E}[\![\mathsf{Max}(P)]\!]\,(T,B) &\triangleq \mathsf{Max}(\mathbb{P}[\![P]\!]\,(T,B), hd)\\
\mathbb{E}[\![\mathsf{Avg}(P)]\!]\,(T,B) &\triangleq \mathsf{Avg}(\mathbb{P}[\![P]\!]\,(T,B), hd)
\end{aligned}$$

The $\mathsf{Count}$ operation counts the number of tuples in the result of $P$, whereas $\mathsf{Sum}$, $\mathsf{Min}$, $\mathsf{Max}$ and $\mathsf{Avg}$ calculate the sum, minimum, maximum, and average of the head elements of $P$ respectively.

Attributes can be used in scalar-expressions. Even more, the underlying components of attributes over a nested type (objectified relationship type) may be accessed. So, if $b$ is an attribute and $p$ a role, we have:

$$\begin{aligned}
\mathbb{E}[\![b]\!]\,(T,B) &\triangleq \begin{cases} \pi_{hd=a,tl=a}\,\epsilon_a\, b & \text{if } b \in B\\ \alpha_{hd=b,tl=b}\,\mathsf{Bind}(T)(b) & \end{cases}\\
\mathbb{E}[\![b.p]\!]\,(T,B) &\triangleq \begin{cases} \pi_{hd=a.p,tl=a.p}\,\epsilon_a\, b & \text{if } b \in B\\ \alpha_{hd=b.p,tl=b.p}\,\mathsf{Bind}(T)(b) & \end{cases}
\end{aligned}$$

Functions can for obvious reasons be used in scalar-expressions as well. Let $E_1, \ldots, E_n$ be scalar-expressions, then we have:

$$\mathbb{E}[\![f(E_1,\ldots,E_n)]\!]\,(T,B) \triangleq f(\mathbb{E}[\![E_1]\!]\,(T,B),\ldots,\mathbb{E}[\![E_n]\!]\,(T,B))$$

For infix functions like $+$ we allow $P+Q$ as an abbreviation for $+(P,Q)$, but we prefer to do this on the verbalisation level of ConQuer-92 rather than on the path-expression level.

As stated before, scalar-expressions are special path-expressions, so we have for a scalar-expression $E$:

$$\mathbb{P}[\![E]\!]\,(T,B) \triangleq \begin{cases} \alpha_{tl=hd}\,\epsilon_{hd}\,E & \text{if } \mathsf{Attr}[\![E]\!] - B = \varnothing \\ \alpha_{tl=hd}\,\alpha_{hd=\mathbb{E}[\![E]\!]\,(T,B)}(\bowtie_{a\in\mathsf{Attr}[\![E]\!]\,-\,B}\ \mathsf{Bind}(T)(a)) & \text{otherwise} \end{cases}$$

$$\mathbb{L}[\![E]\!]\,(T) \triangleq \{\langle a,b\rangle \mid a \sim r \wedge b \sim r \wedge r \in R\}$$

where $R$ is the set of resulting types from scalar-expression $E$. The latter set $R$ can be calculated in a conventional way (like in any other programming language), using the typing in $T$ as a base. As an example, let $P$ be a path-expression resulting in:

| $hd$ | $tl$ |
|---|---|
| 1 | 9 |
| 2 | 10 |
| 8 | 12 |
| 8 | 12 |

then the result of the expression $1 + \mathsf{Avg}(P)$ would result in $4.75$. When interpreted as a path-expression, this would lead to:

| $hd$ | $tl$ |
|---|---|
| 4.75 | 4.75 |

The idea of the coercion from scalar-expressions to path-expressions is to apply it as late as possible; i.e. evaluate a scalar-expression $E$ on the relational algebra level as much as possible as a scalar-expression.

Functions can also be applied on path-expressions. In such a case, the function is applied to each tuple separately. Let $P_1, \ldots, P_n$ be path-expressions (which are not *all* scalar-expressions), $a_1, \ldots, a_n$ be fresh attributes, and $f$ a function symbol, then we have:

$$\mathbb{P}[\![f(P_1,\ldots,P_n)]\!]\,(T,B) \triangleq$$

$$\delta_{a_1,\ldots,a_n}\,\alpha_{hd=f(a_1,\ldots,a_n)}\left(\bowtie_{1\leq i<n}\rho_{a_i=hd}\,\delta_{tl}\,\mathbb{P}[\![P_i]\!]\,(T,B) \bowtie \rho_{a_n=hd}\,\mathbb{P}[\![P_n]\!]\,(T,B)\right)$$

$$\mathbb{L}[\![f(P_1,\ldots,P_n)]\!]\,(T) \triangleq \{\langle v,w\rangle \mid v \sim r \wedge w \sim r \wedge r \in R\}$$

where $R$ is the set of resulting types of function $f$

Note that this path-expression returns the tail of $P_n$ as tail. If the types of the parameters to function $f$ are also known, a stricter type check of the provided path-expressions can be done. For instance, if $t_i$ is the type for the parameter at position $i$, then we must have: $\exists_{x\in\pi_1\mathbb{L}[\![P_i]\!]\,(T)}\,[x \sim t_i]$. As an illustration, let $P$ and $Q$ lead to:

| $hd$ | $x$ | $tl$ |
|---|---|---|
| 1 | $l$ | 2 |
| 3 | $m$ | 5 |

| $hd$ | $y$ | $tl$ |
|---|---|---|
| 1 | $s$ | 3 |
| 4 | $t$ | 5 |

the $P + Q$ will result in:

| $hd$ | $x$ | $y$ | $tl$ |
|---|---|---|---|
| 2 | $l$ | $s$ | 3 |
| 5 | $l$ | $t$ | 5 |
| 4 | $m$ | $s$ | 3 |
| 7 | $m$ | $t$ | 5 |

The limitation in the above definition that not all $P_i$ are scalar-expressions avoids an ambiguity in parsing path-expressions. A scalar-expression of the form $f(P_1, \ldots, P_n)$ is a scalar-expression iff $P_1, \ldots, P_n$ are *all* scalar-expressions.

In ConQuer-92 we could now write (using the abbreviation for simply identified object types):

$$\textsf{Length of Room } x \textsf{ in House } y * \textsf{Width of Room } x \textsf{ in House } y$$

which returns the area of a room $x$ in house $y$. A more advanced example, taken from [Hal95], is:

100 * ( THE SUM OF(the NrVoters that voted for a Politician who seeks Seat $s$) +
+ the NrVoters that voted informally for $s$ )
/ the NrVoters that are on roll in $s$

## 6.4 Conditions

Conditions are build from path-expressions and can be used as constraints on the database, for conditions in a select statement (to be introduced below), and for yes-no queries. Conditions, however, are defined as a special class of path-expressions in the same way, and for the same reasons, as path-expressions.

The semantics of conditions are provided by:

$$\mathbb{C} : \texttt{PEConditions} \times (\textsf{Attrs} \times \mathcal{TP}) \times \wp(\textsf{Attrs}) \to \mathcal{RA}$$

Let S be a binary relation on multisets such as $\mathrm{S} \in \{\subset, \subseteq, =, \supseteq, \supset\}$, let L be a logical connector such as $\mathrm{L} \in \{\vee, \underline{\vee}, \wedge, \Rightarrow\}$, let R be a relational operator such as $\mathrm{R} \in \{<, \leq, =, \neq, \geq, >\}$, let $P$ be a path-expression, $E_1, E_2$ scalar-expressions, and $C, C_1, C_2$ conditions, then:

$$\begin{aligned}
\mathbb{C}[\![\textsf{Some}(P)]\!]\,(T,B) &\triangleq \mathbb{P}[\![P]\!]\,(T,B) \neq \varnothing \\
\mathbb{C}[\![P \mathrm{~S~} Q]\!]\,(T,B) &\triangleq \pi_{hd=hd}\,\mathbb{P}[\![P]\!]\,(T,B) \mathrm{~S~} \pi_{hd=hd}\,\mathbb{P}[\![Q]\!]\,(T,B) \\
\mathbb{C}[\![C_1 \mathrm{~L~} C_2]\!]\,(T,B) &\triangleq \mathbb{C}[\![C_1]\!]\,(T,B) \mathrm{~L~} \mathbb{C}[\![C_2]\!]\,(T,B) \\
\mathbb{C}[\![E_1 \mathrm{~R~} E_2]\!]\,(T,B) &\triangleq \mathbb{E}[\![E_1]\!]\,(T,B) \mathrm{~R~} \mathbb{E}[\![E_2]\!]\,(T,B) \\
\mathbb{C}[\![\neg C]\!]\,(T,B) &\triangleq \mathbb{C}[\![C]\!]\,(T,B) = \varnothing
\end{aligned}$$

We also define exclusion of path-expressions $P_1$ and $P_2$ as: $P_1 \bigotimes P_2 \triangleq \neg\,\mathsf{Some}(\mathsf{Fr}(P) \cap \mathsf{Fr}(Q))$.

Using conditions we can also introduce a selection mechanism for path-expressions. This selection mechanism works similar to the selection mechanism for relational algebra expressions. It is defined as:

$$\begin{aligned}
\mathbb{P}[\![\mathsf{Where}(P,C)]\!]\,(T,B) &\triangleq \sigma_{\mathbb{P}[\![C]\!]\,(T,B\,\cup\,\mathsf{Attr}[\![C]\!])}\left(\mathbb{P}[\![P]\!]\,(T,B \cup \mathsf{Attr}[\![C]\!]) \bowtie_{v \in \mathsf{Attr}[\![C]\!] - B - \pi_1\,\mathbb{T}[\![P]\!]} \mathsf{Bind}(T)(v)\right)\\
\mathbb{L}[\![\mathsf{Where}(P,C)]\!]\,(T) &\triangleq \mathbb{L}[\![P]\!]\,(T)
\end{aligned}$$

The extra joins are needed to bind any free attributes in $C$. Note that even if a attribute in $C$ is bound in $C$, we still need to do the join as they are needed to evaluate the condition in the first place. The attributes contained in $B$ are the ones which are already bound by the 'calling' environment in the case of a subquery. It means that the path-expression is to be used in a projection expression or a condition, and evaluated in an environment where the given attribute has already received a value. Note: the above path-expression only makes sense when: $\mathsf{dom}(T) \subseteq \mathsf{Attr}[\![C]\!] - \mathsf{Attr}[\![P]\!]$, i.e. all attributes used in the condition must be typed.

As an example of the semantics of this operation, let $P$ yield:

| $hd$ | $x$ | $tl$ |
|---|---|---|
| 1 | 3 | 5 |
| 6 | 9 | 8 |

the the expression $\mathsf{Where}(P, tl > x)$ would lead to:

| $hd$ | $x$ | $tl$ |
|---|---|---|
| 1 | 3 | 5 |

A concrete example of the use of this operation in ConQuer-92 would be:

Person who earns a Salary $x$ AND ALSO works for a Company $c$
WHERE $x >$ THE AVERAGE Salary of a Person who works for $c$

Two abbreviations based on the Where operations are:

$$\begin{aligned}
\mathsf{Where}(P_1, C_1; \ldots; P_{l-1}, C_{l-1}; P_l) &\triangleq \mathsf{Where}(P_1, C_1) \cap \ldots \cap \mathsf{Where}(P_{l-1}, C_{l-1})\\
&\cap\ \ \mathsf{Where}(P_l, \neg C_1 \wedge \ldots \wedge \neg C_{l-1})\\
\mathsf{Where}(P_1, C_1; \ldots; P_l, C_l) &\triangleq \mathsf{Where}(P_1, C_1) \cap \ldots \cap \mathsf{Where}(P_l, C_l)
\end{aligned}$$

As each condition is a path-expression we also need an implicit coercion to path-expressions. Let $C$ be a condition, then:

$$\begin{aligned}
\mathbb{P}[\![C]\!]\,(T,B) &\triangleq \mathbb{P}[\![\mathsf{Where}(true, C)]\!]\,(T,B)\\
\mathbb{L}[\![C]\!]\,(T) &\triangleq \left\{\langle u, v\rangle \,\middle|\, u \sim \mathsf{Bool} \wedge v \sim \mathsf{Bool}\right\}
\end{aligned}$$

where Bool denotes the value type for boolean values, and *true* is a constant denotation (and thus a path expression). For example, if $P$ is a path-expression yielding a non-empty result, then the condition $\mathsf{Some}(P)$ is true. Interpreted as a path expression, this condition would then lead to the table:

| *hd* | *tl* |
|---|---|
| *true* | *true* |

Two other simple forms of these operations are of course IF .. THEN .. ELSE .. and IF .. THEN ... An example in ConQuer-92 of the use of this operation (taken from [Hal95]) is:

$$\begin{cases} \text{'P'} & \text{IF } b = 100 \text{ OR } r = 100 \text{ OR } y = 100 \\ \text{'T'} & \text{IF } b > 0 \text{ AND } b > 0 \text{ AND } y > 0 \\ \text{'S'} & \text{OTHERWISE} \end{cases}$$

$$\begin{array}{ll} \text{WHERE} & \text{SOME Color } c \text{ includes blue in Portion: } b \text{ AND} \\ & \text{SOME Color } c \text{ includes red in Portion: } r \text{ AND} \\ & y = 100 - b - r \end{array}$$

Note that the colon symbol after Portion signifies that the value of attribute $b$ is an instance of the value type used to identify Portion. The value of attribute $c$ on the other hand is an instance of the non-value type Color.

What's needed now is a diagram showing the implicit and explicit transitions between path-expressions, scalar-expressions and conditions.

## 6.5 Gathering information

Normally, once a path-expression has been specified one wants a diverse set of information to be returned for the resulting instances. For example, one would like to say:

LIST Budget VIA $g$, Firstname of, Surname given to EACH Person working for Group $g$ part of Department: 'CS'

For this purpose, the confluence operation is used. Since SQL-92 is not able to deal with nested relations, we have changed the definition slightly as opposed to the one used in [HPW93]. If $a_1, \ldots, a_n$ are attributes, $Q_1, \ldots, Q_n$ path-expressions, and $x_1, \ldots, x_n \in \mathsf{Sch} \cdot \mathbb{P}[\![P]\!]$, then the new definition is:

$$\begin{aligned} &\mathbb{P}[\![\, [a_1 : Q_1 : x_1, \ldots, a_n : Q_n : x_n; P] \,]\!]\,(T,B) \triangleq \\ &\quad \pi_X(\mathbb{P}[\![P]\!]\,(T,B) \underset{1 \le i \le n}{\sqsupset\!\bowtie} \alpha_{a_i = hd, x_i = tl}\, \mathbb{P}[\![Q_i]\!]\,(T,B)) \\ &\mathbb{L}[\![\, [a_1 : Q_1 : x_1, \ldots, a_n : Q_n : x_n; P] \,]\!]\,(T) \triangleq \mathbb{L}[\![P]\!]\,(T) \end{aligned}$$

where

$$X = \mathsf{DefMap}\left(\{a_1, \ldots, a_n\} \cup \mathsf{Sch} \cdot \mathbb{P}[\![P]\!]\,(T,B) \cup \bigcup_{1 \le i \le n} \mathsf{Sch} \cdot \mathbb{P}[\![Q_i]\!]\,(T,B)\right)$$

Each time no $x_i$ is provided, $x_i = hd$ will be used as a default. The $x_i$'s are used to connect the $Q_i$'s path-expressions to the $P$. The path $P$ is the main query, whereas the $Q_i$'s are used to 'gather' the required information. The gathered information is merged into the result of the path-expression using attributes $a_i$.

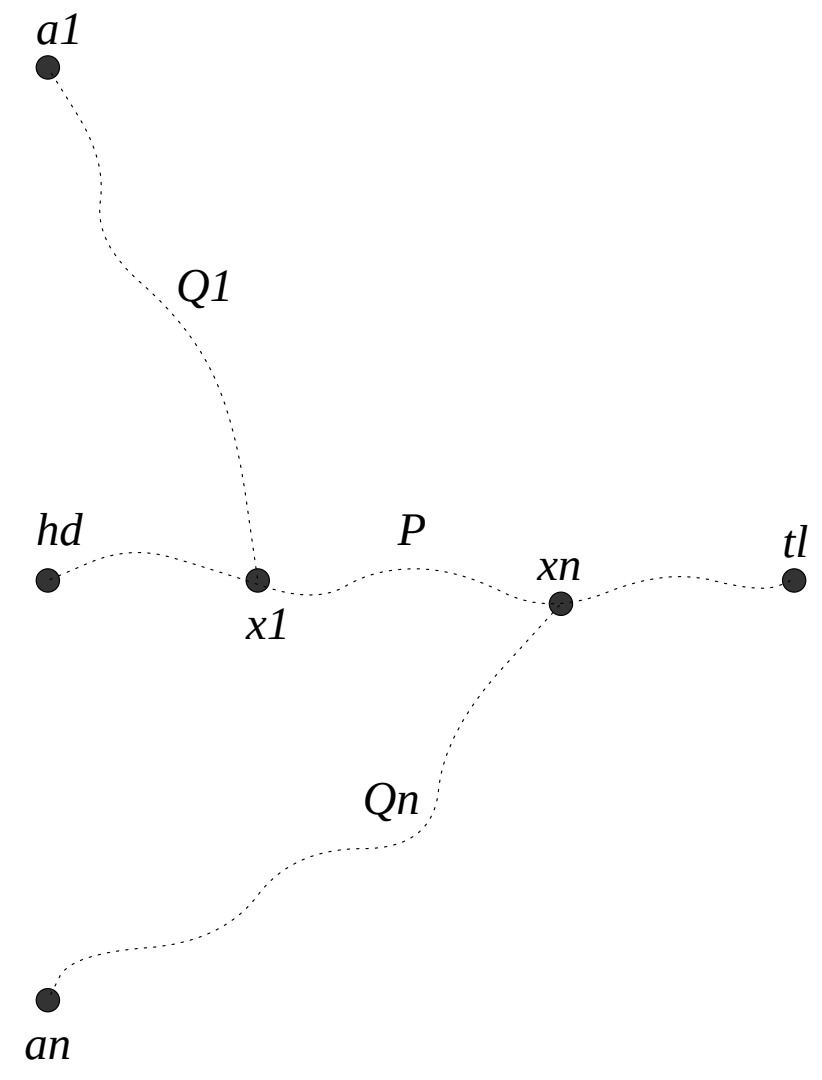

Figure 5: The anatomy of a confluence operation

We do not provide an elaborated example for the confluence operation, but rather limit ourselves to the illustration of what happens. This has been depicted in figure 5. The path-expression $P$ is the base of the query, and the $Q_1$ to $Q_n$ are all pieces of information we are interested in regarding the basic query $P$. The $Q_i$'s are path expressions themselves, and the $x_i$'s provide the connection points between the $Q_i$'s and the query $P$. The final result has to be represented in a table, and the attribute names $a_i$ provide the names of the columns containing the extra information selected by the $Q_i$'s

## 6.6 Group functions

SQL allows for the grouping of relations based on a given set of attribute names by means of the GROUP BY construction. On the result of such a grouping a number of scalar operations can be performed. In the path-expression language we allow for similar constructs. Let $a$ be a new attribute, then:

$$\begin{aligned} \mathbb{P}[\![\mathsf{GCount}(P, X)]\!]\,(T, B) &\triangleq \pi_{hd=a,tl=a}\,\pi_{a=|a|}\,\varphi_X\,\mathbb{P}[\![P]\!]\,(T, B) \\ \mathbb{L}[\![\mathsf{GCount}(P, X)]\!]\,(T) &\triangleq \left\{\langle u, v\rangle \;\middle|\; u \sim \mathsf{Natno} \wedge v \sim \mathsf{Natno}\right\} \end{aligned}$$

$$\mathbb{P}[\![\mathsf{GDsCount}(P,X)]\!]\,(T,B) \triangleq \pi_{hd=a,tl=a}\,\pi_{a=|\mathsf{Ds}(a)|}\,\varphi_X\,\mathbb{P}[\![P]\!]\,(T,B)$$
$$\mathbb{L}[\![\mathsf{GDsCount}(P,X)]\!]\,(T) \triangleq \left\{\langle u,v\rangle \;\middle|\; u \sim \mathsf{Natno} \wedge v \sim \mathsf{Natno}\right\}$$

where $b \in \mathsf{Sch}\cdot\mathbb{P}[\![P]\!]\,(T,B) - X$ is an arbitrarily chosen attribute. Let $a \in \mathsf{Sch}\cdot\mathbb{P}[\![P]\!]\,(T,B)$, then we can also define.

$$\mathbb{P}[\![\mathsf{GSum}(P,X,a)]\!]\,(T,B) \triangleq \pi_{hd=a,tl=a}\,\pi_{a=\mathrm{sum}(a)}\,\varphi_X\,\mathbb{P}[\![P]\!]\,(T,B)$$
$$\mathbb{P}[\![\mathsf{GDsSum}(P,X,a)]\!]\,(T,B) \triangleq \pi_{hd=a,tl=a}\,\pi_{a=\mathrm{sum}(\mathsf{Ds}(a))}\,\varphi_X\,\mathbb{P}[\![P]\!]\,(T,B)$$
$$\mathbb{P}[\![\mathsf{GMin}(P,X,a)]\!]\,(T,B) \triangleq \pi_{hd=a,tl=a}\,\pi_{a=\min(a)}\,\varphi_X\,\mathbb{P}[\![P]\!]\,(T,B)$$
$$\mathbb{P}[\![\mathsf{GMax}(P,X,a)]\!]\,(T,B) \triangleq \pi_{hd=a,tl=a}\,\pi_{a=\max(a)}\,\varphi_X\,\mathbb{P}[\![P]\!]\,(T,B)$$
$$\mathbb{P}[\![\mathsf{GAvg}(P,X,a)]\!]\,(T,B) \triangleq \pi_{hd=a,tl=a}\,\pi_{a=\mathrm{sum}(a)/|a|}\,\varphi_X\,\mathbb{P}[\![P]\!]\,(T,B)$$

For each $F \in \{\mathsf{GSum}, \mathsf{GMin}, \mathsf{GMax}, \mathsf{GAvg}\}$ we have:

$$\mathbb{L}[\![F(P,X,a)]\!]\,(T) \triangleq \left\{\langle u,v\rangle \;\middle|\; u \sim T(a) \wedge v \sim T(a)\right\}$$

The grouping functions in the path-expression language operate in the same way as in SQL. Therefore we will not provide an elaborated example of their semantics.

## 6.7 Sub-expressions

Sub-expressions are a simple, yet neat, way of introducing limitations on (linear) path-expressions. The sub-expression concept is basically an abbreviation which is introduced as:

$$[Q_1,\ldots,Q_n] \triangleq \mathsf{Ds}(\mathsf{Fr}(Q_1)\cap\ldots\cap\mathsf{Fr}(Q_n))$$

This allows us to write $P\circ[Q_1,\ldots,Q_n]\circ R$. The *line* of $P \circ R$ is not disturbed by the sub-expression. The sub-expression operates like a filter. When verbalising, the sub-expression corresponds to a limitation that is put between parenthesis in natural language. As we will be using parenthesis for disambiguation purposes, we suggest the use of the symbol [ and ] to designate a sub-expression. As an example consider:

The Person who is a Coworker [during the Year: 1994] of the Company: "Asymetrix"

which corresponds to the intuitive formulation:

> Deep thought. O ye smartest of computers. Shiniest machine of all. Please,
>
> list the person who is a coworker, during the year 1994, of
>
> the company named 'Asymetrix'

## 6.8 Denotations

For denotations we introduce an extra set of abbreviations that allow us to write compact denotations of instances, in particular for compositely identified types. Before introducing these abbreviations we first need to introduce the syntactic category of path expression denotations (`PEDenotations`). The following two rules are all rules to build these denotations:

1. if $p \in$ `PathExpr`, then $p \in$ `PEDenotations`

   Any normal path-expression can be used to identify a simply identified object type.

2. if $a \in$ Attrs, then $!a \in$ `PEDenotations`

   Sometimes we want to introduce attribute names that represent the abstract instance rather than the concrete values. For example, in Person: $!x$, the $x$ will be a variable of type Person, whereas in Person: $x$ will be a variable of type PersonName (presuming a person is identified by a single name).

3. if $d_1, \ldots, d_1 \in$ `PEDenotations` then $(d_1, \ldots, d_1) \in$ `PEDenotations`.

   For the compositely identified object types, we can simply combine existing instance denotations by making a list of them. Naturally, the order is dictated by the order specified by the reference scheme for the type at hand.

We can now formally introduce the denotations into the path-expression language. We actually do not have to introduce extra language constructs; the denotations are simply a standard abbreviation mechanism.

For value types we have the simplest form of denotation. If $x \in \mathcal{VL} \wedge p \in$ `InfDiscr`, then we have:

$$x : p \triangleq x \circ p$$

For the introduction of abstract attributes we have the following abbreviation. If $x \in \mathcal{TP} \wedge a \in$ Attrs then:

$$x : \,!a \triangleq x \circ a$$

Each object type with a reference scheme defined for it (which could be a simple one:one reference schema for the simply identified object types), the following abbreviation is introduced. For $\mathsf{Idf}(x) = [\langle p_1, q_1 \rangle, \ldots, \langle p_l, q_l \rangle] \wedge d_1, \ldots, d_l \in$ `PEDenotations`, we have:

$$x : (d_1, \ldots, d_l) \triangleq x \circ [\, p_1 \circ q_1^{\leftarrow} \circ \mathsf{Player}(q_1) : d_1, \ldots, p_l \circ q_l^{\leftarrow} \circ \mathsf{Player}(q_l) : d_l \,]$$

If $l = 1$, the parenthesis may be omitted leading to $x : d$; so for instance to Person: $x$, or Person: 'Erik'. Finally for relationship types, we have when $\mathsf{Idf}(x) = [p_1, \ldots, p_l] \wedge d_1, \ldots, d_l \in$ `Denotations`:

$$x : (d_1, \ldots, d_l) \triangleq x \circ [\, p_1^{\leftarrow} \circ \mathbb{D}[\![d_1]\!], \ldots, p_l^{\leftarrow} \circ \mathbb{D}[\![d_l]\!] \,]$$

Again, if $l = 0$, the parenthesis may be omitted.

## 6.9 Macro mechanism

In ConQuer-92, we allow for macro definitions of path-expressions. Currently, however, we do not support (mutually) recursive macro definitions. In a next version recursive definitions will be allowed, however, since the SQL-92 standard does not allow for recursive queries, it does not make sense to support recursive definitions yet. For recursive macros the advanced concept of transitive closure of fix-point is required, and most current SQL implementations are not even capable of supporting SQL-92, let alone SQL-3.

The macros are presumed to be provided by a function:

$$\mathsf{Macros} : \mathsf{Names} \times \mathsf{Attrs}^{+} \rightarrow \texttt{PathExpr}$$

A macro definition is denoted as:

$$\alpha(a_1, \ldots, a_n) \ ::= \ E$$

where $a_i$'s are attributes. An example macro definition in ConQuer-92 would be:

SubTotal(*inv, itm*) ::=
  (Quantity of InvoiceLine: *inv, itm*) * (Unitprice of InvoiceLine: *inv, itm*)

Macros are integrated into the language by the three rules below. Three rules are needed as we allow for macro definitions of conditions, scalar-expressions, and path-expressions in general. Let $\mathsf{Macros}(\alpha, [a_1, \ldots, a_n]) = Q$ be a macro definition, then the path-expression language can be extended according to the following three rules:

1. If $Q \in \texttt{PEScalarExpr}$ and $\mathsf{Attr}[\![Q]\!] = \{a_1, \ldots, a_n\}$, then we have for scalar-expressions $E_1, \ldots, E_n$:

$$\alpha(E_1, \ldots, E_n) \ \triangleq \ Q|^{a_1, \ldots, a_n}_{E_1, \ldots, E_n}$$

2. If $Q \in \texttt{PEConditions}$ and $\mathsf{Attr}[\![Q]\!] = \{a_1, \ldots, a_n\}$, then we have for conditions $C_1, \ldots, C_n$:

$$\alpha(C_1, \ldots, C_n) \ \triangleq \ Q|^{a_1, \ldots, a_n}_{C_1, \ldots, C_n}$$

3. If $Q \in (\texttt{PathExpr} - \texttt{PEScalarExpr} - \texttt{PEConditions})$ (in other words, $Q$ is a 'normal' path-expression), and $\mathsf{Attr}[\![Q]\!] = \{a_1, \ldots, a_n\}$, and $b_1, \ldots, b_n$ are fresh attributes, and $P_1, \ldots, P_n$ are path-expressions, then we have:

$$\alpha(P_1, \ldots, P_n) \ \triangleq \ Q|^{a_1, \ldots, a_n}_{Y_1, \ldots, Y_n}$$

   where $Y_i = (b_i \circ P_i)$.

As an example, $\mathsf{SubTotal}(L_1, L_2)$ would lead to:

$$(\mathsf{Quantity\ of\ InvoiceLine}\colon\ b_1 \circ L_1,\ b_2 \circ L_2) * (\mathsf{Unitprice\ of\ InvoiceLine}\colon\ b_1 \circ L_1,\ b_2 \circ L_2)$$

This example also illustrates why, in the case of a 'normal' path-expression, we need to introduce the fresh attribute names $b_i$'s. If an instance of $L_1$ is selected as an invoice number, then it is now enforced that this number is the same for the quantity and unit price parts of the path-expression.

An expression of the form $Q|^{a_1,\ldots,a_n}_{b_1,\ldots,b_n}$ results in an expressions $Q'$ in which all occurences of $a_i$ have been replaced by $b_i$ (for each $1 \le i \le n$). If there are $i < j$ such that $a_i = a_j$, then $a_i$ is replaced by $b_i$ only!

## 6.10 The typing function

At the beginning of this section we discussed the importance of the typing relation, which determines of which type(s) the instances of attributes may be. In this subsection we look at how we can derive this typing information by examining the parse tree of a path-expression.

Given a path-expression $P$, the typing function $\mathbb{T}[\![P]\!]$ should search the parse tree of $P$ for the following patterns:

| Pattern | Typing |
|---|---|
| $x \circ a$ | $\langle a, x \rangle$ |
| $a \circ x$ | $\langle a, x \rangle$ |
| $p \circ a$ | $\langle a, \mathsf{Rel}(p) \rangle$ |
| $a \circ p$ | $\langle a, \mathsf{Player}(p) \rangle$ |
| $p^{\leftarrow} \circ a$ | $\langle a, \mathsf{Player}(p) \rangle$ |
| $a \circ p^{\leftarrow}$ | $\langle a, \mathsf{Rel}(p) \rangle$ |

where $p \in \mathcal{RO}$, $x \in \mathcal{TP}$ and $a \in \mathsf{Attrs}$. These patterns are typical ways to bind attributes in path-expressions to an underlying domain. In the right hand side of the column we have provided the derived typing for the attribute.

Since it is not hard to write an algorithm to search a parse tree of a path expression for the above patterns, we do not provide a more detailed formalisation.

## 6.11 Derivation Rules

In InfoModeler there is a clear need for the ability to define formalised derivation rules which can be translated to an SQL statement. In this subsection we introduce two classes of derivation rules.

The first class deals with the definition of derived relationship types. If $f$ is a (derivable) relationship type with $\mathsf{Roles}(f) = \{p_1, \dots, p_n\}$, $P$ a path-expression and $a_1, \dots, a_n$ are attributes, then:

$$f(p_1 : a_1, \dots, p_n : a_n) ::= P$$

is a derivation rule for relationship type $f$ with semantics (in the context of a population $\mathsf{Pop}$):

$$\mathsf{Pop}'(f) \triangleq \mathsf{Val}[\![\pi_{p_1=a_1,\dots,p_n=a_n} \mathbb{P}[\![P]\!](T, \varnothing)]\!](\mathsf{Pop}, \varnothing)$$

where typing $T$ is defined by:

$$T = \mathbb{T}[\![P]\!] \cup \{\langle a_i, \mathsf{Player}(p_i)\rangle \mid 1 \leq i \leq n\}$$

For obvious reasons, the typing requirement:

$$\langle a, x\rangle, \langle a, y\rangle \in T \Rightarrow x \sim y$$

should hold.

The population $\mathsf{Pop}$ is the existing population of the database, whereas $\mathsf{Pop}'$ is the updated population after applying the derivation rule. An example of what this will look like in ConQuer-92 is:

A Product $p$ has a taxed price of MoneyAmt $a$ IFF
MoneyAmt $a = 1.5$ * the MoneyAmt that is the ex tax price of a Product $p$

The second class of derivation rules are concerned with ordinary (non relationship type) derivable object types. If $t \in \mathcal{TP} - \mathcal{RL}$ is a type, and $P$ a path-expression, then:

$$t ::= P$$

is a derivation rule for type $t$ with semantics:

$$\mathsf{Pop}'(t) \triangleq \{i(hd) \mid i \in \mathsf{Val}[\![\mathbb{P}[\![P]\!](T, B)]\!](\mathsf{Pop}, \varnothing)\}$$

where typing $T$ is derived by:

$$T = \mathbb{T}[\![P]\!] \cup \{\langle a, t\rangle\}$$

with the typing requirement:

$$\langle b, x\rangle, \langle b, y\rangle \in T \Rightarrow x \sim y$$

In ConQuer-92 this would for instance lead to:

EACH Town-or-village IS a Community that has a Population <= 100000

A special use of these derivation rules is of course for subtype definitions.

### 6.12 Constraints

Path-expressions, in particular that subset of the path-expressions that allows us to formulate conditions, can be used to denote textual constraints. If $P \in \texttt{PathExpr}$ is a path expression and Pop a population, then it can be evaluated as a constraint by requiring:

$$\mathsf{Val}[\![\mathbb{P}[\![P]\!]\,(\mathbb{T}[\![P]\!],\varnothing)]\!]\,(\mathsf{Pop},\varnothing) \neq \varnothing$$

The condition $P$ can now be interpreted as a constraint on population $P$, which necessarily needs to return a non-empty result.

### 6.13 Ordering the result of a path expression

The final aspect of the path-expression language we need to discuss is the ability to order results. Ordering is not a part of the path-expressions themselves, as ordering results in merely a presentation issue from a query language point of view. Therefore we provide a small extension of the path-expressions language, which is not a path-expression itself. In the next section, where we introduce ConQuer-92 itself, we will discuss this in some more detail.

The result of a path-expression $P$ can be sorted by the following operation:

$$\Omega(P, a_1 : o_1; \ldots; a_n : o_n)$$

where $a_i \in \mathsf{Attrs}$ and $o_i \in \{\texttt{Asc}, \texttt{Desc}\}$. The semantics of these functions will not be provided in full detail here, but their definition is rather obvious

# 7 Syntax and Semantics of ConQuer-92

We have now finally reached the stage where we can start working on more readable versions of path-expressions. To this end we first introduce naming functions for the concepts from the conceptual schema used to build path-expressions. These functions are the additional requirements on the meta-model with respect to verbalisations. This means that the factbase structure must be extended accordingly.

Similarly to the path-expression level, the ConQuer level is initially specified using abstract syntax. In appendix B the actual concrete syntax of ConQuer-92 is given. The relation between the ConQuer level and the path-expressions level is that ConQuer provides the textual surface structure of the path-expressions while the path-expressions represent the deeper structures.

## 7.1 Named concepts

Most of the concepts used to build ORM schemas will receive a name of some form. The well known ones are types and roles. With respect to the conceptual schema, path-expressions only require names for these two concepts and combinations thereof.

It should be noted that a role name is *not* the same as the predicate. A role name described the role played by the player in the relationship type. For the types and roles we introduce: $\mathsf{TNm} : \mathcal{OB} \to \Sigma^+$ giving the names of types, and $\mathsf{PNm} : \mathcal{RO} \rightarrowtail \Sigma^+$ giving the names of roles. The type names must be unique, so we should have:

$$\mathsf{TNm}(x) = \mathsf{TNm}(y) \Rightarrow x = y$$

However, this does not have to hold for role names. Nevertheless, their names must be unique within one relationship type:

$$\mathsf{PNm}(x) = \mathsf{PNm}(y) \Rightarrow (x = y \vee \mathsf{Rel}(x) \neq \mathsf{Rel}(y))$$

Note that not every role must necessarily receive a name. Roles have two additional ways of receiving a name. Especially when dealing with objectified relationship types, it turns out to be useful to have reversed role names, i.e. connecting the relationship type to the player rather than the other way around. These reverse names are provided by: $\mathsf{RNm} : \mathcal{RO} \rightarrowtail \Sigma^+$. We will use the role names for the verbalisation of a role entry, while a reversed role name is used for a role exit. These reverse names must be unique within one fact type as well:

$$\mathsf{RNm}(x) = \mathsf{RNm}(y) \Rightarrow (x = y \vee \mathsf{Rel}(x) \neq \mathsf{Rel}(y))$$

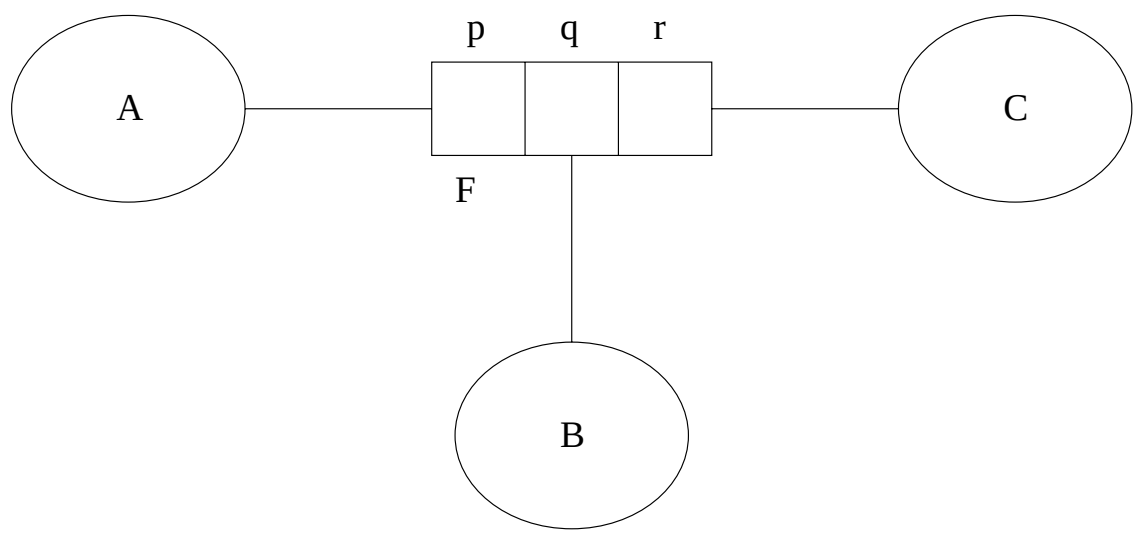

Figure 6: Providing names for schema concepts

As an example, consider the schema depicted in figure 6. For this schema we could have:

| | | |
|---|---|---|
| $\mathsf{TNm}(A) =$ President | $\mathsf{PNm}(p) =$ has | $\mathsf{RNm}(p) =$ are of |
| $\mathsf{TNm}(B) =$ Result | $\mathsf{PNm}(q) =$ are in | $\mathsf{RNm}(q) =$ has as |
| $\mathsf{TNm}(C) =$ Election | $\mathsf{PNm}(r) =$ has | $\mathsf{RNm}(r) =$ is for |
| $\mathsf{TNm}(F) =$ Election-result | | |

This will allow us to verbalise $A \circ p \circ F \circ q^{\leftarrow} \circ B$ as: President has Election-result has as Result. This is already much better readable than an SQL version would be. However, we will add extra names to improve further on such verbalisations.

To allow for more elegant verbalisations we allow a user to provide prefixes and postfixes to type names. The prefixes and postfixes provide the glue to better connect partial verbalisations. So an object type Person might receive as a prefix the and as a postfix who, leading to the Person who .... However, in some cases one would like to express the fact that the person is not determined, i.e. one would like to say a Person who .... To this end we allow for two kinds of prefixes and postfixes. One for the determined class and one for the undetermined class. In a later stage these classes can be extended with plural and singular cases as well. For the moment the prefixes and postfixes are provided by: $\mathsf{Pre} : \mathcal{TP} \times \{\mathtt{undetermined}, \mathtt{determined}\} \to \Sigma^+$, and $\mathsf{Post} : \mathcal{TP} \to \Sigma^+$. Note that in the English language the determined and undetermined postfix will usually be the same. However, in other languages this might not be the case. For our running example the prefixes and postfixes could be:

$$\begin{array}{ll} \mathsf{Pre}(A, \mathtt{undetermined}) = \mathsf{some} & \mathsf{Post}(A) = \mathsf{who} \\ \mathsf{Pre}(A, \mathtt{determined}) = \mathsf{the} & \\ \mathsf{Pre}(B, \mathtt{undetermined}) = \mathsf{some} & \mathsf{Post}(B) = \mathsf{which} \\ \mathsf{Pre}(B, \mathtt{determined}) = \mathsf{the} & \\ \mathsf{Pre}(C, \mathtt{undetermined}) = \mathsf{some} & \mathsf{Post}(C) = \mathsf{that} \\ \mathsf{Pre}(C, \mathtt{determined}) = \mathsf{the} & \\ \mathsf{Pre}(F, \mathtt{undetermined}) = \mathsf{some} & \mathsf{Post}(F) = \mathsf{that} \\ \mathsf{Pre}(F, \mathtt{determined}) = \mathsf{the} & \end{array}$$

For $A \circ p \circ F \circ q^{\leftarrow} \circ B$ we could now have: A president who has an Election-result that has as a Result This verbalisation can be improved even further. Most connections between two object types will involve two roles only. In our case, $A \circ p \circ F \circ q^{\leftarrow} \circ B$ object type $A$ and $B$ are connected via roles $p$ and $q$ (of relationship type $F$). For this frequently occurring class of connections the mix-fix predicate verbalisations can be used. The binary case as given above is the most commonly occurring case, but these verbalisations are also useful for ternaries and quarternaries. Formally, these mix-fix predicates are provided by the relation:

$$\mathsf{MFix} \subseteq \mathcal{RL} \times (\Sigma^*)^+ \times \mathsf{Roles}^+$$

For this naming function we must have:

$$\mathsf{MFix}(f, [\alpha_1, \ldots, \alpha_k], [p_1, \ldots, p_l]) \Rightarrow k + 1 = l$$

requiring that the number of 'gaps' in the verbalisation corresponds to the number of roles. Since a relationship type can have multiple mix-fix predicate verbalisations associated MFix is a relationship.

In the case of a relationship with an arity of higher than one we can also have mix-fix predicates spanning only part of the relationship type. It is obvious that these names allow us to formulate elegant paths through relationship types, including ones of higher arity.

For our running example we could have the following mix-fix predicate verbalisations:

$$\begin{array}{l}
\textsf{MFix}(F, [\textsf{has participated in}, \textsf{leading to}], [p, q, r]) \\
\textsf{MFix}(F, [\textsf{has received a}, \textsf{from the participation in}], [p, r, q]) \\
\textsf{MFix}(F, [\textsf{of the participation by}, \textsf{in}], [q, p, r]) \\
\textsf{MFix}(F, [\textsf{of the participation in}, \textsf{by}], [q, r, p]) \\
\textsf{MFix}(F, [\textsf{has participation of}, \textsf{leading to}], [r, p, q]) \\
\textsf{MFix}(F, [\textsf{has lead to}, \textsf{for the participation of}], [r, q, p]) \\
\textsf{MFix}(F, [\textsf{has participated in an election leading in}], [p, q]) \\
\textsf{MFix}(F, [\textsf{has participated in}], [p, r]) \\
\textsf{MFix}(F, [\textsf{of the participation in an election by}], [q, p]) \\
\textsf{MFix}(F, [\textsf{of the participation in}], [q, r]) \\
\textsf{MFix}(F, [\textsf{has participation of}], [r, p]) \\
\textsf{MFix}(F, [\textsf{has lead to}], [r, p])
\end{array}$$

It goes without saying that not all combinations of roles within one relationship type have to be named by the users. Only for role-sequences that are used frequently a mix-fix predicate verbalisation makes sense. Our running example could now be verbalised as:

Some person has participated in an election leading to some Result which is in our opinion reasonably readable.

## 7.2 Basic information descriptors

In this subsection we define the syntactic category of information descriptors. The information descriptors in ConQuer-92 correspond directly to the path-expressions on the path-expression level. The semantics of ConQuer-92 information descriptors is defined by a function:

$$\mathbb{I} : \texttt{InfDiscr} \rightarrow \texttt{PathExpr}$$

This function does not take any extra parameters besides the ConQuer information descriptor since this translation is just a matter of syntactic re-writing.

The names for types, and their prefixes lead to:

**[P1]** $\textsf{TNm}(x) = n \quad \vdash \quad \mathbb{I}[\![n]\!] = x$

**[P2]** $\textsf{Pre}(x, c) = r \wedge \textsf{TNm}(x) = n \quad \vdash \quad \mathbb{I}[\![r\ n]\!] = x$

In the example we have $\textsf{TNm}(A) = \textsf{President}$, so we have $\mathbb{I}[\![\textsf{President}]\!] = A$. Furthermore, since $\textsf{Pre}(A, \texttt{undetermined}) = \textsf{some}$ and $\textsf{Pre}(A, \texttt{determined}) = \textsf{the}$, we have

$$\mathbb{I}[\![\textsf{some Person}]\!] = \mathbb{I}[\![\textsf{the Person}]\!] = x$$

Note that in parsing we do not care at all about the determined or undetermined cases for pre- and post-fixes (hence the free variable $c$). In section 8, where we consider verbalisations, we will indeed take this into consideration.

Attributes which are used as variables in ConQuer-92 information descriptors receive a name via the function $\mathsf{VNm} : \mathsf{Attrs} \rightarrowtail \Sigma$. Named attributes must have a unique name, so this function must adhere to:

$$\mathsf{VNm}(x) = \mathsf{VNm}(y) \Rightarrow x = y$$

Since ConQuer does have a limited form of scoping (the Path function is the only operation leading to scoping of variables), we could reduce this claim that for all attributes within a scope the above uniqueness of variable names must hold.

For using variables in the context of types we have the following rules:

**[P3]** $\mathsf{TNm}(x) = n \wedge \mathsf{VNm}(a) = m \quad \vdash \quad \mathbb{I}[\![n\ m]\!] = x \circ a$

**[P4]** $\mathsf{Pre}(x, c) = r \wedge \mathsf{TNm}(x) = n \wedge \mathsf{VNm}(a) = m \quad \vdash \quad \mathbb{I}[\![r\ n\ m]\!] = x \circ a$

If $a$ is an attribute with name $p$, then for our running example we have:

$$\mathbb{I}[\![\mathsf{some\ Person}]\!] = \mathbb{I}[\![\mathsf{the\ Person}]\!] = \mathbb{I}[\![\mathsf{Person}\ p]\!] = A \circ a$$

The names given to roles and reversed roles lead to the following four rules:

**[P5]** $\mathsf{PNm}(p) = n \quad \vdash \quad \mathbb{I}[\![n]\!] = p$

**[P6]** $\mathsf{PNm}(p) = n \wedge \mathsf{Post}(\mathsf{Player}(p)) = o \quad \vdash \quad \mathbb{I}[\![o\ n]\!] = p$

**[P7]** $\mathsf{RNm}(p) = n \quad \vdash \quad \mathbb{I}[\![n]\!] = p^{\leftarrow}$

**[P8]** $\mathsf{RNm}(p) = n \wedge \mathsf{Post}(\mathsf{Rel}(p)) = o \quad \vdash \quad \mathbb{I}[\![o\ n]\!] = p^{\leftarrow}$

For the running example we therefore have:

$$\mathbb{I}[\![\mathsf{has}]\!] = \mathbb{I}[\![\mathsf{who\ has}]\!] = p \text{ and } \mathbb{I}[\![\mathsf{that\ are\ of}]\!] = \mathbb{I}[\![\mathsf{are\ of}]\!] = p^{\leftarrow}$$

## 7.3 Complex information descriptors

In this section we introduce the operations that can be used to build more complex information descriptors. The most important construct is concatenation:

**[P9]** $I, J \in \texttt{InfDiscr} \quad \vdash \quad \mathbb{I}[\![I \; J]\!] = \mathbb{I}[\![I]\!] \circ \mathbb{I}[\![J]\!]$

This construct allows us to concatenate:

$$\begin{array}{lcl} \mathbb{I}[\![\textsf{a Person}]\!] & = & A \\ \mathbb{I}[\![\textsf{who has}]\!] & = & p \\ \mathbb{I}[\![\textsf{an Election-result}]\!] & = & F \\ \mathbb{I}[\![\textsf{that has as}]\!] & = & q^{\leftarrow} \\ \mathbb{I}[\![\textsf{a Result}]\!] & = & B \end{array}$$

to: $\mathbb{I}[\![\textsf{a Person who has an Election-result that has as a Result}]\!] = A \circ p \circ F \circ q^{\leftarrow} \circ B$.

To explicitly cater for subtypes and their selective effects on the result of a path expression, we introduce the IS construct:

**[P10]** $\mathbb{I}[\![I]\!] = P \circ xE \wedge \mathbb{I}[\![J]\!] = yF \circ Q \wedge x \sim y \quad \vdash \quad \mathbb{I}[\![I \; \textsf{IS} \; J]\!] = P \circ xE \circ yF \circ Q$

where $E$ and $F$ may be denotations or empty. As an example, let object type Person have two (overlapping) subtypes: Employee and Student. For Employees we store their salary, and for Persons we store their hobbies. Then we could for instance have:

Some Salary earned by some Employee who IS a Student who has as Hobby: "Cycling"

This expression has a differing semantics from:

Some Salary earned by some Employee who has as Hobby

as it does not limit the Employees to Persons.

The mix-fix predicate verbalisations are integrated into ConQuer-92 by the following definition:

**[P11]** $I_2, \ldots, I_{l-1} \in \texttt{InfDiscr} \wedge \textsf{MFix}(r, [\alpha_1, \ldots, \alpha_{l-1}], [p_1, \ldots, p_l]) \quad \vdash$
$$\mathbb{I}[\![\alpha_1 \; I_2 \; \ldots \; I_{l-1} \; \alpha_{l-1}]\!] = \langle p_1, p_2 : \mathbb{I}[\![I_2]\!], \ldots, p_{l-1} : \mathbb{I}[\![I_{l-1}]\!], p_l \rangle$$

to cater for any post-fix for the player of role $p_1$, we also introduce:

**[P12]** $I_2, \ldots, I_{l-1} \in \texttt{InfDiscr} \wedge \textsf{MFix}(r, [\alpha_1, \ldots, \alpha_{l-1}], [p_1, \ldots, p_l]) \wedge \textsf{Post}(p) = o \quad \vdash$
$$\mathbb{I}[\![o \; \alpha_1 \; I_2 \; \ldots \; I_{l-1} \; \alpha_{l-1}]\!] = \langle p_1, p_2 : \mathbb{I}[\![I_2]\!], \ldots, p_{l-1} : \mathbb{I}[\![I_{l-1}]\!], p_l \rangle$$

The path resulting from a mix-fix predicate verbalisation always leads from the head of $p_1$ to the tail of $p_l^{\leftarrow}$, so from $\textsf{Player}(p_1)$ to $\textsf{Rel}(p_l)$.

Path reversal on the information descriptor level is verbalised as:

**[P13]** $I \in \texttt{InfDiscr} \quad \vdash \quad \mathbb{I}[\![\textsf{THE REVERSE OF } I]\!] = \mathbb{I}[\![I]\!]^{\leftarrow}$

The front elements of an information descriptor result from:

**[P14]** $I \in \texttt{InfDiscr} \quad \vdash \quad \mathbb{I}[\![\textsf{ONLY } I]\!] = \textsf{Fr}\,\mathbb{I}[\![I]\!]$

A multiset result can be coerced to a set by:

**[P15]** $I \in \texttt{InfDiscr} \quad \vdash \quad \mathbb{I}[\![\textsf{DISTINCT } I]\!] = \textsf{Ds}\,\mathbb{I}[\![I]\!]$

All binary operations of the path-expression level are present on the information descriptor level. To this end we first define the relation SetOper as:

| ConQuer-92 Operation | Path Expression Operator |
|---|---|
| UNITED WITH | $\cup$ |
| INTERSECTED WITH | $\cap$ |
| MINUS | $-$ |
| WHICH ARE ALL IN | $\subseteq$ |
| THAT INCLUDES ALL | $\supseteq$ |
| MATCHING ALL | $\equiv$ |
| MISSING | $\underline{\otimes}$ |
| WITH | $\times$ |

These operations are now integrated in ConQuer-92 by:

**[P16]** $I, J \in \texttt{InfDiscr} \wedge \textsf{SetOper}(o, O) \quad \vdash \quad \mathbb{I}[\![I \; o \; J]\!] = \mathbb{I}[\![p]\!] \; O \; \mathbb{I}[\![q]\!]$

The union, intersection, and set difference operations will usually be to restrictive since the intersection is applied to at least the head and tail combinations. So if $P$ and $Q$ result in:

$P$:

| hd | tl |
|---|---|
| 1 | 2 |
| 2 | 3 |
| 2 | 4 |
| 1 | 8 |
| 3 | 4 |

$Q$:

| hd | tl |
|---|---|
| 1 | 2 |
| 2 | 9 |
| 8 | 3 |
| 3 | 1 |

would lead for $P$ INTERSECTED WITH $Q$ to:

| hd | tl |
|---|---|
| 1 | 2 |

In general we want such operations to be applied to the head elements of paths only. Therefore we introduce the FrSetOper operations:

| ConQuer-92 Operation | Path Expression Operator |
|---|---|
| OR OTHERWISE | $\rlap{\cup}{f}$ |
| AND ALSO | $\rlap{\cap}{f}$ |
| BUT NOT | $\rlap{-}{f}$ |

For these operations we have:

**[P17]** $I, J \in \texttt{InfDiscr} \wedge \mathsf{FrSetOper}(o, O) \quad \vdash \quad \mathbb{I}[\![I\ o\ J]\!] = \mathbb{I}[\![I]\!]\ O\ \mathbb{I}[\![J]\!]$

For $P$ AND ALSO $Q$ we would have:

| hd | tl |
|---|---|
| 1 | 1 |
| 2 | 2 |
| 3 | 3 |

The values in the result of information descriptors can be compared to each other using relational operators like $<, \leq, \geq$. For this purpose we introduce the ValueComp operations, which is identified as:

| ConQuer-92 Operation | Path Expression Operator |
|---|---|
| $=$ | $=$ |
| IS EQUAL TO | $=$ |
| $<>$ | $\neq$ |
| IS NOT EQUAL TO | $\neq$ |
| $<$ | $<$ |
| IS LESS THAN | $<$ |
| $<=$ | $\leq$ |
| IS LESS THAN OR EQUAL TO | $\leq$ |
| $>$ | $>$ |
| IS GREATER THAN | $>$ |
| $>=$ | $\geq$ |
| IS GREATER THAN OR EQUAL TO | $\geq$ |

For this class of operations we have:

**[P18]** $I, J \in \texttt{InfDiscr} \wedge \mathsf{ValueComp}(o, O) \quad \vdash \quad \mathbb{I}[\![I\ o\ J]\!] = (\mathsf{TlCoerce}\,\mathbb{I}[\![I]\!])\ O\ (\mathsf{HdCoerce}\,\mathbb{I}[\![J]\!])$

We are now able to write:

a Person who earns a Salary $>$ Salary that is earned by Person: 'Erik'

Finally, we introduce an *inline* projection operation on information descriptors that allows us to do projections of existing information descriptor, while still treating the result as an information descriptor.

**[P19]** $I \in \texttt{InfDiscr} \land \forall_{1 \leq i \leq l} [\mathsf{VNm}(a_i) = n_i] \quad \vdash$
$$\mathbb{I}[\![\mathsf{THE\ PATH\ FROM}\, n_1\, \mathsf{VIA}\, n_2, \ldots, n_{l-1}\, \mathsf{TO}\, n_l\, \mathsf{OF}\, I]\!] = \mathsf{Path}(\mathbb{I}[\![I]\!], a_1, \ldots, a_n)$$

Note that $n$ must be larger than 1, and that if $n = 2$ the VIA subclause should be dropped. Examples are:

THE PATH FROM $c$ VIA $s$ TO $p$ OF
Person $p$
( who works for a Company $c$ that is based in the Country: 'USA'
AND ALSO
who earns a Salary $s$)

leading to an information descriptor (with an underlying path-expression) that has as head column $c$ and tail column $p$.

## 7.4 Scalar expressions

On this level we obviously also have scalar-expressions. They are provided by the set `Expressions`, and the semantics are provided as: $\mathbb{X} : \texttt{Expressions} \rightarrow \texttt{PEScalarExpr}$. Again, this is just a syntactic re-write function.

The first class of functions on expressions is contained in Function, which is defined by:

| ConQuer-92 Function | Function |
|---|---|
| THE COUNT OF | Count |
| THE SUM OF | Sum |
| THE MINIMUM | Min |
| THE MAXIMUM | Max |
| THE AVERAGE | Avg |

**[P20]** $I \in \texttt{InfDiscr} \land \mathsf{Function}(f, F) \quad \vdash \quad \mathbb{X}[\![f\ I]\!] = F(\mathsf{HdCoerce}\, \mathbb{I}[\![I]\!])$

An example would be:

THE SUM OF Salary earned by a Person working for the Company: 'Asymetrix'

**[P21]** $c \in \cup\mathcal{D} \quad \vdash \quad \mathbb{X}[\![c]\!] = c$

Formally there should be a difference between a constant of the path-expression level, which is an abstract mathematical object, and its verbalisation on the information descriptor level. It is only for pragmatic reasons that we ignore this distinction in our current formalisation.

**[P22]** $\mathsf{VNm}(a) = m \quad \vdash \quad \mathbb{X}[\![m]\!] = a$

**[P23]** $\mathsf{VNm}(a) = m \wedge \mathsf{RNm}(p) = r \quad \vdash \quad \mathbb{X}[\![m.r]\!] = a.p$

Note that these last two rules only make sense when the variable $m$ is properly bound to a type already.

Functions and operations are introduced for ConQuer-92 expressions as:

**[P24]** $E_1, \ldots, E_l \in \texttt{Expressions} \wedge f$ is a function symbol $\vdash$
$\mathbb{X}[\![f\,(\,E_1\,,\ldots,E_l\,)]\!] = f(\mathbb{X}[\![E_1]\!], \ldots, \mathbb{X}[\![E_l]\!])$

**[P25]** $E_1, E_2 \in \texttt{Expressions} \wedge$ $f$ is an infix operator symbol $\vdash$
$\mathbb{X}[\![E_1\ f\ E_2]\!] = f(\mathbb{X}[\![E_1]\!], \mathbb{X}[\![E_2]\!])$

Similarly to the path expression level, each expression is also an information descriptor:

**[P26]** $E \in \texttt{Expressions} \quad \vdash \quad \mathbb{I}[\![E]\!] = \mathbb{X}[\![E]\!]$

Functions can be applied to path-expressions. On the information descriptor level this leads to the following verbalisation.

**[P27]** $I_1, \ldots, I_l \in \texttt{InfDiscr} \wedge \{\mathbb{I}[\![I_1]\!], \ldots, \mathbb{I}[\![I_l]\!]\} \not\subseteq \texttt{Expressions} \wedge f$ is a function symbol $\vdash$
$\mathbb{I}[\![f\,(\,I_1\,,\ldots,I_l\,)]\!] = f(\mathsf{HdCoerce}\,\mathbb{I}[\![I_1]\!], \ldots, \mathsf{HdCoerce}\,\mathbb{I}[\![I_l]\!])$

**[P28]** $f(I_1, I_2) \in \texttt{InfDiscr} \wedge f$ is an infix operator symbol $\vdash \quad \mathbb{I}[\![I_1\ f\ I_2]\!] = \mathbb{I}[\![f(I_1, I_2)]\!]$

Some simple examples are:

(Salary which is earned by a Person who works for the Company: 'Asymetrix')
* (the ExchangeRate for Currency: 'AUD')

Cosinus(Distance from the Planet: 'Earth' to the Planet: 'Venus')

## 7.5 Conditions

We now define the set of condition: $\mathtt{Conditions}$ for ConQuer-92. Its semantics are provided by the relation $\mathbb{C} : \mathtt{Conditions} \rightarrow \mathtt{PEConditions}$. We start with comparison operations of expressions. The value comparison operations we used on information descriptors ($\mathsf{ValueComp}$) can be used in conditions as well to compare the values of expressions:

**[P29]** $E_1, E_2 \in \mathtt{Expressions} \wedge \mathsf{ValueComp}(r, R) \quad \vdash \quad \mathbb{B}[\![E_1 \; r \; E_2]\!] = \mathbb{X}[\![E_1]\!] \; R \; \mathbb{X}[\![E_2]\!]$

An example would be:

(THE MAXIMUM Salary of a Person who works for the Company: Asymetrix) $>$ 100000

For building information descriptors can be compared as well using a variety of set comparison operators. Let SetComp be filled by:

| ConQuer-92 Operator | Path Expression Operator |
|---|---|
| EQUALS | $=$ |
| $=$ | $=$ |
| DOES NOT EQUAL | $\neq$ |
| $<>$ | $\neq$ |
| IS DISJOINT FROM | $\otimes$ |
| IS A SUBSET OF | $\subset$ |
| $<$ | $\subset$ |
| IS A SUBSET OF OR EQUAL TO | $\subseteq$ |
| $<=$ | $\subseteq$ |
| IS A SUPERSET OF | $\supset$ |
| $>$ | $\supset$ |
| IS A SUPERSET OF OR EQUAL TO | $\supseteq$ |
| $>=$ | $\supseteq$ |

For information descriptors we now have:

**[P30]** $I_1, I_2 \in \mathtt{InfDiscr} \wedge \mathsf{SetComp}(r, R) \quad \vdash \quad \mathbb{B}[\![I_1 \; r \; I_2]\!] = \mathbb{I}[\![I_1]\!] \; R \; \mathbb{I}[\![I_2]\!]$

As an illustration consider:

Person who has watched the Movie: 'Star Trek Generations'
IS A SUBSET OF
Person who works for the Department: 'Computer Science'

Conditions themselves can be combined in a number of ways. For obvious reasons, these combining operations are based on the operations from logic. Let LogicConn be defined as:

| ConQuer-92 Operator | Condition Operator |
|---|---|
| AND | $\wedge$ |
| & | $\wedge$ |
| OR | $\vee$ |
| | | $\vee$ |
| EXCLUSIVE OR | $\underline{\vee}$ |
| || | $\underline{\vee}$ |
| IMPLIES | $\Rightarrow$ |
| => | $\Rightarrow$ |
| IFF | $\Longleftrightarrow$ |
| <=> | $\Longleftrightarrow$ |

We now have:

**[P31]** $C_1, C_2 \in \texttt{Conditions} \wedge \mathsf{LogicConn}(r, R) \quad \vdash \quad \mathbb{B}[\![C_1\ r\ C_2]\!] = \mathbb{B}[\![C_1]\!]\ R\ \mathbb{B}[\![C_2]\!]$

**[P32]** $C \in \texttt{Conditions} \quad \vdash \quad \mathbb{B}[\![\mathsf{NOT}\ C]\!] = \neg\,\mathbb{B}[\![C]\!]$

**[P33]** $C \in \texttt{Conditions} \quad \vdash \quad \mathbb{C}[\![\tilde{\ }C]\!] = \neg\,\mathbb{B}[\![C]\!]$

Conditions can be used, similarly as in SQL, be used for selections (restrictions). In the case of ConQuer-92 selections can be used on information descriptors. The selection operation is verbalised as:

**[P34]** $C \in \texttt{Conditions} \wedge I \in \texttt{InfDiscr} \quad \vdash \quad \mathbb{I}[\![I\ \mathsf{WHERE}\ C]\!] = \mathsf{Where}(\mathbb{I}[\![I]\!], \mathbb{B}[\![C]\!])$

Besides the simple WHERE clause, we have a number of other verbalisations based on this operation. Other verbalisations of the selection statement are:

**[P35]** $C \in \texttt{Conditions} \wedge I \in \texttt{InfDiscr} \quad \vdash \quad \mathbb{I}[\![\mathsf{IF}\ C\ \mathsf{THEN}\ I]\!] = \mathsf{Where}(\mathbb{I}[\![I]\!], \mathbb{I}[\![C]\!])$

**[P36]** $C \in \texttt{Conditions} \wedge I, J \in \texttt{InfDiscr} \quad \vdash$
$\mathbb{I}[\![\mathsf{IF}\,c\,\mathsf{THEN}\,p\,\mathsf{ELSE}\,q]\!] = \mathsf{Where}(\mathbb{I}[\![I]\!], \mathbb{B}[\![C]\!]; \mathbb{I}[\![J]\!])$

**[P37]** $C_1, \ldots, C_l \in \texttt{Conditions} \wedge I_1, \ldots, I_l \in \texttt{InfDiscr} \wedge \quad \vdash$
$\mathbb{I}[\![I_1\ \mathsf{IF}\,C_1\,; \ldots I_{l-1}\ \mathsf{IF}\,C_{l-1}\,; I_l\ \mathsf{OTHERWISE}]\!] =$
$\mathsf{Where}(\mathbb{I}[\![I_1]\!], \mathbb{B}[\![C_1]\!]; \ldots; \mathbb{I}[\![I_{l-1}]\!], \mathbb{B}[\![C_{l-1}]\!]; \mathbb{I}[\![I_l]\!])$

**[P38]** $C_1, \ldots, C_l \in \texttt{Conditions} \wedge I_1, \ldots, I_l \in \texttt{InfDiscr} \wedge \quad \vdash$
$\mathbb{I}[\![p_1\ \mathsf{IF}\,c_1\,; \ldots p_l\ \mathsf{IF}\,c_l]\!] = \mathsf{Where}(\mathbb{I}[\![I_1]\!], \mathbb{B}[\![C_1]\!]; \ldots; \mathbb{I}[\![I_l]\!], \mathbb{B}[\![C_l]\!])$

**[P39]** $C \in \texttt{Conditions} \wedge \forall_{1 \le i \le l} [\mathsf{VNm}(a_i) = n_i] \vdash$
$\mathbb{I}[\![\mathsf{SELECT}\, n_1, \ldots, n_l\, \mathsf{WHERE}\ C\ ]\!] = \mathsf{Path}(\mathbb{I}[\![C]\!], a_1, \ldots, a_l)$

Some illustrative examples of the use of the WHERE operations were already provided in the previous section.

Since on the path-expression level every condition is a path-expression, on the ConQuer-92 level every condition is an information descriptor:

**[P40]** $C \in \texttt{Conditions} \vdash \mathbb{I}[\![C]\!] = \mathbb{B}[\![C]\!]$

Conversely, information descriptors can be coerced to conditions by:

**[P41]** $P \in \texttt{InfDiscr} \vdash \mathbb{B}[\![\mathsf{SOME}\, P]\!] = \mathsf{Some}(\mathbb{I}[\![P]\!])$

## 7.6 Gathering information

The verbalisation of the confluence operation is not a beauty. In practice we shall prefer to use a graphical representation on screen. Furthermore, we also provide a version of the confluence operation that is integrated with the to be introduced LIST statement.

**[P42]** $I_1, \ldots, I_l, J \in \texttt{InfDiscr} \wedge \forall_{1 \le i \le l} [\mathsf{VNm}(a_i) = n_i \wedge \mathsf{VNm}(b_i) = m_i] \vdash$
$\mathbb{I}[\![I_1\ as(n_1)\ via(m_1), \ldots, I_l\ as(n_l)\ via(m_l)\, \mathsf{EACH}\, J]\!] =$
$[a_1 : \mathbb{I}[\![I_1]\!] : b_1, \ldots, a_l : \mathbb{I}[\![I_l]\!] : b_l ; \mathbb{I}[\![J]\!]]$

where:

$$as(n) = \textbf{if}\ n = \epsilon\ \textbf{then}\ \epsilon\ \textbf{else}\ \mathsf{AS}\, n\ \textbf{fi}$$
$$via(n) = \textbf{if}\ n = \epsilon\ \textbf{then}\ \epsilon\ \textbf{else}\ \mathsf{VIA}\, n\ \textbf{fi}$$

## 7.7 Group functions

We introduce the grouping related operations in two groups. The first two are intended for counting the entire result of a grouped information descriptor, and the second group is used to perform arithmetic operations on the grouping results.

The functions on groupings are verbalised as follows:

**[P43]** $I \in \texttt{InfDiscr} \wedge \forall_{1 \le i \le l} [\mathsf{VNm}(a_i) = n_i] \vdash$
$\mathbb{I}[\![\mathsf{THE\ COUNT\ OF}\, I\, \mathsf{GROUPED\ BY}\, n_1, \ldots, n_l]\!] = \mathsf{GCount}(\mathbb{I}[\![I]\!], \{a_1, \ldots, a_l\})$

**[P44]** $I \in \texttt{InfDiscr} \wedge \forall_{1 \le i \le l} [\mathsf{VNm}(a_i) = n_i] \quad \vdash$
$\mathbb{I}[\![\mathsf{THE\ DISTINCT\ COUNT\ OF}\, I\, \mathsf{GROUPED\ BY}\, n_1, \ldots, n_l]\!] =$
$\mathsf{GDsCount}(\mathbb{I}[\![I]\!], \{a_1, \ldots, a_l\})$

For the other grouping functions we introduce the relation GroupFunction as:

| ConQuer-92 Operator | Condition Operator |
|---|---|
| THE SUM OF | GSum |
| THE DISTINCT SUM OF | GDsSum |
| THE MIMIMUM OF | GMin |
| THE MAXIMUM OF | GMax |
| THE AVERAGE OF | GAvg |

These operations are integrated in the language by:

**[P45]** $I \in \texttt{InfDiscr} \wedge \mathsf{GroupFunction}(f, F) \forall_{1 \le i \le l} [\mathsf{VNm}(a_i) = n_i] \quad \vdash$
$\mathbb{I}[\![f\ I\, \mathsf{GROUPED\ BY}\, n_1, \ldots, n_l]\!] = F(\mathbb{I}[\![I]\!], \{a_1, \ldots, a_l\}, hd)$

**[P46]** $I \in \texttt{InfDiscr} \wedge \mathsf{GroupFunction}(f, F) \forall_{1 \le i \le l} [\mathsf{VNm}(a_i) = n_i] \wedge \mathsf{VNm}(b) = m \quad \vdash$
$\mathbb{I}[\![f\ m\, \mathsf{IN}\, I\, \mathsf{GROUPED\ BY}\, n_1, \ldots, n_l]\!] = F(\mathbb{I}[\![I]\!], \{a_1, \ldots, a_l\}, hd)$

## 7.8 Sub-expressions

The sub-expression concept is introduced on the information descriptor level as:

**[P47]** $I_1, \ldots I_l \in \texttt{InfDiscr} \quad \vdash \quad \mathbb{I}[\![\, [I_1, \ldots, I_l]\, ]\!] = [\ \mathbb{I}[\![I_1]\!], \ldots, \mathbb{I}[\![I_l]\!]\ ]$

As we will be using parenthesis for disambiguation purposes, we suggest the use of the symbol [ and ] to designate a sub-expression, even on the information descriptor level. An example was already given in the previous section.

## 7.9 Denotations

Now it is time to define a special class of constructs that can be used to denote instances of types. The mechanics of the denotations has already been discussed on the path-expression level. This class is represented as the set `Denotations`, and its semantics are provided by: $\mathbb{D} : \mathcal{TP} \times \texttt{Denotations} \rightarrow \texttt{PathExpr}$. The semantics defining rules are:

**[P48]** $I \in \texttt{InfDiscr} \quad \vdash \quad \mathbb{D}[\![I]\!] = \mathbb{I}[\![I]\!]$

**[P49]** $\mathsf{VNm}(a) = n \quad \vdash \quad \mathbb{D}[\![!n]\!] = !a$

**[P50]** $D_1, \ldots, D_l \in \texttt{Denotations} \quad \vdash \quad \mathbb{D}[\![(D_1, \ldots, D_l)]\!] = (\mathbb{D}[\![D_1]\!], \ldots, \mathbb{D}[\![D_l]\!])$

The denotations are now integrated in the class of information descriptors by:

**[P51]** $\mathsf{TNm}(x) = n \wedge D \in \texttt{Denotations} \quad \vdash \quad \mathbb{I}[\![n : D]\!] = x : \mathbb{D}[\![D]\!]$

**[P52]** $\mathsf{Pre}(x, c) = r \wedge \mathsf{TNm}(x) = n \wedge D \in \texttt{Denotations} \quad \vdash \quad \mathbb{I}[\![r\ n : D]\!] = x : \mathbb{D}[\![D]\!]$

## 7.10 Macros

Each macro definition results in extension of the ConQuer-92 language. On the path-expression level, macros where introduced by the Macros function. On the ConQuer level, the language extensions resulting from these definitions are provided as:

**[P53]** $\mathsf{Macros}(f, [a_1, \ldots, a_l]) \in \texttt{PEScalarExpr} \quad \vdash \quad \mathbb{X}[\![f(e_1, \ldots, e_l)]\!] = f(\mathbb{X}[\![e_1]\!], \ldots, \mathbb{X}[\![e_l]\!])$

**[P54]** $\mathsf{Macros}(f, [a_1, \ldots, a_l]) \in \texttt{PEConditions} \quad \vdash \quad \mathbb{B}[\![f(c_1, \ldots, c_l)]\!] = f(\mathbb{B}[\![c_1]\!], \ldots, \mathbb{X}[\![c_l]\!])$

**[P55]** $\mathsf{Macros}(f, [a_1, \ldots, a_l]) \in \texttt{PathExpr} \quad \vdash \quad \mathbb{I}[\![f(p_1, \ldots, p_l)]\!] = f(\mathbb{B}[\![p_1]\!], \ldots, \mathbb{X}[\![p_l]\!])$

These definitions imply that macros are, from a syntactical point of view, treated as if they were function symbols. As a result, in the grammar for ConQuer as provided in the appendix we will not see any explicit notions for the macros; they are part of the possible functions that are defined on ConQuer expressions.

In this section we do not revisit the derivation rules. Derivable types can be used in path-expressions and ConQuer expressions just as any other type. When a ConQuer query (and its underlying path-expressions) needs to be evaluated, then the derivation rule for the derivable type needs to be substituted in the existing expression.

## 7.11 Parsing and ambiguities

Since this section defined the semantics of ConQuer-92 in terms of an abstract syntax, we will not be confronted with ambiguities. However, one ConQuer-92 expression might have more than one possible parse tree in terms of the syntax provided in the appendix (ignoring issues of commutative or associative operations like AND ALSO). These ambiguities will usually result from multiply used role names (like has, of, etc). The different parse trees for one ConQuer expression correspond to different path-expressions, which have a different semantics.

At the moment it is not certain whether users of the InfoAssistant tool will actually type in complete ConQuer expressions themselves. If this is not going to be the case, then we will never have to actually parse ConQuer expressions, which would avoid any of these ambiguities all together. Nonetheless, when a further integration is made between NUQL and ConQuer, the notion of ambiguities due to alternative parse trees will become even larger. In the remainder of this subsection we discuss a way in which our knowledge about the schema combined with the typing within the resulting path-expressions can help us in deciding between alternative parse trees.

If $p$ is a ConQuer-92 expression (or a NUQL expression for that matter), and $T_1$ is a parse tree for $p$ and $T_2$ is a different parse tree for $p$, then to $p$ we can either associate path expression $\mathbb{I}[\![T_1]\!]$ or $\mathbb{I}[\![T_2]\!]$. If tree $T_1$ and $T_2$ are different, than this means that we have an ambiguity on our hands! Using $\mathbb{L}$ we can sometimes dismiss one (or both) of the alternatives. If $\mathbb{L}[\![\mathbb{I}[\![T_1]\!]]\!] = \varnothing$ then $T_2$ is the likely alternative, and vice versa. The reason for dismissing $T_1$ when $\mathbb{L}[\![\mathbb{I}[\![T_1]\!]]\!] = \varnothing$ lies in the fact that the interpretation of expression $p$ as parse tree $T_1$ would lead to a path-expression which is structurally empty. In other words, by looking at the schema (the relatedness of types), and looking at how the types are connected in the path-expressions (using $\mathbb{L}$), it can be proven that the path-expression $\mathbb{I}[\![T_1]\!]$ will always lead to an empty result for any population of the underlying schema.

If all alternatives lead to a structurally empty result, then ConQuer (or NUQL) expression $p$ is a structurally empty query formulation, and should be marked by the system as an incorrect query. If a parsing ambiguity cannot be solved this way, the alternative interpretations need to re-verbalised and shown to the user. This latter process is particularly useful as a feedback mechanism for translation of NUQL expressions to path-expressions, as it allows us to show the possible interpretations of a NUQL expressions in terms that are still close to natural language (the ConQuer expressions).

## 7.12 Normalisation

Let $I$ be an information descriptor. In the path expression $\mathbb{I}[\![I]\!]$ certain patterns will occur that can be optimised with respect to their verbalisation (to be discussed in the next section), as well as the mapping to SQL. Let $P$ be a path expression (resulting from parsing an information descriptor), and $T$ be its typing (so $T = \mathbb{T}[\![P]\!]$). In the parse tree of $P$ we would now like to replace the following patterns:

1. replace $p \circ f \circ q^{\leftarrow}$ where $\mathsf{Rel}(p) = \mathsf{Rel}(q) = f$ by $\langle p, q \rangle$

   Note: $\langle p, q \rangle$ is the notation for a binary mix-fix predicate!

2. replace $p \circ q^{\leftarrow}$ where $\mathsf{Rel}(p) = \mathsf{Rel}(q) = f$ by $\langle p, q \rangle$

3. replace $x \circ E$ where $x \in \mathcal{VL} \wedge E \in \texttt{PEScalarExpr}$ by $x : E$.

Note: if $x \circ E$ occurs while $x$ is not a value type, then this is not a semantically sensible path expression. $E$ will always return a value type instances, so concatenating an expression to a non-value type always returns empty.

4. Replace for each pair $\langle \Theta, \varphi \rangle \in \{\langle \cup, \Cup \rangle, \langle \cap, \Cap \rangle, \langle -, \not- \rangle\}$ the expression $\mathsf{Fr}(Q_1)\Theta\,\mathsf{Fr}(Q_2)$ by the expression $Q_1 \varphi Q_2$.

Note that $p \circ A \circ q^{\leftarrow}$ where $\mathsf{Rel}(p) = \mathsf{Rel}(q) \sim A$ is *not* replaced. The $A$ could be a subtype of $\mathsf{Rel}(q)$, and removal of $A$ from the path-expression would change its semantics.

## 7.13 Listing results

The result of an information descriptor (via the path-expression and relational algebra expression) is a bag. In most real life applications some order on this bag is required. Therefore we should allow for sorting on top of ConQuer-92 information descriptors. Note: projection of the results is an integrated part of the information descriptors. These filters should be added *on top* of ConQuer expressions, and should not be a part of it.

The last syntactic category we introduce is therefore the list specification (`ListSpec`) class. Its semantics are expressed by the function $\mathbb{Q}$. We do not provide a formal semantics, as all what these operations do is take the bag resulting from an information descriptor and order the results. If $I \in$ `InfDiscr`, then we have the following possible list specifications:

**[P56]** $\mathbb{Q}[\![\mathsf{LIST}\ I]\!] = \mathbb{I}[\![I]\!]$

listing the results without enforcing an order

**[P57]** $\mathbb{Q}[\![\mathsf{LIST}\ I\ \mathsf{ORDERED\ ASCENDING}]\!] = \Omega(\mathbb{I}[\![I]\!], hd : \mathtt{Asc})$

listing the results in $I$ ordered in ascending order for the heads of $I$,

**[P58]** $\mathbb{Q}[\![\mathsf{LIST}\ I\ \mathsf{ORDERED\ DESCENDING}]\!] = \Omega(\mathbb{I}[\![I]\!], hd : \mathtt{Desc})$

listing the results in $I$ ordered in descending order for the heads of $I$. However, sometimes users may want to order on other columns in the result than the head column. Let $o_1, \ldots, o_l \in \{\mathsf{ASCENDING}, \mathsf{DESCENDING}\}$, and $I \in$ `InfDiscr` and $v_1, \ldots, v_l$ each be a variables in $I$ where $\forall_{1 \leq i \leq l}\,[\mathsf{VNm}(a_i) = v_i]$, then the general format of the LIST statement is:

**[P59]** $\mathbb{Q}[\![\mathsf{LIST}\ I\ \mathsf{ORDERED\ WITH}\ o_1\ v_1, \ldots, o_l\ v_l]\!] = \Omega(\mathbb{I}[\![I]\!], ord(o_1) : a_1; \ldots; ord(o_l) : a_l)$

where $ord(o) = \textbf{if } o = \texttt{Asc} \textbf{ then } \texttt{Desc} \textbf{ fi}$. Note: we presume that there are two standard variables: $\mathsf{VNm}(hd) = \mathsf{HEAD}$ and $\mathsf{VNm}(tl) = \mathsf{TAIL}$.

The columns that are actually printed by the list statement could still be non-value types. For instance, LIST Costumers of Company: 'Asymetrix' would lead to a list of the abstract instances of costumers rather than their costumer nr (presuming they are identified through such a number). It should be clear that the LIST statement must replace the abstract costumer instances by the concretised costumer nrs. It will also be these last nrs that would be used in the ordering operations.

Formally we presume that we have a function $\mathsf{Denote} : \Omega \times \mathsf{POP} \rightarrow (\cup \mathcal{D})^+$ which expresses each instance in terms of some set of values. We should of course have:

$$i \in \cup \mathcal{D} \Rightarrow \mathsf{Denote}(i, p) = [i]$$

The exact definition of this function depends on the conceptual schema, the reference schemas, and the current population. This function can then be used to denote the instances resulting from the list statement. When mapping path expressions to SQL, this function becomes implicit since the abstract instances are never stored, but rather their denotations in terms of concrete values.

Finally, one may sometimes like to do a final projection on the columns in the table resulting from an information descriptor. Therefore we extend the LIST statement with an optional projection clause. This would allow us to formulate:

LIST HEAD, $s/1000$ OF Person who works for the Company: 'Asymetrix' AND ALSO earns a Salary: $s$

In this case, the table resulting from

Person who works for the Company: 'Asymetrix' AND ALSO earns a Salary: $s$

is projected on Person and $s/1000$ using a normal projection operation as introduced in section 5. In the projection list, we are allowed to use any scalar expression, where the variables HEAD and TAIL are used as standard variable names for the head ($hd$) and tail ($tl$) columns of the table resulting from the information descriptor.

# 8 Verbalisation Rules

This section discusses verbalisation rules for path-expressions. These rules should be interpreted as a standardised verbalisation format for information descriptors.

Sometimes path-expressions will be generated automatically, e.g. when doing a point to point query. In such a case, a verbalisation of the path-expression needs to be build from scratch.

When a user enters an information descriptor manually then the user may have used a verbalisation that can be improved upon. In that case, the information descriptor specified by the user needs to be interpreted as a path-expression, and then re-verbalised by the system.

When entering queries using NUQL, we may be confronted with a situation where one NUQL query may have more than one interpretation in terms of a path-expression. If this occurs, the different interpretations have to be verbalised using the rules stated in this section, and shown to the user as alternatives.

For the verbalisation of path-expression we introduce the function: $\mathsf{PVerb} : \mathtt{PathExpr} \times \wp(\mathcal{TP}) \times \wp(\mathsf{Attr} \times \mathcal{TP}) \to \Sigma^{+}$. Again, we use the style of denotational semantics, so we write: $\mathsf{PVerb}[\![P]\!]\,(T, L)$.

The parameters constituting the environment of the verbalisation function, $L$ and $T$ provide some extra information used to make better verbalisations. The $L$ provides the maximum set of types of the path expression (if any) directly to the left of the current path expression. With to the left we mean here, with respect to concatenation $\circ$. So if $x$ and $y$ are types, in $Q \circ x \circ P$ this means that $\{x\}$ is the set of types directly to the left of $P$ and in the case of $((Q \circ x) \cup (R \circ y)) \circ P$ this is $\{x, y\}$. Below we will see that this information is required to prevent ambiguities in the verbalisations. We do not provide a formalisation of a way to obtain $L$; it is simply a matter of analysing the parse tree of tha path-expression and find it's direct neighbours. Note, and this is a very important case, that when we consider a path-expression without a left context, then $L = \mathcal{TP}$ must be used. For example, if $p$ is a role-entry and $p$ is the complete path expression we want to verbalise, then the left context is empty while $L = \mathcal{TP}$.

The $T$ is simply the typing function for the attributes (variables) as we have seen before when defining the semantics of ConQuer. Note that we presume that the input of this is a path-expression (or rather a parse tree thereof) that has been normalised conform the rules introduced in subsection 7.12.

## 8.1 Empty path expressions

Empty path expressions will only occur in the context of query by navigation. They are verbalised as follows:

**[V1]** $\vdash \quad \mathsf{PVerb}[\![\epsilon]\!]\,(T, L) = \mathsf{start}$

## 8.2 Types and denotations

For simple occurrences of types in path expressions we have:

**[V2]** $\mathsf{TNm}(x) = n \wedge \mathsf{Pre}(x, \mathtt{undetermined}) = r \quad \vdash \quad \mathsf{PVerb}[\![x]\!]\,(T, L) = r\ n$

From this rule follows that in general we will prefer to use the articles for the undetermined case. For example: A Human or Ein Mensch rather than The Human Dieser Mensch. However, when a concrete instance of the type is given, our preference changes to the determined case: The Person: 'T.A. Halpin'. Therefore, instance denotations are verbalised by:

**[V3]** $\mathsf{TNm}(x) = n \land \mathsf{Pre}(x, \mathtt{determined}) = r \land d \in \mathtt{PEDenotations} \quad \vdash$
$\mathsf{PVerb}[\![x : d]\!]\,(T, L) = r\ n : \mathsf{DenVerb}[\![D]\!]\,(T, L)$

The denotations themselves are verbalised by the function $\mathsf{DenVerb}[\![P]\!]\,(T, L)$ which has a similar signature as the PVerb function.

**[V4]** $P \in \mathtt{PathExpr} \quad \vdash \quad \mathsf{DenVerb}[\![P]\!]\,(T, L) = \mathsf{PVerb}[\![P]\!]\,(T, L)$

**[V5]** $a \in \mathsf{Attrs} \land \mathsf{VNm}(a) = n \quad \vdash \quad \mathsf{DenVerb}[\![!a]\!]\,(T, L) = \ !n$

**[V6]** $d_1, \ldots, d_l \in \mathtt{PEDenotations} \land l > 0 \quad \vdash$
$\mathsf{DenVerb}[\![(d_1, \ldots, d_l)]\!]\,(T, L) = (\mathsf{DenVerb}[\![d_1]\!]\,(T, L), \ldots, \mathsf{DenVerb}[\![d_l]\!]\,(T, L))$

**[V7]** $d \in \mathtt{PEDenotations} \quad \vdash \quad \mathsf{DenVerb}[\![(d)]\!]\,(T, L) = \mathsf{DenVerb}[\![d]\!]\,(T, L)$

Note due to the normalisation rules specified in subsection 7.12 we know that a pattern $x \circ e$ where $x$ is a (value) type and $e$ a scalar-expression will have been changed to $x : e$.

## 8.3 Concatenation

For $\circ$ there are six separate verbalisation rules. When verbalising a path-expression, we should try to apply them in the order given here.

The first rule is concerned with the verbalisation of a role-entry.

**[V8]** $\mathsf{PNm}(p) = n \land \neg\exists_{q \neq p}\,[\mathsf{PNm}(p) = \mathsf{PNm}(q) \land \mathsf{Player}(q) \in L \land \mathsf{Rel}(q) \in \pi_1 \mathbb{L}[\![Y]\!]\,(T)] \quad \vdash$
$\mathsf{PVerb}[\![p \circ Y]\!]\,(T, L) = n\ \mathsf{PVerb}[\![Y]\!]\,(T, \pi_2 \mathbb{L}[\![p]\!]\,(T))$

The difficulty with a role entry is that a role name $\mathsf{PNm}(p)$ does not have to be unique within a schema. For example, involved in is by far the most popular role name. (Note: has is not a role name but a binary mix-fix predicate verbalisation; they will be dealt with in the next subsection.) When verbalising a path-expression, however, we must make sure that we refer to a unique role. This is why we need to check the uniqueness of the role name within the context of the role-entry in the path-expression. This is done by:

$$\neg\exists_{q \neq p}\left[\mathsf{PNm}(p) = \mathsf{PNm}(q) \land \mathsf{Player}(q) \in L \land \mathsf{Rel}(q) \in \pi_1 \mathbb{L}[\![Y]\!]\,(T)\right]$$

This clause checks whether there is another role with the same name ($\mathsf{PNm}(p) = \mathsf{PNm}(q)$) such that the context of the role entry cannot distinguish between them.

For a role-exit, the same considerations apply:

**[V9]** $\mathsf{RNm}(p) = n \land \neg\exists_{q \neq p}\left[\mathsf{PNm}(p) = \mathsf{PNm}(q) \land \mathsf{Rel}(q) \in L \land \mathsf{Player}(q) \in \pi_1 \mathbb{L}[\![Y]\!]\,(T)\right] \quad \vdash$
$\mathsf{PVerb}[\![p^{\leftarrow} \circ Y]\!]\,(T, L) = n \;\; \mathsf{PVerb}[\![Y]\!]\,(T, \pi_2 \mathbb{L}[\![p^{\leftarrow}]\!]\,(T))$

If a role name, or a reverse role name, are not unique within their context, we are forced to make these names unique in some way. We know that role names and reverse role names are unique within the context of one fact type. So if a role name, or reversed role name, is not unique we are *forced* to suffix this name with the fact type name. This leads to the rules:

**[V10]** $\mathsf{PNm}(p) = n \land \mathsf{TNm}(\mathsf{Rel}(p) = m \quad \vdash$
$\mathsf{PVerb}[\![p]\!]\,(T, L) = n.m$

**[V11]** $\mathsf{RNm}(p) = n \land \mathsf{TNm}(\mathsf{Rel}(p) = m \quad \vdash$
$\mathsf{PVerb}[\![p^{\leftarrow}]\!]\,(T, L) = n.m$

As the above verbalisations are independent of their context in a path-expression, they are not defined in the context of a concatenation ($\circ Y$). In this case, the last rule for concatenation (see below) can be used.

When verbalising a concatenation of two types directly following each other we need to introduce the connecting word IS. This is captured by the rule:

**[V12]** $x, y \in \mathcal{TP} \land x \sim y \quad \vdash \quad \mathsf{PVerb}[\![xE \circ yF]\!]\,(T, L) = \mathsf{PVerb}[\![xE]\!]\,(T, L) = n \; \mathsf{IS} \; \mathsf{PVerb}[\![yF]\!]\,(T, \{x\})$

where $E$ and $F$ may be denotations or empty.

Finally, the most generic rule is given below, which simply concatenates verbalisations:

**[V13]** $X, Y \in \mathtt{PathExpr} \quad \vdash$
$\mathsf{PVerb}[\![X \circ Y]\!]\,(T, L) = \mathsf{PVerb}[\![X]\!]\,(T, L) \;\; \mathsf{PVerb}[\![Y]\!]\,(T, \pi_2 \mathbb{L}[\![X]\!]\,(T))$

## 8.4 Mix-fix predicate verbalisations

The next class of verbalisation rules we introduce deal with mix fix predicates. Three rules are introduced which should, again, be applied in the order in which they are introduced. If a mix-fix predicate verbalisation is unique within its path-expression context, then the following rule can be applied:

**[V14]** $P_2, \ldots, P_{l-1} \in \mathtt{PathExpr} \wedge \mathsf{MFix}(r, [\alpha_1, \ldots, \alpha_{l-1}], [p_1, \ldots, p_l]) \wedge o = \mathsf{Post}(\mathsf{Player}(p_1)) \wedge$
$\neg\exists_{[q_1,\ldots,q_l] \neq [p_1,\ldots,p_l],s} [\forall_{1 \leq i \leq l} [\mathsf{Player}(q_i) \in T_i] \wedge \mathsf{MFix}(s, [\alpha_1, \ldots, \alpha_{l-1}], [q_1, \ldots, q_l])] \quad \vdash$
$\mathsf{PVerb}[\![\langle p_1, p_2 : P_2, \ldots, p_{l-1} : P_{l-1}, p_l \rangle]\!] (T, L) =$
$o\, \alpha_1\ \mathsf{PVerb}[\![P_2]\!] (T, \{\mathsf{Player}(p_2)\})\, \alpha_2 \ldots \alpha_{l-1}\ \mathsf{PVerb}[\![P_l]\!] (T, \{\mathsf{Player}(p_l)\})$

where $T_1 = T, T_2 = \pi_1 \mathbb{L}[\![P_2]\!] (T), \ldots T_l = \pi_1 \mathbb{L}[\![P_l]\!] (T)$. The uniqueness requirement in this case is provided as:

$$\neg\exists_{[q_1,\ldots,q_l] \neq [p_1,\ldots,p_l],s} [\forall_{1 \leq i \leq l} [\mathsf{Player}(q_i) \in T_i] \wedge \mathsf{MFix}(s, [\alpha_1, \ldots, \alpha_{l-1}], [q_1, \ldots, q_l])]$$

When a mix-fix predicate verbalisation is not unique within its context, we are forced to make the verbalisation unique by adding a suffix to the verbalisation. For this case we have:

**[V15]** $P_2, \ldots, P_{l-1} \in \mathtt{PathExpr} \wedge \mathsf{MFix}(r, [\alpha_1, \ldots, \alpha_{l-1}], [p_1, \ldots, p_l]) \wedge o = \mathsf{Post}(\mathsf{Player}(p_1)) \wedge \mathsf{TNm}(r) = n \quad \vdash$
$\mathsf{PVerb}[\![\langle p_1, p_2 : P_2, \ldots, p_{l-1} : P_{l-1}, p_l \rangle]\!] (T, L) =$
$o\, \alpha_1.n\ \mathsf{PVerb}[\![P_2]\!] (T, \{\mathsf{Player}(p_2)\})\, \alpha_2 \ldots \alpha_{l-1}\ \mathsf{PVerb}[\![P_l]\!] (T, \{\mathsf{Player}(p_l)\})$

which suffixes the first part of the verbalisation ($\alpha_1$) with the name of the fact type ($\alpha_1.n$). Usually, this latter rule will not be needed. Even for the most commonly used verbalisation has, the context will usually provide a good disambiguation.

If $l = 2$, then $\langle p, q \rangle$ may be the result of a normalisation of $p \circ q^{\leftarrow}$. For these combinations no mix-fix predicate might have been defined. For these (hopefully rare) cases we introduce:

**[V16]** $\mathsf{PNm}(p) = n \wedge \mathsf{RNm}(q) = m \wedge \mathsf{TNm}(\mathsf{Rel}(p)) = f \wedge \mathsf{Post}(\mathsf{Player}(p)) = o \quad \vdash$
$\mathsf{PVerb}[\![\langle p, q \rangle]\!] (T, L) = o\ n\ f\ m$

Note that $n\ f\ m$ is always unique by itself.

## 8.5 Unary operations

The remaining verbalisation rules are now simple straightforward rules which basically define the inverse of the semantics function for ConQuer as defined in the previous section. We will not encounter any additional complications due to possible ambiguities.

For path reversal we have:

**[V17]** $P \in \texttt{PathExpr} \quad \vdash \quad \textsf{PVerb}[\![P^{\leftarrow}]\!]\,(T, L) = \textsf{THE REVERSE OF } \textsf{PVerb}[\![P]\!]\,(T, L)$

The verbalisation of the unary operations is provided by UnOp:

| Verbalisation | Path Expression |
|---|---|
| DISTINCT | Ds |
| ONLY | Fr |
| $\epsilon$ | HdCoerce |
| $\epsilon$ | TlCoerce |

The verbalisation rule is then:

**[V18]** $P \in \texttt{PathExpr} \wedge \textsf{UnOp}(o, O) \quad \vdash \quad \textsf{PVerb}[\![O\ P]\!]\,(T, L) = o\ \textsf{PVerb}[\![P]\!]\,(T, L)$

## 8.6 Binary operations

For the set based operations we defined two classes. The first class operates only on the front elements. These are the FrSetOper operations:

| Verbalisation | Path Expression Operator |
|---|---|
| OR OTHERWISE | $\Cup$ |
| AND ALSO | $\Cap$ |
| BUT NOT | $\not\!-$ |

For these operations we have:

**[V19]** $P, Q \in \texttt{PathExpr} \wedge \textsf{FrSetOper}(o, O) \quad \vdash$
$\textsf{PVerb}[\![P\ O\ Q]\!]\,(T, L) = \textsf{PVerb}[\![P]\!]\,(T, L)\ o\ \textsf{PVerb}[\![Q]\!]\,(T, L)$

Note that the above rule has presidence to the one below. This prevents us from getting

ONLY $P$ UNITED WITH ONLY $Q$

as a verbalisation where $P$ OR OTHERWISE $Q$ would have been more appropriate.

The set of set operations operating on the entire path was SetOper:

| Verbalisation | Path Expression Operator |
|---|---|
| UNITED WITH | $\cup$ |
| INTERSECTED WITH | $\cap$ |
| MINUS | $-$ |
| WHICH ARE ALL IN | $\underline{\subseteq}$ |
| THAT INCLUDES ALL | $\underline{\supseteq}$ |
| MATCHING ALL | $\underline{\equiv}$ |
| MISSING | $\underline{\otimes}$ |
| WITH | $\times$ |

with verbalisation rule:

**[V20]** $P, Q \in \texttt{PathExpr} \wedge \mathsf{BinOp}(o, O) \quad \vdash$

$$\mathsf{PVerb}[\![P\ O\ Q]\!]\,(T, L) = \mathsf{PVerb}[\![P]\!]\,(T, L)\ o\ \mathsf{PVerb}[\![Q]\!]\,(T, L)$$

The value comparison ValueComp were defined as:

| Verbalisation | Path Expression Operator |
|---|---|
| $=$ | $=$ |
| IS EQUAL TO | $=$ |
| $<>$ | $\neq$ |
| IS NOT EQUAL TO | $\neq$ |
| $<$ | $<$ |
| IS LESS THAN | $<$ |
| $<=$ | $\leq$ |
| IS LESS THAN OR EQUAL TO | $\leq$ |
| $>$ | $>$ |
| IS GREATER THAN | $>$ |
| $>=$ | $\geq$ |
| IS GREATER THAN OR EQUAL TO | $\geq$ |

The verbalisation rule is then:

**[V21]** $P, Q \in \texttt{PathExpr} \wedge \mathsf{ValueComp}(o, O) \quad \vdash$

$$\mathsf{PVerb}[\![P\ O\ Q]\!]\,(T, L) = \mathsf{PVerb}[\![P]\!]\,(T, L)\ o\ \mathsf{PVerb}[\![Q]\!]\,(T, L)$$

## 8.7 Path re-shuffling

The path-reshuffler is verbalised by:

**[V22]** $P \in \texttt{PEScalarExpr} \wedge \forall_{1 \leq i \leq l}\,[\mathsf{VNm}(a_i) = n_i] \quad \vdash$

$$\mathsf{PVerb}[\![\mathsf{Path}(P, a_1, \ldots, a_n)]\!]\,(T, L) =$$
$$\text{THE PATH FROM } n_1 \text{ VIA } n_2, \ldots, n_{l-1} \text{ TO } n_l \text{ OF } \mathsf{PVerb}[\![P]\!]\,(T, L)$$

## 8.8 Functions

For functions and operators we simply have:

**[V23]** $P_1, \ldots, P_l \in \texttt{PathExpr} \wedge f$ is a function symbol $\vdash$

$$\mathsf{PVerb}[\![f(P_1, \ldots, P_l)]\!]\,(T, L) = f\,(\,\mathsf{PVerb}[\![E_1]\!]\,(T, L), \ldots, \mathsf{PVerb}[\![E_l]\!]\,(T, L)\,)$$

**[V24]** $P_1, P_2 \in \texttt{PathExpr} \wedge\ f$ is an infix operator symbol $\vdash$

$$\mathsf{PVerb}[\![f(P_1, P_2)]\!]\,(T, L) = \mathsf{PVerb}[\![P_1]\!]\,(T, L) f\,\mathsf{PVerb}[\![P_2]\!]\,(T, L)$$

## 8.9 Selection

For the selection statements we have four rules. They should be tried to be applied in the order of specification. The four rules will deal with selection statements of increasing complexity.

**[V25]** $P \in \texttt{PathExpr} \wedge C \in \texttt{PEConditions} \quad \vdash$

$$\mathsf{PVerb}[\![\mathsf{Where}(P, C)]\!]\,(T, L) = \mathsf{PVerb}[\![P]\!]\,(T, L)\ \mathsf{WHERE}\ \ \mathsf{CondVerb}[\![C]\!]\,(T, L)$$

**[V26]** $P, Q \in \texttt{PathExpr} \wedge C \in \texttt{PEConditions} \quad \vdash$

$$\mathsf{PVerb}[\![\mathsf{Where}(P, C; Q)]\!]\,(T, L) =$$
$$\mathsf{IF}\ \mathsf{CondVerb}[\![C]\!]\,(T, L)\ \mathsf{THEN}\ \mathsf{PVerb}[\![P]\!]\,(T, L)\ \mathsf{ELSE}\ \mathsf{PVerb}[\![Q]\!]\,(T, L)$$

**[V27]** $P_1, \ldots, P_l, Q \in \texttt{PathExpr} \wedge C_1, \ldots, C_l \in \texttt{PEConditions} \quad \vdash$

$$\mathsf{PVerb}[\![\mathsf{Where}(P_1, C_1; \ldots; P_l, C_l; Q)]\!]\,(T, L) =$$
$$\mathsf{PVerb}[\![P_1]\!]\,(T, L)\ \mathsf{IF}\ \mathsf{PVerb}[\![C_1]\!]\,(T, L); \ldots;$$
$$\mathsf{PVerb}[\![P_l]\!]\,(T, L)\ \mathsf{IF}\ \mathsf{PVerb}[\![C_l]\!]\,(T, L); \mathsf{PVerb}[\![Q]\!]\,(T, L)$$

**[V28]** $P_1, \ldots, P_l, Q \in \texttt{PathExpr} \wedge C_1, \ldots, C_l \in \texttt{PEConditions} \quad \vdash$

$$\mathsf{PVerb}[\![\mathsf{Where}(P_1, C_1; \ldots; P_l, C_l]\!]\,(T, L) =$$
$$\mathsf{PVerb}[\![P_1]\!]\,(T, L)\ \mathsf{IF}\ \mathsf{PVerb}[\![C_1]\!]\,(T, L); \ldots;$$
$$\mathsf{PVerb}[\![P_l]\!]\,(T, L)\ \mathsf{IF}\ \mathsf{PVerb}[\![C_l]\!]\,(T, L)$$

Note that when a user enters an information descriptor they may choose to use any of the above given verbalisations. The system should then normalise this when re-displaying the path-expression on the screen.

## 8.10 Group functions

Grouping functions are again simply the inverse of the semantics function for ConQuer. So we have:

**[V29]** $P \in \texttt{PathExpr} \wedge \forall_{1 \leq i \leq l}\,[\mathsf{VNm}(a_i) = n_i] \quad \vdash$

$$\mathsf{PVerb}[\![\mathsf{GCount}(P, \{a_1, \ldots, a_l\})]\!]\,(T, L) =$$
$$\mathsf{THE\ COUNT\ OF}\ \mathsf{PVerb}[\![P\ ]\!]\,(T, L)\ \mathsf{GROUPED\ BY}\ n_1, \ldots, n_l$$

**[V30]** $P \in \texttt{PathExpr} \wedge \forall_{1 \leq i \leq l}\,[\mathsf{VNm}(a_i) = n_i] \quad \vdash$

$$\mathsf{PVerb}[\![\mathsf{GDsCount}(P, \{a_1, \ldots, a_l\})]\!]\,(T, L) =$$
$$\mathsf{THE\ DISTINCT\ COUNT\ OF}\ \mathsf{PVerb}[\![P]\!]\,(T, L)\ \mathsf{GROUPED\ BY}\ n_1, \ldots, n_l$$

For the other grouping functions we introduce the relation GroupFunction as:

| Verbalisation | Condition Operator |
|---|---|
| THE SUM OF | GSum |
| THE DISTINCT SUM OF | GDsSum |
| THE MIMIMUM OF | GMin |
| THE MAXIMUM OF | GMax |
| THE AVERAGE OF | GAvg |

For this class of operations we have the following two verbalisation rules:

**[V31]** $P \in \texttt{PathExpr} \wedge \mathsf{GroupFunction}(f, F) \wedge \forall_{1 \leq i \leq l} [\mathsf{VNm}(a_i) = n_i] \quad \vdash$
$\mathsf{PVerb}[\![F(P, \{a_1, \ldots, a_l\}, hd)]\!]\,(T, L) =$
$f\ \mathsf{PVerb}[\![P]\!]\,(T, L)$ GROUPED BY $n_1, \ldots, n_l$

**[V32]** $P \in \texttt{PathExpr} \wedge \mathsf{GroupFunction}(f, F) \wedge \mathsf{VNm}(x) = m \wedge \forall_{1 \leq i \leq l} [\mathsf{VNm}(a_i) = n_i] \quad \vdash$
$\mathsf{PVerb}[\![F(P, \{a_1, \ldots, a_l\}, x)]\!]\,(T, L) =$
$f\ m$ IN $\mathsf{PVerb}[\![P]\!]\,(T, L)$ GROUPED BY $n_1, \ldots, n_l$

## 8.11 Gathering information

The confluence operation is verbalised as:

**[V33]** $P_1, \ldots, P_l, Q \in \texttt{PathExpr} \wedge \forall_{1 \leq i \leq l} [\mathsf{VNm}(a_i) = n_i \wedge \mathsf{VNm}(b_i) = m_i] \quad \vdash$
$\mathsf{PVerb}[\![[a_1 : P_1 : b_1, \ldots, a_l : P_l : b_l; Q]]\!]\,(T, L) =$
$\mathsf{PVerb}[\![P_1]\!]\,(T, L)\ as(n_1)\ via(m_1)$
$, \ldots, \mathsf{PVerb}[\![P_l]\!]\,(T, L)\ as(n_l)\ via(m_l)$ EACH $\mathsf{PVerb}[\![Q]\!]\,(T, L)$

where:

$$as(n) = \textbf{if } n = \epsilon \textbf{ then } \epsilon \textbf{ else } \mathsf{AS}\, n \textbf{ fi}$$
$$via(n) = \textbf{if } n = \epsilon \textbf{ then } \epsilon \textbf{ else } \mathsf{VIA}\, n \textbf{ fi}$$

## 8.12 Sub-expressions

Sub-expressions hardly need any verbalisation:

**[V34]** $P_1, \ldots, P_l \in \texttt{PathExpr} \quad \vdash$
$\mathsf{PVerb}[\![[P_1, \ldots, P_l]]\!]\,(T, L) = [\,\mathsf{PVerb}[\![P_1]\!]\,(T, L), \ldots, \mathsf{PVerb}[\![P_l]\!]\,(T, L)]$

## 8.13 Scalar expressions

Scalar expressions are verbalised using $\mathsf{ScalVerb}[\![S]\!]\,(T,L)$. We define the Function class of operations by:

| Verbalisation | Function |
|---|---|
| THE COUNT OF | Count |
| THE SUM OF | Sum |
| THE MINIMUM OF | Min |
| THE MAXIMUM OF | Max |
| THE AVERAGE OF | Avg |

The verbalisation rule then becomes:

**[V35]** $P \in \mathtt{PathExpr} \wedge \mathsf{Function}(f, F) \quad \vdash \quad \mathsf{ScalVerb}[\![F(P)]\!]\,(T,L) = f\ \mathsf{PVerb}[\![P]\!]\,(T,L)$

Constants are verbalised as themselves:

**[V36]** $c \in \cup\mathcal{D} \quad \vdash \quad \mathsf{ScalVerb}[\![c]\!]\,(T,L) = c$

For variables we have:

**[V37]** $\mathsf{VNm}(a) = m \quad \vdash \quad \mathsf{ScalVerb}[\![a]\!]\,(T,L) = m$

**[V38]** $\mathsf{VNm}(a) = m \wedge \mathsf{RNm}(p) = r \quad \vdash \quad \mathsf{ScalVerb}[\![a.p]\!]\,(T,L) = m.r$

For functions and operators we have:

**[V39]** $E_1, \ldots, E_l \in \mathtt{PEScalarExpr} \wedge f$ is a function symbol $\quad \vdash$
$\mathsf{ScalVerb}[\![f(E_1, \ldots, E_l)]\!]\,(T,L) = f\,(\,\mathsf{ScalVerb}[\![E_1]\!]\,(T,L), \ldots, \mathsf{ScalVerb}[\![E_l]\!]\,(T,L)\,)$

**[V40]** $E_1, E_2 \in \mathtt{PEScalarExpr} \wedge f$ is an infix operator symbol $\quad \vdash$
$\mathsf{ScalVerb}[\![f(E_1, E_2)]\!]\,(T,L) = \mathsf{ScalVerb}[\![P_1]\!]\,(T,L) f\ \mathsf{ScalVerb}[\![P_2]\!]\,(T,L)$

As scalar expressions can be use as path expressions we have:

**[V41]** $E \in \mathtt{PEScalarExpr} \quad \vdash \quad \mathsf{PVerb}[\![E]\!]\,(T,L) = \mathsf{ScalVerb}[\![E]\!]\,(T,L)$

## 8.14 Conditions

The verbalisation of conditions is provided by $\mathsf{CondVerb}[\![C]\!]\,(T, L)$. The value comparison operations we used on information descriptors (ValueComp) can be used in conditions as well:

**[V42]** $E_1, E_2 \in \texttt{PEScalarExpr} \wedge \mathsf{ValueComp}(r, R) \quad \vdash$

$$\mathsf{CondVerb}[\![E_1\ R\ E_2]\!]\,(T, L) = \mathsf{ScalVerb}[\![E_1]\!]\,(T, L)\ r\ \ \mathsf{ScalVerb}[\![E_2]\!]\,(T, L)$$

Information descriptors can be compared as well using a variety of comparison operators. Let SetComp be filled by:

| Verbalisation | Path Expression Operator |
|---|---|
| EQUALS | $=$ |
| = | $=$ |
| DOES NOT EQUAL | $\neq$ |
| <> | $\neq$ |
| IS DISJOINT FROM | $\otimes$ |
| IS A SUBSET OF | $\subset$ |
| < | $\subset$ |
| IS A SUBSET OF OR EQUAL TO | $\subseteq$ |
| <= | $\subseteq$ |
| IS A SUPERSET OF | $\supset$ |
| > | $\supset$ |
| IS A SUPERSET OF OR EQUAL TO | $\supseteq$ |
| >= | $\supseteq$ |

For information descriptors we now have:

**[V43]** $P_1, P_2 \in \texttt{PathExpr} \wedge \mathsf{SetComp}(r, R) \quad \vdash$

$$\mathsf{CondVerb}[\![P_1\ R\ P_2]\!]\,(T, L) = \mathsf{PVerb}[\![P_1]\!]\,(T, L)\ r\ \ \mathsf{PVerb}[\![P_2]\!]\,(T, L)$$

Conditions can be combined in a number of ways. Let LogicConn be defined as:

| Verbalisation | Condition Operator |
|---|---|
| AND | $\wedge$ |
| & | $\wedge$ |
| OR | $\vee$ |
| \| | $\vee$ |
| EXCLUSIVE OR | $\veebar$ |
| \|\| | $\veebar$ |
| IMPLIES | $\Rightarrow$ |
| => | $\Rightarrow$ |
| IFF | $\Longleftrightarrow$ |
| <=> | $\Longleftrightarrow$ |

We now have:

**[V44]** $C_1, C_2 \in \texttt{PEConditions} \wedge \mathsf{LogicConn}(r, R) \quad \vdash$
$$\mathsf{CondVerb}[\![C_1 \; R \; C_2]\!]\,(T, L) = \mathsf{CondVerb}[\![C_1]\!]\,(T, L)\; r\;\; \mathsf{CondVerb}[\![C_2]\!]\,(T, L)$$

**[V45]** $C \in \texttt{PEConditions} \quad \vdash \quad \mathsf{CondVerb}[\![\neg C]\!]\,(T, L) = \mathsf{NOT}\;\; \mathsf{CondVerb}[\![C]\!]\,(T, L)$

**[V46]** $C \in \texttt{PEConditions} \quad \vdash \quad \mathsf{CondVerb}[\![\neg C]\!]\,(T, L) = \tilde{}\,\mathsf{CondVerb}[\![C]\!]\,(T, L)$

# 9 Miscelaneous

In this section we discuss two remaining issues. Firstly, an example parsing and translation process is provided. We show for a given ConQuer-92 expression its parse tree, and the way in which it can be stored (a syntax tree).

The final part of this section discusses nine phases in which ConQuer can be gradually be introduced in the InfoModeler product.

## 9.1 Example parsing process

Consider the example we have discussed earlier:

Person who earns Salary x AND ALSO works for a Company c
WHERE x ¿ THE AVERAGE Salary of a Person who works for c

The parse tree resulting from this expression is shown in figures 7, 8, and 9.

We could store a ConQuer-92 expression as an entire parse tree, but it is usual to store parse trees in a more condensed format in the form of a syntax tree. In appendix B were the ConQuer-92 grammar is provided, we also provide a record structure for each key syntactical category. If we now presume the following namings of types and mix-fix predicate names in the underlying domain:

| | |
|---|---|
| $\mathsf{TNm}(x) = \mathsf{Person}$ | $\mathsf{MFix}(F, [\mathsf{earns}], [p_1, p_2])$ |
| $\mathsf{TNm}(y) = \mathsf{Salary}$ | $\mathsf{MFix}(F, [\mathsf{of}], [p_2, p_1])$ |
| $\mathsf{TNm}(z) = \mathsf{Company}$ | $\mathsf{MFix}(G, [\mathsf{works\ for}], [q_1, q_2])$ |

where $\mathsf{Player}(p_1) = x$, $\mathsf{Player}(p_2) = y$, $\mathsf{Player}(q_1) = x$ and $\mathsf{Player}(q_2) = z$, then the

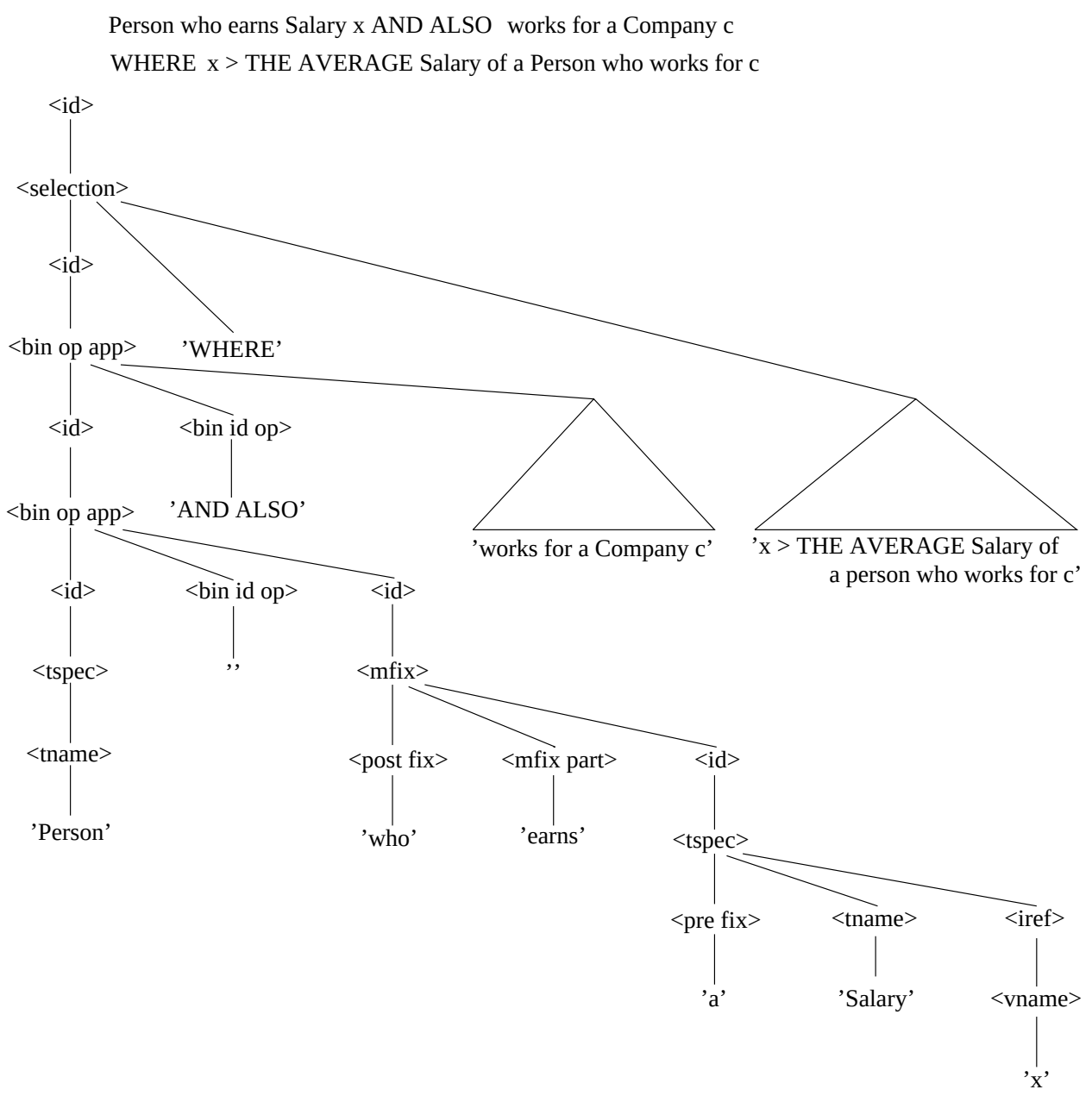

Figure 7:

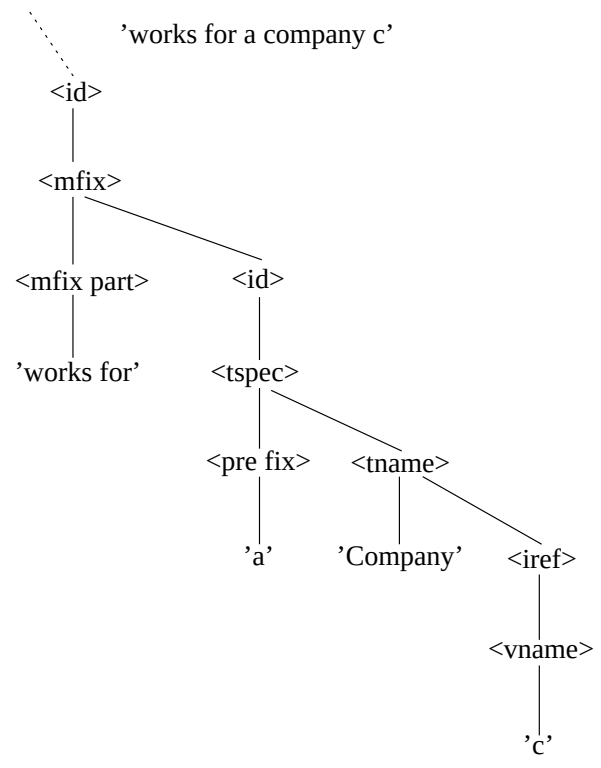

Figure 8:

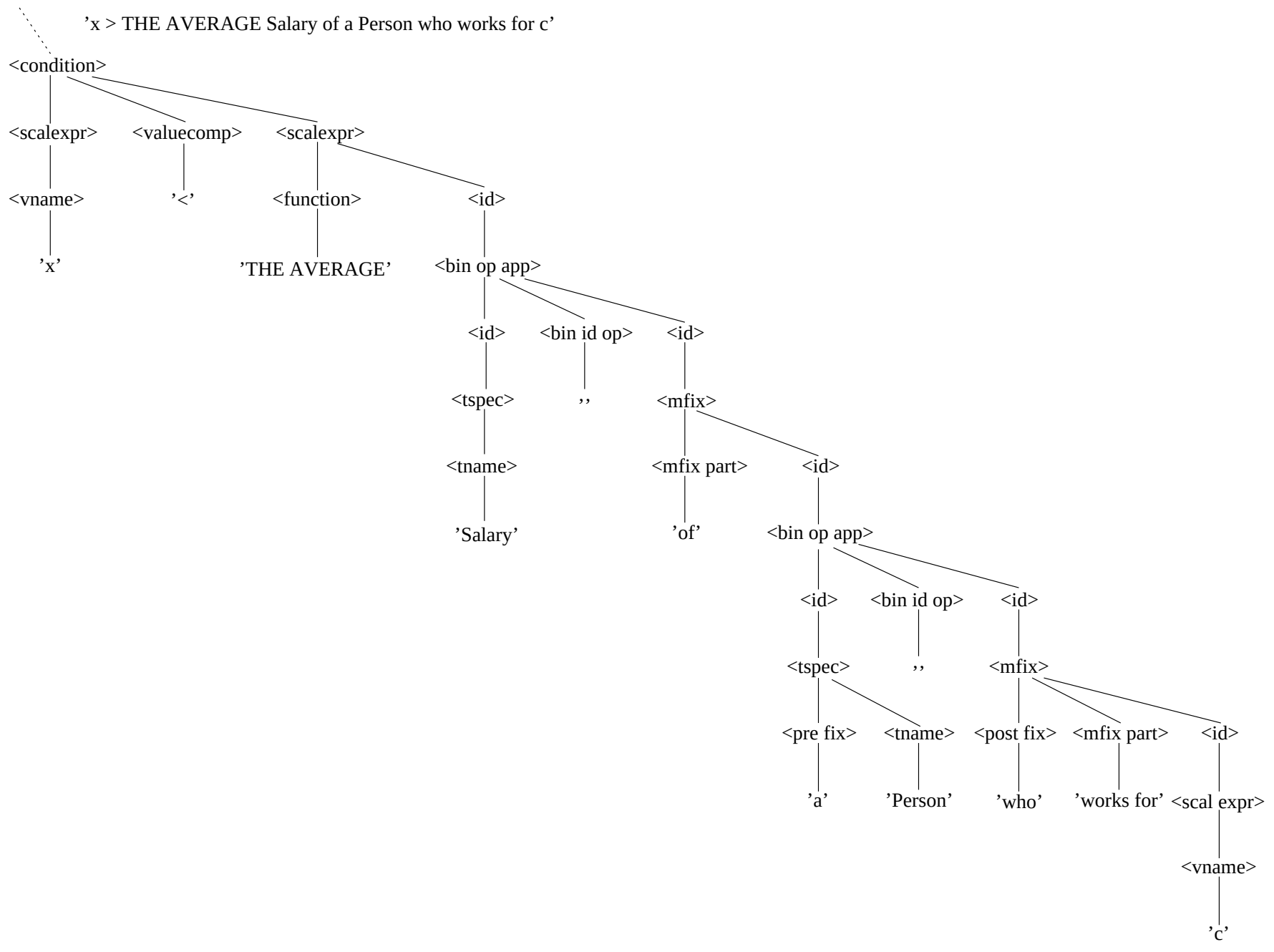
'x > THE AVERAGE Salary of a Person who works for c'
<condition>
<scalexpr>
<valuecomp>
<scalexpr>
<vname>
'<'
<function>
<id>
'x'
'THE AVERAGE'
<bin op app>
<id>
<bin id op>
<id>
<tspec>
' '
<mfix>
<tname>
<mfix part>
<id>
'Salary'
'of'
<bin op app>
<id>
<bin id op>
<id>
<tspec>
' '
<mfix>
<pre fix>
<tname>
<post fix>
<mfix part>
<id>
'a'
'Person'
'who'
'works for'
<scal expr>
<vname>
'c'

Figure 9:

above parse tree would lead to the following set of records:

00 SELECTION($\{\langle 01, 08\rangle\}$, NULL)
01 BINARY_OP_APPLIC("AND ALSO", $02, 06$)
02 BINARY_OP_APPLIC("", $03, 04$)
03 TYPE_SPEC(NULL, "Person", $x$, NULL, NULL)
04 MFIX("who", $p_1$, $\{\langle$"earns", $p_2$, $05\rangle\}$)
05 TYPE_SPEC("a", "Salary", $y$, "c", NULL)
06 MFIX(NULL, $q_1$, $\{\langle$"works for", $q_2$, $07\rangle\}$)
07 TYPE_SPEC("a", "Company", $z$, "c", NULL)
08 SCAL_EXPR_COMP($09, 10$, " $<$ ")
09 VAR_NAME("x", NULL)
10 COERCE_FUNCTION("THE AVERAGE", $\{11\}$)
11 BINARY_OP_APPLIC("", $12, 13$)
12 TYPE_SPEC(NULL, "Salary", $y$, NULL, NULL)
13 MFIX(NULL, $p_2$, $\{\langle$"of", $p_1$, $14\rangle\}$)
14 BINARY_OP_APPLIC("", $15, 16$)
15 TYPE_SPEC("a", "Person", $x$, NULL, NULL)
16 MFIX("who", $q_1$, $\{\langle$"works for", $q_2$, $17\rangle\}$)
17 VAR_NAME("c", NULL)

This shows how we can actually store ConQuer-92 expressions. The record structures have been set up in such a way that the ConQuer-92 expressions and their underlying path-expressions can be stored in one single format.

For example, using the definition of the semantics of ConQuer-92, we can now simply rewrite to a path-expression. This results in:

$$\begin{aligned}&\mathsf{Where}(\mathsf{Fr}(x \circ \langle p_1, p_2\rangle \circ y \circ \mathsf{x}) \cap \mathsf{Fr}(\langle q_1, q_2\rangle \circ z \circ \mathsf{c}),\\ &\quad \mathsf{x} > \mathsf{HdCoerce}(\mathsf{Avg}(y \circ \langle p_2, p_1\rangle \circ x \circ \langle q_1, q_2\rangle \circ \mathsf{c}))\end{aligned}$$

Operationally it means that we do not have to store the path-expressions separately.

## 9.2 Order of implementation

It is the aim of the InfoModeler team to release the first version of InfoModeler as soon as possible, and then gradually increase the support of the ConQuer-92 language. Below I have provided a list of phases. In the initial phases some non-terminals are listed that will not yet be implemented in that phase. It means that in the grammar for that phase, the mentioned non-terminal (and al other non-terminals solely used to define it) is removed. I have tried to evenly divide the extra work involved between each step as much as possible (based on my intuition), while also making sure the resulting languages are still sensible.

**Phase 1:** *Not yet implemented:* `shuffler`, `group accounting`, `sub expression`, `selection`, `confluence`, denotations in `type specification`s limited to simple constants and variables only, set comparisons in the `condition`s.

**Phase 2:** *Not yet implemented:* `group accounting`, `selection`, `confluence`, denotations in `type specification`s limited to simple constants and variables only, set comparisons in the `condition`s.

**Phase 3:** *Not yet implemented:* denotations in `type specification`s limited to simple constants and variables only, set comparisons in the `condition`s.

**Phase 4:** ConQuer-92 is complete.

**Phase 5:** Tighter integration with NUQL.

**Phase 6:** Extensions of ConQuer-92.

# 10 Conclusions

In this report we have given a complete formal definition of the ConQuer-92 conceptual query language. From a practical (= Asymetrix) point of view, this language is now ready to be implemented in the InfoAssistant QueryTool. The next step is therefore providing a compiler from ConQuer-92 to SQL-92.

For the longer term, more should be done to improve the quality of verbalisations and make a closer integration of ConQuer-92 and NUQL. One could imagine a non-ambiguous subset of ConQuer-92 which is used to verbalise path-expressions in a normalised form (basically what happens already), and a more liberal ConQuer-92 language from a user's point of view. So a user is allowed to enter more natural language like sentences which are then transformed by the system to a normalised format (while resolving ambiguities). Furthermore, when SQL-3 becomes readily available ConQuer-92 can be extended with forms of recursion.

From a theoretical (= Universities) point of view, the extensions to the original LISA-D provided by ConQuer-92 will have to be incorporated into next versions of the strictly academic LISA-D language. LISA-D is expected to see more extensions in the next years. Our list of wishes includes support for uncertainty and relevance making a closer integration with expert systems and information retrieval systems feasible. For example:

LIST The Authors of a Book ABOUT { 'Pollution', 'Rivers' }

Furthermore, incorporation of more linguistic principles into the LISA-D language may lead to better verbalisations.

# A Grammar of Path Expressions

In this appendix we define the concrete syntax of path expressions. This syntax will dictate the form and shape of the data structures used to store queries. Although queries are presented to the user and entered by users in terms of ConQuer-92 expressions, they will be stored as a combined form of path-expression parse trees and ConQuer-92 parse trees.

As we are now dealing with concrete syntax as opposed to abstract syntax as defined in section 6, the definitions given here do include parenthesis to allow for disambiguation of the expressions.

## A.1 Atoms

We distinguish a number of elementary parts of path-expression parse tree. The first ones are concerned with schema elements. For each $x \in \mathcal{TP}$, $p \in \mathcal{RO}$, and $a \in \mathsf{Attrs}$ we have:

`<type>      ::=` $x$

`<role>      ::=` $p$

`<attribute> ::=` $a$

Each constant is also an element of the path-expression language. So, if $c$ is a constant, then:

`<constant>  ::=` $c$

Any function symbol $f$ can be used in a path-expression, so we have:

`<function>  ::=` $f$

In the language of path-expressions a number of operators can be used. These operations are:

<binary path operator> ::= <path concatenator> | <product> | <path set comparitor>
| <value comparitor> | <set operator>

<path concatenator> ::= $'\circ'$

<path selector> ::= $'\mathsf{Where}'$

<product> ::= $'\times'$

<value comparitor> ::= $'<'$ | $'\leq'$ | $'='$ | $'\neq'$ | $'\geq'$ | $'>'$

<set operator> ::= $'\cup'$ | $'\cap'$ | $'-'$ | $'\cup\!\!\!\!f'$ | $'\cap\!\!\!\!f'$

| ′$\not f$′

<path set comparitor> ::= ′$\underline{\subseteq}$′ | ′$\underline{\supseteq}$′ | ′$\underline{\equiv}$′ | ′$\underline{\otimes}$′

<unary path operator> ::= ′Ds′ | ′Fr′ | ′TlCoerce′ | ′HdCoerce′

<group function> ::= ′GSum′ | ′GDsSum′ | ′GMin′ | ′GMin′ | ′GMax′ | ′GAvg′

<group counter> ::= ′GCount′ | ′GDsCount′

<logical connector> ::= ′$\vee$′ | ′$\underline{\vee}$′ | ′$\wedge$′ | ′$\Rightarrow$′

<path function> ::= ′Count′ | ′Sum′ | ′Min′ | ′Max′ | ′Avg′

<exists quantifier> ::= ′Some′

<path shuffler> ::= ′Path′

<path reverser> ::= ′$\leftarrow$′

<set comparitor> ::= ′$\subset$′ | ′$\subseteq$′ | ′$=$′ | ′$\neq$′ | ′$\#$′ | ′$\supseteq$′ | ′$\supset$′

<order operator> ::= ′$\Omega$′

## A.2 Linear Path Expressions

The first real class of path-expressions we introduce are the linear path expressions:

```
<linear path expression> ::=
   <type>                                                                  |
   <role>                                                                  |
   <linear path expression> <path reverser>                                |
   <linear path expression> <path concatenator> <linear path expression>
```

## A.3 Path Expressions

The path-expressions in general are introduced below. Note that `<path reverser>` is re-introduced for path-expressions in general. This is needed as a reversed linear path expression is still a linear path-expression, and hence needs to be part of definition of the linear path-expressions. Nonetheless, any path-expression in general can be reversed.

```
   <path expression> ::=
      <linear path expression>
      <instance denotation>
      <path expression> <path reverser>
      <unary path operator> <path expression>
      <path expression> <binary path operator> <path expres-
sion>                    |
      <path shuffler>  '(' <path expression> ',' <attribute list> ')'
      <function>       '(' <path expression list> ')'
      <path selector>  '(' <option sequence> [ ';' <option> ] ')'
      <group counter>  '(' <path expression> ',' <attribute set> ')'
      <group function> '(' <path expression> ',' <attribute set> ',' <at-
tribute> ')' |
      '<' <role> ',' <role to path list>, ',' <role> '>'
      <path confluence>
      <sub expression>
      '(' <path expression> ')'
      <scalar expression>
      <condition>
```

Note that any scalar expression or condition can simply be interpreted as a path-expression, and are therefore part of the path-expressions.

```
      <instance denotation> ::=
         <type> ':' <denotation>

         <denotation> ::=
            <path expression> |
            '!' <attribute>   |
            '(' <denotation list> ')'

            <denotation list> ::=
               <denotation> [ {',' <denotation>}... ]

      <attribute list> ::=
         <attribute> [ {',' <attribute>}... ]

      <attribute set> ::=
         '{' <attribute list> '}'

      <role to path list> ::=
         <role> ':' <path expression> [ {',' <role> ':' <path ex-
pression>}... ]
```

```
<path confluence> ::=
   '[' <confluence element list> ']'

   <confluence element list> ::=
      <confluence element> [ {',' <confluence element>}... ]

      <confluence element> ::=
         <attribute> ':' <path expression> ':' <attribute>

<sub expression> ::=
   '[' <path expression list> ']'

   <path expression list> ::=
      <path expression> [ {',' <path expression>}... ]

<option sequence> ::=
   <option> [ {';' <option>}... ]

   <option> ::=
      <path expression> ';' <condition>
```

## A.4 Scalar expression

The next syntactic category we introduce are the scalar expressions.

```
<scalar expression> ::=
   <constant>                                  |
   <path function> '(' <path expression> ')'   |
   <attribute>                                 |
   <attribute> '.' <role>                      |
   <function> '(' <scalar expression list> ')' |
   '(' <scalar expression> ')'

   <scalar expression list> ::=
      <scalar expression> [ {',' <scalar expression>}... ]
```

## A.5 Conditions

The conditions are defined by:

```
<condition> ::=
```

```
      <exists quantifier> <path expression>                        |
      <path expression>   <set comparator>    <path expres-
sion>   |
      <scalar expression> <value comparator>  <scalar expres-
sion> |
      <condition>         <logical connector> <condition>          |
      <negation>          <condition>                              |
      '(' <condition> ')'
```

### A.6 Queries

Finally, the set of path-expression queries are identified by the following definitions:

```
   <path expression query> ::=
      <path expression> |
      <order operator> '(' <path expression> ',' <order cri-
terion sequence> ')'

      <order criterion sequence> ::=
         <order criterion> [ {';' <order criterion>}... ]

      <order criterion> ::=
         <attribute name> ': Asc'  |
         <attribute name> ': Desc'
```

## B Grammar of ConQuer-92

In this appendix we define the concrete syntax of ConQuer-92. As we are now dealing with concrete syntax as opposed to abstract syntax as defined earlier, the definitions given here do include parenthesis to allow for disambiguation of the expressions. In this appendix we shall also discuss hierarchical record structures which can be used to store the ConQuer-92 expressions in a hybrid form between ConQuer-92 syntax trees and path-expression syntax trees.

### B.1 Atoms

We distinguish a number of elementary parts of ConQuer-92 parse trees. The first ones are concerned with schema elements. For each $x \in \mathsf{ran}(\mathsf{TNm})$, $p \in \mathsf{ran}(\mathsf{PNm})$, $r \in \mathsf{ran}(\mathsf{RNm})$, $v \in \mathsf{ran}(\mathsf{VNm})$, $a \in \mathsf{ran}(\mathsf{Pre})$, and $b \in \mathsf{ran}(\mathsf{Post})$ we have:

```
   <type name>         ::= x
```

```
<role name>         ::= p

<reverse role name> ::= q

<variable name>     ::= v

<prefix>            ::= a

<postfix>           ::= b
```

Each partial mix fix verbalisation leads to a mix fix part. So, if MFix$(r, A, R)$ and $a \in A$ we have:

```
<mixfix verb part>  ::= a
```

For each constant $c$ we have:

```
<constant>          ::= c
```

For each function name $f$ of an arithmetic function we have:

```
<function name>     ::= f
```

For each binary arithmetic operation (each of which is also a function name!) $o$ we have:

```
<bin operator>      ::= o
```

The predefined operators used in ConQuer-92 are divided in the following classes:

```
<concat>               ::= ϵ

<distinctor>           ::= 'DISTINCT'

<fronts selector>      ::= 'ONLY'

<function>             ::= 'THE COUNT OF' | 'THE SUM OF' | 'THE MINIMUM'
                         | 'THE MAXIMUM' | 'THE AVERAGE'

<group function>       ::= <function> | 'THE DISTINCT COUNT OF'
| 'THE DISTINCT SUM OF'

<logical connector>    ::= 'AND' | '&' | 'OR' | '|' | 'EXCLUSIVE OR'
                         | '||' | 'IMPLIES' | '=>' | 'IFF' | '<=>'

<path reverser>        ::= 'THE REVERSE OF'

<set comparitor>       ::= 'EQUALS' | '=' | 'DOES NOT EQUAL'
| '<>'
                         | 'IS DISJOINT FROM' | 'EXCLUDES' |
'IS A PROPER SUBSET OF'
                         | '<' | 'IS A SUBSET OF' | '<='
                         | 'IS A PROPER SUPERSET OF' | '>' |
'IS A SUPERSET OF'
```

```
                          | '>='

   <set operation>       ::= 'UNITED WITH' | 'INTERSECTED WITH'
| 'MINUS'
                          | 'WHICH ARE ALL IN' | 'THAT INCLUDES ALL'
| 'MATCHING ALL'
                          | 'MISSING' | 'WITH' | 'OR OTHERWISE'
|'AND ALSO'
                          | 'BUT NOT'

   <subtype selector>    ::= 'IS'

   <value comparitor>    ::= '=' | 'IS EQUAL TO' | '<>' | 'IS NOT EQUAL TO'
                          | '<' | 'IS LESS THAN' | '<=' | 'IS LESS THAN OR EQUAL TO'
                          | '>' | 'IS GREATER THAN' | '>='
                          | 'IS GREATER THAN OR EQUAL TO'
```

Note the definition of <concat>. In an information descriptor the concatenation operator is empty. For example, the Person and who works are information descriptors. When concatenating we get the Person who works, i.e. no operation is inserted.

For our convenience we introduce the following aggregate classes:

```
<binary information descriptor operator> ::=
   <set operation>       |
   <value comparitor>    |
   <subtype selector>    |
   <bin operator>        |
   <concat>

<unary information descriptor operator> ::=
   <path reverser>   |
   <fronts selector> |
   <distinctor>
```

## B.2 Information descriptors

The information descriptors are introduced as one large potpouri of options. Information descriptors will be stored in case specific records. So for each of the non-terminals in the definition of information descriptors we will define one specific record. When actually implementing these record structures in C++, one might want to introduce a general superclass for information descriptors with as subclasses the specific records for each options.

```
<information descriptor> ::=
   <type specification>             |
   <role reference>                 |
   <constant occurrence>            |
   <mixfix predicate verbalisation> |
   <unary operation application>    |
   <binary operation application>   |
   <shuffler>                       |
   <function or macro application>  |
   <selection>                      |
   <confluence>                     |
   <group accounting>               |
   <sub expression>                 |
   '(' <information descriptor> ')' |
   <scalar expression>              |
   <condition>
```

As we will introduce specialised records for each of the above syntactical classes, the data structure for information descriptors is a disjunction of the records of the options.

```
INF_DESCR ::=
  TYPE_SPEC | ROLE_REF | CONST | MFIX |
  UNARY_OP_APPLIC | BINARY_OP_APPLIC|
  SHUFFLE | FUNCTION | CONFLUENCE |
  GROUP_ACCT | SUB_EXPR | SCALAR_EXPR|
  CONDITION
```

The first kind of information descriptors deals with type specifications.

```
<type specification> ::=
   [ <prefix> ] <type name> [ <instance reference> ]

   <instance reference> ::=
      <variable name>  |
      ':' <denotation>

      <denotation> ::=
         <information descriptor> |
         '!' <variable name>      |
         '(' <denotation list> ')'

      <denotation list> ::=
         <denotation> [ { ',' <denotation> }... ]
```

For type specification we have the following record structure:

TYPE_SPEC ( PreFix: STRING OP, TypeName: STRING, Type: *TYPE,
InstRefVarName: STRING OP, InstRefDenot: DENOT OP )

In our notation we use OP to indicate an optional attribute, and REP for a repetitive attribute. The record refers to a record for denotations. We need to introduce this separate redord since denotations can be recursively defined. So we have:

DENOT ( InfDescr: INF_DESCR OP, VarName: STRING, Denot: DENOT OP REP )

We can refer to roles in two ways corresponding to role entries, and role exits. This leads to:

```
<role reference> ::=
   [ <postfix> ] <role name> |
   <reverse role name>
```

Role references are stored in the following record:

ROLE_REF ( PostFix: STRING OP, RoleKind: KIND, RoleName: STRING, Role: *ROLE )

where KIND = $\{\texttt{entry}, \texttt{exit}\}$.

Constant occurrences lead to the following non-terminal and record structure:

```
<constant occurrence> ::=
   <constant>
```

CONST ( Constant: STRING )

A mix-fix predicate verbalisation basically consist of an optional postfix for the preceding type (if any), followed by a non-empty sequence of mix-fix predicate parts and information descriptors. The syntax is therefore:

```
<mixfix predicate verbalisation> ::=
   [ <postfix> ] [ { <mixfix verb part> <information de-
scriptor> }... ]
      <mixfix verb part> <information descriptor>
```

A mix-fix predicate verbalisation can be stored in the following record structure:

MFIX ( PostFix: STRING OP, StartRole: ROLE*, ( MixFix: STRING, Role: *ROLE, InfDescr: INF_DESCR ) REP )

Both unary and binary operations for information descriptors lead to the following relative easy definitions of the syntax and records:

```
      <unary operation application> ::=
         <unary information descriptor operator> <informa-
tion descriptor>

      <binary operation application> ::=
         <information descriptor> <binary information descrip-
tor operator>
            <information descriptor>
```

UNARY_OP_APPLIC ( UnaryOp: UNARY_OP, InfDescr: INF_DESCR )

BINARY_OP_APPLIC ( BinaryOp: BINARY_OP, LeftInfDescr: INF_DESCR, RightInfDescr: INF_DESCR )

Note: UNARY_OP and BINARY_OP are enumeration types consisting of the `binary information descriptor operator` and `unary information descriptor operator` respectively.

The Harlem Shuffle operation involves a non empty list of variable names and an information descriptor; the information descriptor to be re-shuffled. So we have:

```
      <shuffler> ::=
         'THE PATH FROM' <variable name>
            [ 'VIA' <variable name list> ]
               'TO' <variable name> <information descriptor>

         <variable name list> ::=
            <variable name> [ {',' <variable name>}... ]
```

We store an occurrence of the shuffle operation as:

SHUFFLE ( VarName: STRING REP, InfDescr: INF_DESCR )

We would like to stress here again that macros for information descriptors are treated as ordinary functions. So as a syntactic category they are treated indistinctively. This leads to:

```
      <function or macro application> ::=
         <function name> '(' <information descriptor list> ')'
```

The record structure is now:

FUNCTION ( FunctionName: STRING, InfDesc: INF_DESCR REP )

Although selections can have different syntactical representations, they are basically a sequence of conditions and information descriptor together with an optional default (if none of the conditions yields true). When verbalising a selection, the preferences as discussed in section 8 should be used. <selection> ::=

'IF' <condition> 'THEN' <information descriptor> |
'IF' <condition> 'THEN' <information descriptor> 'ELSE'
<information descriptor> |
<information descriptor> 'WHERE' <condition> |
<alternatives sequence> 'OTHERWISE' <information descriptor> |
<alternatives sequence>

<alternatives sequence> ::=
<alternative> [ {';' <alternative>}... ]

<alternative> ::=
<information descriptor> 'IF' <condition>

The selection operation is stored in the following record:

SELECTION ( ( InfDescr: INF_DESCR, Cond: CONDITION ) REP, Default: INF_DESCR OP)

In a confluence operation a number of information descriptors is provided selecting aspects we are interested in starting out from an existing information descriptor. Variable names needed to be introduced to link the information descriptors yielding the aspects we are interested in to the existing information descriptor (VIA), and to provide names for the resulting columns (AS). The syntax of this construct is captured as:

<confluence> ::=
<confluence element list> 'EACH' <information descriptor>

<confluence element list> ::=
<confluence element> [ {',' <confluence element>}... ]

<confluence element> ::=
<information descriptor> [ 'AS' <variable name> ]
[ 'VIA' <variable name> ] This operation is stored in the following format:

CONFLUENCE ( BaseInfDescr: INF_DESCR, ( AspectInfDescr: INF_DESCR, AsVarName: STRING OP, ViaVarName STRING OP ) REP )

Group accounting requires evolves around three syntactic classes. Firstly, a variable selecting which column in the grouped information descriptor the group accounting function needs to be applied is needed. Secondly, the actual information descriptor that needs to be grouped is required. Thirdly, the variable names that need to be grouped are needed. These observations lead to the following definition of the group accounting operation:

```
<group accounting> ::=
   <group function> [ <variable> 'IN' ] <information descriptor>
      'GROUPED BY' <variable name list>
```

The record structure for this operation is:

GROUP_ACCT ( GrFuncName: GR_FUNC_OP, VarName: STRING OP, InfDescr: INF_DESCR,
( GroupingVarName: STRING REP ) )

Finally, sub expressions are simply a list of information descriptors enclosed by square brackets:

```
<sub expression> ::=
   '[' <information descriptor list> ']'

   <information descriptor list> ::=
      <information descriptor> [ {, <information descriptor>}... ]
```

We store this as:

SUB_EXPR ( InfDescr: STRING REP )

## B.3 Scalar expression

The scalar expressions of ConQuer-92 are build from five main classes We will not introduce special syntactic categories for these classes of scalar expressions. We will, however, discuss the record structures needed to store these classes separately.

```
<scalar expression> ::=
   <constant>                                                |
   <function> '(' <information descriptor> ')'               |
   <variable name> [ '.' <role name> ]                       |
   <function or macro name> '(' <scalar expression list> ')' |
   <scalar expression> <bin operator> <scalar expression>    |
   '(' <scalar expression> ')'
```

```
<scalar expression list> ::=
   <scalar expression> [ {',' <scalar expression>}... ]
```

The record structure for scalar expressions is a disjunction of five alternatives:

```
SCALAR_EXPR ::=
  SC_CONST | COERCE_FUNCTION | VAR_NAME |
  SC_FUNCTION | SC_BINARY_OP_APPLIC
```

The alternatives are provided as:

```
SC_CONST ( Constant: STRING )

COERCE_FUNCTION ( Function: STRING, InfDescr: INF_DESCR )

VAR_NAME ( VarName: STRING, RoleName: STRING OP )

SC_FUNCTION ( Function: STRING, ScalarExpr: SCALAR_EXP REP )

SC_BINARY_OP_APPLIC ( BinaryOp: SC_BIN_OP, LeftScalarExpr: SCALAR_EXP,
                      RightScalarExpr: SCALAR_EXP )
```

## B.4 Conditions

The conditions consist of six main classes. The syntax is provided as:

```
<condition> ::=
  'SOME' <information descriptor>                                  |
   <information descriptor> <set comparitor>    <infor-
mation descriptor> |
   <scalar expression>      <value comparitor>  <scalar ex-
pression>      |
   <condition>              <logical connector> <condi-
tion>            |
   <function or macro name> '(' <condition list> ')'               |
   <negated condition>                                             |
   '(' <condition> ')'

   <negated condition> ::=
      'NOT' <condition> |
      '~' <condition>
```

The record structure for conditions is identified by:

CONDITION :=
  CONDITIONER | INF_DESCR_COMP | SCAL_EXPR_COMP
  BIN_COND_OPER | COND_FUNCTION | NEGATION

The alternatives are defined by:

CONDITIONER ( InfDescr: INF_DESCR )

INF_DESCR_COMP ( LeftInfDescr: INF_DESCR, RightInfDescr: INF_DESCR, SetComp: SET_COMP )

SCAL_EXPR_COMP ( LeftScalExpr: SCALAR_EXPR, RightInfDescr: SCALAR_EXPR, ValueComp: VALUE_COMP )

BIN_COND_COMP ( LeftCond: CONDITION, RightCond: CONDITION, LogConn: LOGIC_CONN )

COND_FUNCTION ( Function: STRING, Condition: CONDITION REP )

NEGATION ( Negation: CONDITION )

Note: SET_COMP, VALUE_COMP and LOGIC_CONN are enumeration types containing all `set comparitor`, `value comparitor` and `logical connectors` respectively.

## B.5 List statement

The list statement is defined by the following definitions:

```
<list statement> ::=
   'LIST' [ <scalar expression list> 'FROM' ]
      <information descriptor> [ <order specification> ]

   <order specification> ::=
      'ORDERED' <order> |
      'ORDERED WITH' <order list>

      <order> ::=
         'ASCENDING' | 'DESCENDING'

      <order list> ::=
         <order item> [ {',' <order item>}... ]
```

```
<order item> ::=
   <variable name> <order> |
   'HEAD'  <order>         |
   'TAIL'  <order>
```

A list statement can be stored in the following record structure:

```
LIST ( ScalExpr: SCALAR_EXPR REP OP, InfDescr: INF_DESCR,
       ( VarName: STRING, Ord: ORDER ) REP OP)
```

Even though the list statement allows us to use ASCENDING and DESCENDING as an ordering specification without referring to a specific variable name, this absence implicitly refers to the head column of the table resulting from the information descriptor.

# References

[Bar84] H.P. Barendregt. *The Lambda Calculus: Its Syntax and Semantics*, volume 103 of *Studies in Logic and the Foundations of Mathematics*. North-Holland, Amsterdam, The Netherlands, Revised Edition, 1984.

[Hal95] T.A. Halpin. *Conceptual Schema and Relational Database Design*. Prentice-Hall, Sydney, Australia, 2nd edition, 1995.

[HP95] T.A. Halpin and H.A. Proper. Subtyping and Polymorphism in Object-Role Modelling. *Data & Knowledge Engineering*, 15:251–281, 1995.

[HPW93] A.H.M. ter Hofstede, H.A. Proper, and Th.P. van der Weide. Formal definition of a conceptual language for the description and manipulation of information models. *Information Systems*, 18(7):489–523, October 1993.

[HPW94] A.H.M. ter Hofstede, H.A. Proper, and Th.P. van der Weide. A Conceptual Language for the Description and Manipulation of Complex Information Models. In G. Gupta, editor, *Seventeenth Annual Computer Science Conference*, volume 16 of *Australian Computer Science Communications*, pages 157–167, Christchurch, New Zealand, January 1994. University of Canterbury.

[HPW97] A.H.M. ter Hofstede, H.A. Proper, and Th.P. van der Weide. Exploiting Fact Verbalisation in Conceptual Information Modelling. *Information Systems*, 22(6/7):349–385, September/November 1997.

[Lev79] A. Levy. *Basic Set Theory*. Springer-Verlag, Berlin, Germany, 1979.

[Lew85] A. Lew. *Computer Science: A Mathematical Introduction*. Prentice-Hall, Englewood Cliffs, New Jersey, 1985.

[Par90] H. Partsch. *Specification and Transformation of Programs - a Formal Approach to Software Development*. Springer-Verlag, Berlin, Germany, 1990.

[Pro94a] H.A. Proper. Generating significant examples for conceptual schema validation. Interactive Query Formulation using Query By Navigation 94-4, Asymetrix Research Laboratory, University of Queensland, Australia, 1994.

[Pro94b] H.A. Proper. Interactive query formulation using point to point queries. Confidential Asymetrix Research Report 94-1, Asymetrix Research Laboratory, University of Queensland, Australia, 1994.

[Sto77] J.E. Stoy. *Denotational Semantics: The Scott-Strachey Approach to Programming Language Semantics*. MIT Press, Cambridge, Massachusetts, 1977.